\documentclass[%
preprint,
superscriptaddress,
 amsmath,amssymb,
 aps,pre,longbibliography,
]{revtex4-1}

\usepackage{multirow}
\usepackage{braket}
\usepackage{relsize}
\usepackage{mathrsfs}
\usepackage{amsfonts}
\usepackage{rotating}
\usepackage{url}
\usepackage{color}
\usepackage{mathtools,amssymb}
\usepackage{verbatim}
\usepackage{overpic}
\usepackage{eufrak}
\usepackage{graphicx}
\usepackage{dcolumn}
\usepackage{bm}

\newcommand{\para}{\pi}

\newcommand{\cond}{\mathfrak{I}}

\begin{document}

\title{Absorbing Random Walks Interpolating Between Centrality Measures on Complex Networks}

\author{Aleks J. Gurfinkel}%
 \email{To whom correspondence should be addressed. E-mail: alexander.gurfinkel@gmail.com.} 
 \affiliation{%
 Department of Physics, Florida State University, Tallahassee, FL 32306-4350, USA\\
}%

\author{Per Arne Rikvold}%

\affiliation{%
 Department of Physics, Florida State University, Tallahassee, FL 32306-4350, USA\\
}%
\affiliation{%
PoreLab, NJORD Centre, Department of Physics, University of Oslo, P.O. Box 1048 Blindern, 0316 Oslo, Norway
}%



\date{\today}

\begin{abstract}
Centrality, which quantifies the “importance” of individual nodes,
is among the most essential concepts in modern network theory. As there are
many ways  in which a node can be important, many different centrality measures 
are in use. 
Here, we concentrate on versions of the common {\it betweenness\/} and {\it closeness\/}
centralities. The former measures  the fraction of paths between pairs
of nodes that go through a given node, while the latter measures
an average “inverse distance” between a particular node and all other nodes. Both centralities 
only consider shortest paths ({\it i.e.}, geodesics) between pairs of nodes. Here we
develop a method, based on absorbing Markov chains, that enables
us to continuously interpolate both of these centrality measures away from the geodesic limit and toward a limit where no restriction is placed on the length of the 
paths the walkers can explore. At this second limit, the interpolated betweenness and closeness 
centralities reduce, respectively, to the well-known {\it current betweenness\/} 
and {\it resistance closeness\/} ({\it information\/})  centralities. 
The method is tested numerically on four real networks, revealing complex changes in node 
centrality rankings with respect to the value of the interpolation parameter. Non-monotonic betweenness behaviors are found to characterize nodes that lie close to inter-community boundaries in the studied networks.

\end{abstract}

\maketitle


\section{Introduction\label{sec:intro}}

Modern network theory has evolved 
through a synthesis of mathematical graph theory \cite{EULER1736,EULER1956,bollobas1979graph}
with problems and methods from social sciences 
\cite{MORENO1934,opsahl2010node}
and physics \cite{KIRCHHOFF1847,KIRCHHOFF1958,ALBE02,CALD07,DORO08,newman2010networks,ESTR11}, 
into a powerful paradigm for analysis of 
complex systems consisting of interacting entities.  
Current interdisciplinary applications include modeling of transport in porous media and composites 
\cite{BLUN13,SAVA17}, reaction networks in chemical synthesis \cite{SHAM19}, 
food webs in ecology \cite{ROSS13}, 
transportation and distribution networks \cite{VERM14,xu2014architecture,GURF15}, 
economics and sociology \cite{KUTN19}, the Internet and the World Wide Web \cite{TIRO15},
and many more. 

The focus of the present paper is {\it centrality\/}, which, 
together with the adjacency relationship and the degree distribution, is one of the most basic and widely studied concepts in network theory. Centrality measures are prescriptions for 
quantitatively assigning importance to nodes in complex networks, and the power of the concept stems from the flexibility of characterizing importance in different ways. Applications of centrality 
measures are widespread, ranging from Internet searches (Google's PageRank algorithm \cite{page1999pagerank}) 
to determinations of  proteins  necessary for cell survival \cite{jeong2001lethality}.

In fact, centrality results are not just useful to
identify important nodes: with specific, quantitative information about individual nodes,
a centrality that reproduces this information can reveal
principles inherent in the structure of the network. Along these lines, 
in \cite{xu2014architecture} we investigated the architecture of the
Florida electric power grid. In particular, we found a strong correlation between  the known generating capacities of power plants and the values of a centrality based on 
Estrada's {\it communicability\/} concept 
\cite{estrada2008communicability,estrada2009communicability}.
This centrality has a parameter that controls the (graph) distance over which nodes can 
influence each other. 
Quantification of such correlations between node attributes and network structure requires 
centrality measures with a built-in tuning parameter. 

Here we develop an interpolation scheme connecting members of a set of 
four common centrality measures. 
Two of these measure how ``close" a node is to others ({\it closeness\/}), 
while the other two 
measure a node's tendency to lie on paths connecting pairs of other nodes ({\it betweenness\/}). 
Our main result is a new parametrization, based on absorbing random walks and their   
correspondence to electrical resistor networks 
\cite{Doyle06randomwalks,newman2005measure,brandes2005centrality,klein1993resistance}, 
that interpolates between the two 
betweenness measures and between the two closeness measures, respectively. 
In contrast to the parameter in the communicability centrality, our random-walk 
parameter tunes the centralities' preference for shortest paths (geodesics) versus longer paths. 
Using this parameter, harmonic  closeness centrality \cite{Dekker2005,Rochat2009,newman2010networks} (based on geodesics, as described below)
is deformed into 
a closeness centrality based on the Klein resistance distance \cite{brandes2005centrality,klein1993resistance}. Furthermore, 
using exactly the same parametrized absorbing random walk, the geodesic betweenness 
centrality \cite{Anthonisse1971,Freeman1977} 
is deformed into Newman's random-walk betweenness \cite{newman2005measure}. 
These four measures thus represent a natural class:  \em walker-flow centralities\em. 
Equations defining the four limiting centralities we consider are given in Sec.~\ref{sec:origcentralities}. 

As noted below, our work is not the first concerning interpolations between different 
centrality measures. 
However, to the best of our knowledge, it is the first to interpolate between the four walker-flow centralities both (a) {\it precisely} and (b) using the same parameter for both the closeness and betweenness continua. Furthermore, our interpolation is based on an easy-to-visualize random walk, allowing analysis at both the microscopic (individual walker step) and macroscopic (final centrality weighting) levels. Also, the transition probabilities of the random walk are closely related to the 
physics of lossy power transmission lines, allowing connections to the engineering literature, 
{\it e.g.\/}, \cite{steer2010microwave}. 

We  perform numerical tests of our analytical interpolation formalism on 
four real example networks of moderate size: a network of social interactions in a group of kangaroos \cite{kangadata,grant1973dominance}, 
Zachary's well-known karate club network \cite{zachary}, and two versions of the electrical power grid of the U.S. state of Florida \cite{dale}. 
Two of these networks are weighted and two are unweighted, and they range from strongly 
connected to relatively sparse. 
The numerical centrality rankings of the nodes are found to vary, 
not just between the closeness and betweenness 
measures, but also with the value of the interpolation parameter. The dependence of the 
individual centrality values on the 
interpolation parameter is generally smooth, but not necessarily monotonic. Plateaus and 
extrema in the intermediate parameter range  reveal the existence of nearly degenerate 
paths and the proximity of certain nodes to community boundaries in the network. 

In recent work, Bozzo and Franceschet \cite{bozzo}, and Tizghadam 
and Leon-Garcia \cite{tizghadam}, have found that random-walk betweenness can be written in 
terms of  resistance distances and the closely related pseudo-inverse of the graph Laplacian. Alamgir 
and von Luxburg \cite{alamgir2011phase} present an interpolation between graph distance and 
resistance distance, which is equivalent to an interpolation (different from ours) between closeness 
and resistance closeness.  Avrachenkov {\it et al\/}.\@ 
\cite{avrachenkov2013alpha,avrachenkov2015beta} present two  betweenness-like measures, where 
a parameter tunes the centrality's preference for geodesics. 
However, these do not precisely reduce to the betweenness. 
Kivim{\"a}ki {\it et al\/}.\@ \cite{kivimaki2016two,kivimaki2014developments} introduce the 
randomized-shortest-path (RSP) framework, which assigns Boltzmann weights to all paths in 
the network. Their inverse temperature parameter also tunes the preference for geodesics.  
In \cite{kivimaki2016two}, RSP is used to interpolate between graph distance and 
resistance distance, while in \cite{kivimaki2014developments}  it is used to interpolate 
between random-walk betweenness and a measure similar to standard betweenness centrality. 
In \cite{bavaud2012interpolating}, Bavaud and Guex accomplish a weighting equivalent to 
RSP through the minimization of a free-energy functional. Fran{\c{c}}oisse {\it et al\/}.\@ 
also reach similar results with a  different path-weighting scheme in \cite{franccoisse2017bag}. 
Estrada, Higham, and Hatano \cite{estrada2009communicability} calculate a version of 
betweenness centrality by assigning lower weights to longer paths.  

The remainder of this paper is organized as follows. In Sec.~\ref{sec:notation}, we introduce notations and conventions. In Sec.~\ref{sec:origcentralities}, we discuss well-known centrality measures. In Sec.~\ref{sec:param}, we develop two new parametrized centralities, based on a specific absorbing random walk, that interpolate between (a) closeness and resistance-closeness centralities and (b) betweenness and random-walk betweenness centralities. In Sec.~\ref{results_a}, we  report the behavior of these centralities on four example networks. In Sec.~\ref{sec:degen}, we use the numerical examples to explain how the presence of similarly-long paths leads to specific centrality behaviors. In Sec.~\ref{sec:nonmon}, we provide a model for the non-monotonicity encountered in some of the numerics. 
In Sec.~\ref{sec:conc}, we provide concluding comments and discuss plans for future study. 
Some technical details are addressed in four appendices.

\section{Walker-Flow Centralities } \label{sec:currflow}
\subsection{Notation and conventions}\label{sec:notation}

The most commonly studied centrality measures  can be found in, \em e.g.\/\em, Ch.7 of \cite{newman2010networks},  and many can be written in the  matrix  form: 
\begin{equation}
\label{centrality}
c_i=  \sum_{j} \mathbf{M}_{i j},
\end {equation}
where $c_i$ is the centrality of node $i$. The matrix element $\mathbf{M}_{i j}$ encodes the level of influence that node $j$ exerts on node $i$, and the final centrality is the sum of such influences.
Commonly, centrality measures include a normalization factor to ensure that  $\sum_i c_i=1$. In this paper, to better facilitate the inter-centrality comparisons in Sec.~\ref{sec:results}, we will only deal with {\it unnormalized} centrality measures. 

We concentrate on weighted, undirected networks, so that the adjacency matrix 
$\mathbf{A}$ is symmetric with, the weights of its nonzero elements denoting 
{\it affinities\/} between nodes. In other words: as  $\mathbf{A}_{i j}$ increases, 
nodes $i$ and $j$ become more closely associated
\cite{kivimaki2014developments}. Setting $\mathbf{M}=\mathbf{A}$ results in the weighted 
degree centrality: $c_i = A_i $, where $A_i=\sum_j \mathbf{A}_{i j}$ is the weighted degree 
of node $i$. 
(Unweighted networks will be treated as special cases of weighted networks, in which 
all nonzero $\mathbf{A}_{i j} = 1$.)

Many commonly studied centralities make use of the graph {\it distance\/} between nodes. 
The {\it weighted} graph distance, $d_{i j}$, is the length of the shortest edge path from 
node $i$ to $j$, where the length of a given edge $(a,b)$ is $d_{(a,b)}=(\mathbf{A}_{a b})^{-1}$. 
Defining edge lengths to be inverses of the nonzero weighted adjacency-matrix elements
is standard practice when using affinity weights in the adjacency matrix: 
distance is considered to be the inverse of affinity 
\cite{newman2001scientific,kivimaki2014developments,opsahl2010node}. 
Note that  $d_{a b}$ need not equal $d_{(a,b)}$, since the shortest weighted path from $a$ to $b$ does not necessarily follow the edge $(a,b)$.

\subsection{Centralities based on shortest paths, resistor networks, and random walks }
\label{sec:origcentralities}

\subsubsection{Closeness}

A prominent centrality based on graph distances is Bavelas's original closeness centrality  $c^\mathrm{CLO}_{i} =  (\sum_j d_{ij})^{-1}$ \cite{bavelas1950communication}, here modified for weighted graphs as in \cite{opsahl2010node}.
This closeness measure cannot be put into the form of Eq.~(\ref{centrality}), 
but in \cite{newman2010networks} Newman argues for the superiority of a modified 
closeness:
\noindent
\begin{equation}
\label{eq:HCL}
\mathbf{M}^\mathrm{HCL}_{ij}= d_{ij}^{-1} .
\end{equation}

\noindent
This measure is referred to as the {\it harmonic closeness centrality} (HCL) and is studied in 
\cite{latora2001efficient,Rochat2009,Dekker2005}. The present paper deals primarily with the 
harmonic closeness and measures that can be similarly described in the form 
of Eq.~(\ref{centrality}). Unless otherwise specified, any mention of ``closeness'' will 
refer to the harmonic type. However, the ideas presented here can be 
straightforwardly applied to the standard closeness as well. This is because we  
develop generalizations of the $d_{i j}$ themselves, so  either form, 
$(\sum_j d_{i j})^{-1}$ or $\sum_j d_{i j}^{-1}$, can be calculated.

In \cite{brandes2005centrality}, Brandes and Fleischer define a version of closeness centrality where the graph distance $d_{ij}$ is replaced by Klein and Randi\'c's resistance distance $R^\mathrm{eff}_{ij}$ \cite{klein1993resistance}. 
They prove the resulting  centrality measure equivalent to the \em information centrality \em \cite{stephenson1989rethinking}, whose original definition made no reference to resistor networks. This centrality  is given by $c^\mathrm{INF}_i =1/\sum_jR^\mathrm{eff}_{ij} = 1/\sum_{j}I^{-1}_{ij}$. 
Here, $I_{ij}$ is the current flowing from $i$ to $j$ when a unit potential difference is introduced between those nodes.
(The last equality is true by the definition of resistance distance; $R^\mathrm{eff}_{ij}$ is just the inverse of the current flow  from $i$ to $j$.) 

Alternatively,  starting from the harmonic closeness centrality (HCL), rather than the original closeness,
enables us to work with the centrality matrix $\mathbf{M}$. Following the same substitutions as in \cite{brandes2005centrality}, we obtain

\noindent
\begin{equation}\label{eq:RCC}
\mathbf{M}^\mathrm{RCC}_{ij}=\frac{1}{R^\mathrm{eff}_{ij}}= I_{ij}.
\end{equation}

\noindent
Due to the similarity with HCL, the centrality of Eq.~(\ref{eq:RCC}) can be termed the 
\em resistance-closeness centrality \em (RCC) (or \em harmonic information centrality\/\em). 
Eqs.~(\ref{eq:HCL}-\ref{eq:RCC}) have the same structure; the difference is that  HCL only 
considers geodesics (as captured by the weighted graph distance $d$), while RCC  considers currents  
(encoded by $R^\mathrm{eff}$)  that explore the entire network. In Sec.~\ref{sec:param}, 
we derive a parameter that can interpolate between these two limits. 
This derivation relies  on a useful isomorphism between random walks on networks and 
currents in corresponding resistor networks. Reference \cite{Doyle06randomwalks} describes this isomorphism and provides an equation for $R^\mathrm{eff}$ in terms of random walks.

\subsubsection{Betweenness}
\label{sec:bet}

One of the most common centrality measures, betweenness \cite{newman2010networks,Anthonisse1971,Freeman1977}, is defined as 
\begin{equation}
\label{eq:bet}
\mathbf{M}^\mathrm{BET}_{ij}=\sum_s n_{s ij} /g_{s j}.
\end{equation}
 Here, $g_{sj}$ counts the number of \em shortest \em  paths between node $s$ and node $j$, while $n_{sij}$ counts the number of such paths that pass through $i$. 

As an example of the random-walk/resistor-network isomorphism, in \cite{newman2005measure}, Newman re-frames his random-walk centrality in terms of electrical 
currents $I$ flowing along network edges, each of which has  an equal resistance. 
This {\it current-betweenness centrality} (CBT) can be written similarly to the standard 
betweenness of Eq.~(\ref{eq:bet}) as
\begin{equation}\label{eq:CBT}
\mathbf{M}^\mathrm{CBT}_{ij}=\sum_s I_{s i  j} /I_{s j}.
\end{equation}
Here, $I_{s i  j}$ denotes the current passing through node $i$ when a current $I_{sj}$ is passed into the network at $s$ and flows out of the network at $j$. The notation here is chosen to reveal the similarity to Eq.~(\ref{eq:bet}). (It is necessary to separately denote the current  flowing on an edge from $i$ to $j$, should such an edge exist. We  refer to this edge current as $I_{(i,j)}$, and in general $I_{i j}\neq I_{(i,j)}$.)

The current flow $I_{(a,b)}$ along any network edge $(a,b)$ is determined by Kirchhoff's laws, as applied when edge conductances $C_{k l}$ are  taken to equal $\mathbf{A}_{k l}$ (where we do not allow for self-edges). This  condition  is proven in \cite{Doyle06randomwalks} to be mathematically equivalent to a random walker's transition probabilities being proportional to $\mathbf{A}_{k l}$. Such a process gives the same result as the current-betweenness centrality, and is also described by Eq.~(\ref{eq:CBT}), provided that (a)  $I_{sj}$ is taken to be the number of random walks starting on node $s$ and eventually absorbed at $j$, and (b) $I_{sij}$ is the sum of  the walker currents that flow \em into \em $i$ during this process:  $I_{sij}=\sum_{a:I_{(a,i)}>0}  I_{(a,i)}$. Therefore, current-betweenness centrality can be described by the same random-walk dynamics that leads to the resistance-closeness centrality. 

In Eqs.~(\ref{eq:bet}-\ref{eq:CBT}), the analogy between current-betweenness and standard betweenness centrality is particularly clear. As with the two closeness centralities in Eqs.~(\ref{eq:HCL}-\ref{eq:RCC}), the difference is between a centrality (BET) based only on 
geodesics, as denoted by $n$ and $g$, and a centrality (CBT)  based on currents (or random walks) $I$ that explore  the entire network, not just the shortest path. 
Interpolation between Eqs.~(\ref{eq:bet}-\ref{eq:CBT}) will be achieved with the same parameter 
as between Eqs.~(\ref{eq:HCL}-\ref{eq:RCC}).

\subsection{parametrization for walker-flow centralities \label{sec:param}}
\subsubsection{Random walks that prefer short paths: conditional current }
\label{sec:shortpaths}
We have noted above that the current betweenness and resistance closeness centralities both can be described in terms of walker flows. In this paper, we introduce a method to interpolate continuously between the current betweenness and betweenness on the one hand, and the resistance closeness and the closeness on the other hand. As such, all the centralities in Eqs.~(\ref{eq:HCL}-\ref{eq:CBT}), and their related measures, such as the information centrality, can be viewed as belonging to the same class: {\it walker-flow centralities}.

Given that the discussed walker-flow  centralities can be equivalently described in terms of either resistor networks or random walks, one expects the interpolation parameter to take the form of either (a)  resistances or (b) walker transition rates. These two interpretations are equivalent for our purposes, and we will employ both descriptions interchangeably, since they lead to simpler arguments in different contexts.

We introduce a parameter $\para_D$ that controls the  probability of a walker's death  before reaching the target node $j$. (The detailed definition of $\para_D$ is described in Sec.~\ref{sec:rootgr} below.) This parameter has an interpretation in terms of network scale: the higher $\para_D$, the less of the network will be explored by the walker. This {\it random walk } parameter naturally connects with the resistance closeness  and the current betweenness due to the electrical isomorphism: when  the parameter is  zero we recover the standard random walk, which can be used to calculate those centralities. 

\begin{figure*}
\includegraphics[scale=0.7, trim={0 0cm 0.2cm 0cm},clip]{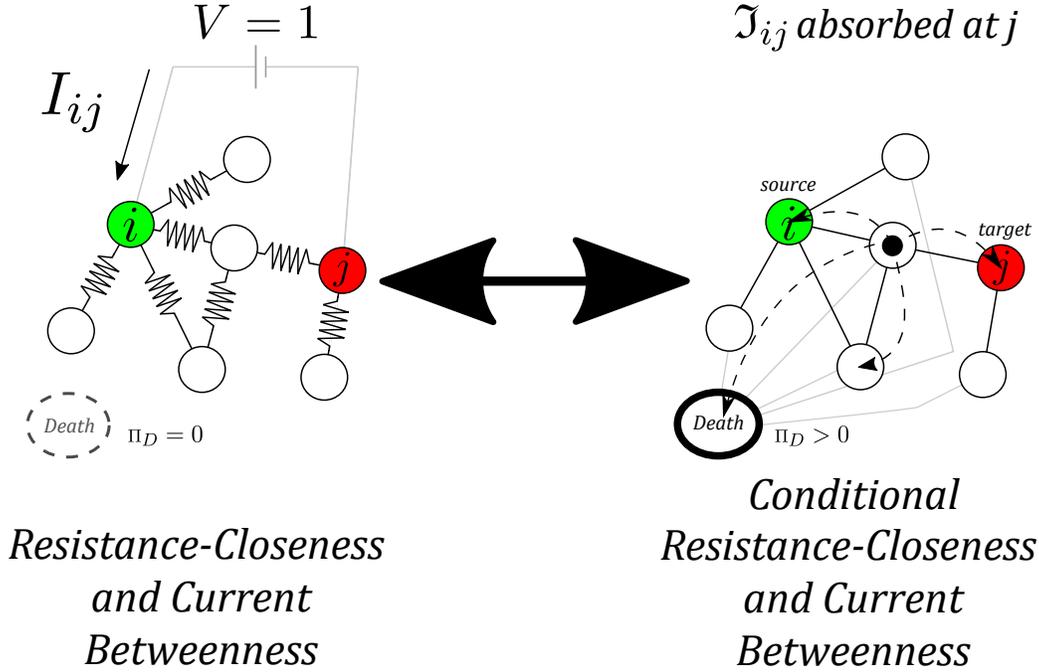}
\caption{\label{fig:gcc} \em The relationship between current-based centralities and centralities using conditional current $\cond$\em.  Even though resistance closeness and betweenness are different centralities, they can be described using the same electrical network (left). Likewise, the conditional resistance closeness and conditional current betweenness are described using the same random-walk process (right). When $\para_D=0$, the conditional current and physical current are identical: $\cond = I$, and the left and right sides of the figure become equivalent, illustrating the isomorphism between random walks and resistor networks. On the left, the network currents are  found according to Kirchhoff's Laws. On the right, $\para_D>0$, so $\cond \neq I$, and network currents are determined by counting edge traversals of random walkers (illustrated by the black disk) that do \em not \em land on the ``death'' node. The walker's transition to the death node is controlled by the parameter $\para_D$, while the transition probabilities to the network nodes are inversely proportional to the weighted degree of the node currently being occupied by the walker. The walker begins on the ``source'' node $i$ and ends on the ``target'' node $j$.}
\end{figure*}

The shortest possible walk from $i$ to $j$ covers the weighted graph distance $d_{i j}$, so increasing $\para_D$ tends to connect the walk with the centralities based on geodesics: betweenness and closeness.  Walkers operating in the high $\para_D$ regime are likely to die before making the full journey to a distant node, but the few walkers that survive will be overwhelmingly likely to have followed geodesics. Therefore, we  restrict our attention to 
walkers that 
   \em do not die\em, leading to a ``conditional current'' $\cond$ of walkers. In Sec.~\ref{sec:CalcCond}, we provide an explicit formula for $\cond$ in terms of absorbing Markov matrices. The conditional current $\cond$, once substituted for the physical current $I$ in Eq.~(\ref{eq:CBT}),  provides a parametrized version of current-betweenness centrality, the \em conditional current-betweenness centrality\/\em. In Sec.~\ref{sec:reff} we  provide a calculation, also based on $\cond$, that parametrizes the resistance-closeness centrality, resulting in the \em conditional resistance-closeness centrality\em.  With the restriction to conditional current, the parametrizations can reduce to the centralities discussed in the previous section at appropriate values of $\para_D$: Conditional current-betweenness centrality reduces to current-betweenness centrality (as $\para_D\to 0$) and betweenness centrality (as $\para_D\to \infty$), while conditional resistance-closeness centrality reduces to resistance-closeness centrality and harmonic closeness centrality in the same limits, respectively.  

Figure \ref{fig:gcc} illustrates the correspondence between resistor-network centralities and the parametrization controlled by $\para_D$ and $\cond$. On the left side 
of the figure, resistance closeness and current betweenness are represented by the same diagram because both of those centralities can be calculated from current flows in the same resistor network. The parametrized forms of both these centralities are represented by the diagram on the right side of the figure.  

The new centralities based on conditional current  have a clear interpretation at intermediate parameter values. While as $\para_D\to\infty$ the walker will only successfully follow geodesic paths, at smaller non-zero values of $\para_D$ the walker will be restricted to short paths that are nearly geodesic. Finally, as $\para_D$ goes to zero, the length of the walker's paths will be completely unrestricted. Thus, $\para_D$ can be said to tune the centralities' preference for geodesic paths.

   \subsubsection{ Specification of the interpolation parameter $\para_D$}
   \label{sec:rootgr}
   
      \begin{figure*}
\includegraphics[scale=0.7, trim={0 0cm 0cm 0cm},clip]{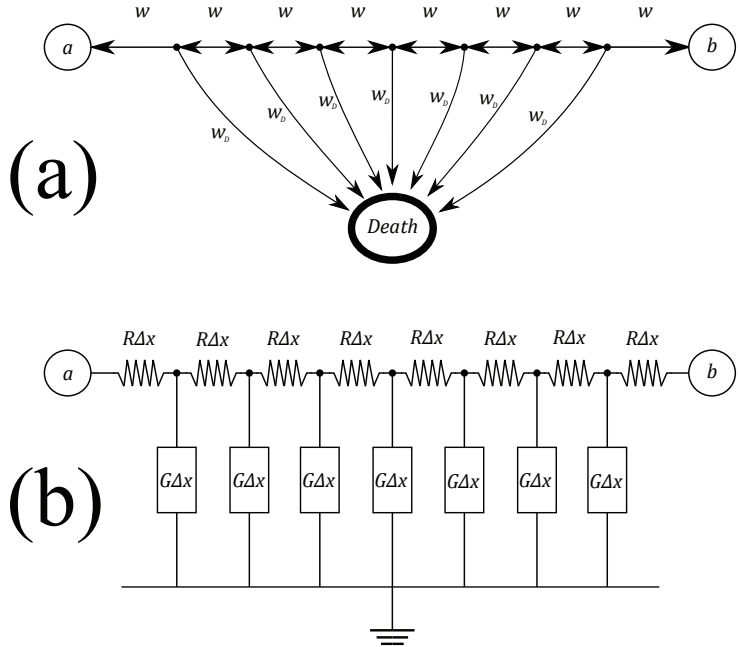}
\caption{\label{fig:longline} \em Weighted network edge from node $a$ to $b$ in (a) random-walk and (b) resistor-network descriptions\/\em. Case (b) is equivalent to a discrete approximation of a  transmission line with constant resistance $R$ and ground conductance $G$ per unit length.  
In this figure, the number of intermediary edges, $n_\mathrm{edge}$, is 8.  
$\Delta x = d_{(a,b)}/n_\mathrm{edge}$.}
\end{figure*}
   
   Even though our two conditional current centralities are different measures, they are both based on the same random-walk dynamics controlled by the same parameter $\para_D$. This random walk must be capable of reproducing the harmonic closeness centrality, which places an important requirement on its transition probabilities. In the case of weighted networks, the harmonic closeness centrality [see Eq.~(\ref{eq:HCL})] relies on the entries of the weighted adjacency matrix 
$\mathbf{A}$. Recall that, in closeness metrics, the inverse $(\mathbf{A}_{a b})^{-1}$ of an edge weight is generally associated with the length $d_{(a,b)}$ of that edge \cite{ opsahl2010node}. Therefore, the random-walk transition probabilities must be sensitive to the weights of edges as well. 

   
To properly incorporate edge lengths (equivalently, inverse weights) into the absorbing random walk, we break each edge into a  number of intermediary edges, connected by fictitious intermediary nodes, and then   take the continuum limit. For example, Fig.~\ref{fig:longline}(a) shows the edge $(a,b)$ broken into $n_\mathrm{edge}=8$ intermediary edges, connected by $n_\mathrm{edge}-1$ fictitious nodes. The weight of the entire edge is given by the adjacency matrix: $w_{(a,b)}=\mathbf{A}_{a b}=d_{(a,b)}^{-1}$. Therefore, an intermediary edge has weight 
\begin{equation}
   \label{eq:intermed}
   w= (d_{(a,b)}/n_\mathrm{edge})^{-1}=w_{(a, b)} * n_\mathrm{edge}. 
\end{equation}

In addition to its two connections along the original edge $(a,b)$, each intermediary node has an edge (weight $w_D$) to the absorbing ``death'' node. The behavior of $w_D$ as $n_\mathrm{edge} \to \infty$ is taken from an analogy with the lossy transmission line model from electrical power engineering \cite{steer2010microwave}. Fig.~\ref{fig:longline}(b) depicts the lossy transmission line model with ground conductance per unit length $G$, line resistance per unit length $R$, and line inductance and ground capacitance set to zero. The correspondence between electrical networks and random walks \cite{Doyle06randomwalks} implies that  edge weights are proportional to conductances in the equivalent electrical network. If we take the proportionality factor to be one \cite{RExplained}, $w_D=G\Delta x$ and $w=(R \Delta x)^{-1}$, where $\Delta x = d_{(a,b)}/n_\mathrm{edge}$.  This means that edge weights (and hence, the adjacency matrix) have units of conductance, while edge lengths have units of resistance.
   
  With intermediary edge weights in terms of $G$, in the continuum limit, we obtain random-walk transition probabilities $p_\nu$ over the edges $\nu$ incident on a given node $a$ (see Appendix \ref{app:cont}):
  \begin{equation}
  \label{eq:tranprob}
p_\nu(a)=\frac{[\sinh(\sqrt{GR}d_\nu)]^{-1}}{N-1-k_a + \sum_\mu [\tanh(\sqrt{GR}d_\mu)]^{-1}}.
  \end{equation}
 Here, the index $\mu$ runs over all edges incident on $a$,  $k_a$ is the \em unweighted \em degree of $a$, $d_\nu$ is the length  of edge $\nu$, and $N$ is the number of nodes in the network. The probability of the walker on $a$ dying before successfully crossing an edge is therefore
  \begin{equation}
  \label{eq:dedprob}
p_D(a)=1- \frac{\sum_\nu[\sinh(\sqrt{GR}d_\nu)]^{-1}}{N-1-k_a + \sum_\mu [\tanh(\sqrt{GR}d_\mu)]^{-1}}.
  \end{equation}

Eqs. (\ref{eq:tranprob}) and (\ref{eq:dedprob}) are parametrized by the combination $\sqrt{GR}$, which has units of inverse length. In the theory of power transmission, $\sqrt{GR}$ is the inverse attenuation length of voltage signals along a lossy power line with negligible inductance and capacitance \cite{steer2010microwave}. For our purposes, $\sqrt{GR}$ is the parameter  that controls the probability $p_D(a)$ of walker death at  node $a$. In the next section, we show that increasing this parameter accomplishes the interpolations from resistance closeness to closeness and from current betweenness to betweenness. Thus, the centrality interpolation parameter is
\begin{equation}
\para_D=\sqrt{GR}.
\end{equation}
Eqs. (\ref{eq:tranprob}) and (\ref{eq:dedprob}) give sensible results for values of $\para_D$ between $0$ and $\infty$. Table \ref{paraDlims} summarizes the limiting values. In the limit $\para_D \to 0$, the  probabilities correctly reduce to those of a standard random walk. 
   
      \begin{table}[t]
      \caption{Walker transition probabilities for different values of $\para_D$ in finite networks. 
        }
      \begin{center}

\label{paraDlims}
    \begin{tabular}{c|c|c|c} 

&$\lim{\para_D \to \infty}$ & $\para_D>0$ & $\lim{\para_D \to 0}$ \\
\hline
&&Eq.~(\ref{eq:tranprob}) & \;(standard random walk)
\\   
$p_\nu(a)$ & 0 &$\frac{[\sinh(\para_D d_\nu)]^{-1}}{ \vphantom{\tilde{E}}N-1-k_a + \sum_\mu [\tanh(\para_D d_\mu)]^{-1}}$ & $\frac{(d_\nu)^{-1}}{\vphantom{\tilde{E}} \sum_\mu (d_\mu)^{-1}}=\frac{w_\nu}{\vphantom{\tilde{E}} \sum_\mu w_\mu}$\\
\hline
&&Eq.~(\ref{eq:dedprob})&
\\
$p_D(a)$ & 1 & $1-\frac{\sum_\nu[\sinh(\para_D d_\nu)]^{-1}}{\vphantom{\tilde{E}} N-1-k_a + \sum_\mu [\tanh(\para_D d_\mu)]^{-1}}$ & 0

    \end{tabular}
  \end{center}

\end{table}

With Eqs.~(\ref{eq:tranprob}) and (\ref{eq:dedprob}) in place, we have completely described a model of the original network as a lossy resistor network. We have focused on the random-walk description of the resistor network to aid in the calculations of the following section. However, we stress that the description in terms of the electrical quantities $R$ and $G$ is equally valid. In fact, such a lossy resistor network would be physically realizable.

\subsubsection{ $\cond$ at extreme values of $\para_D$}

The entries of Table~\ref{paraDlims} are  transition probabilities for a {\it single} walker step. They do not necessarily reflect what will happen in the {\it conditional} random walk,  where we do not count walkers that die before reaching the target. In Sec.~\ref{sec:CalcCond}, we  derive a formula for calculating $\cond$ based on a surviving walker's complete journey, not just a single step. However, we can already understand the behavior of $\cond$ at the limits of small and large $\para_D$. 

\begin{figure}
\hspace{-2em} 
\includegraphics[scale=.85, trim={0cm .3cm 0cm .0cm},clip]{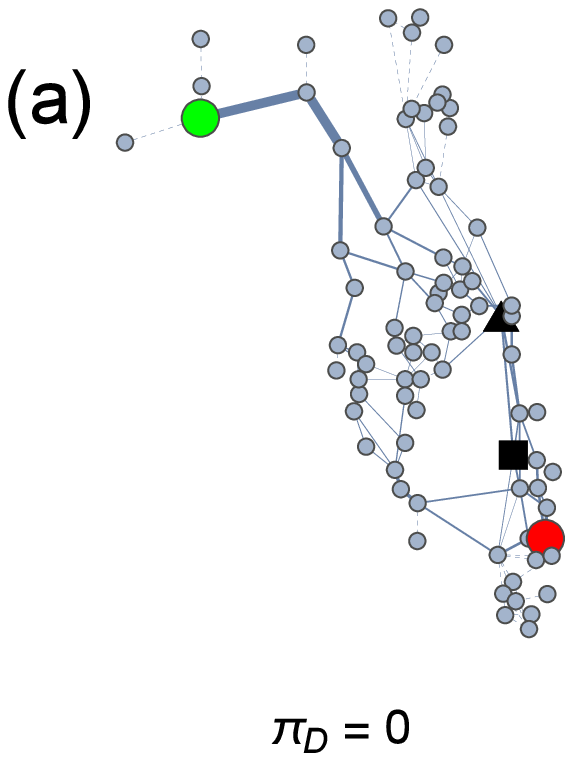}
\hspace{-8em} 
\includegraphics[scale=.85, trim={0cm .3cm 0cm .0cm},clip]{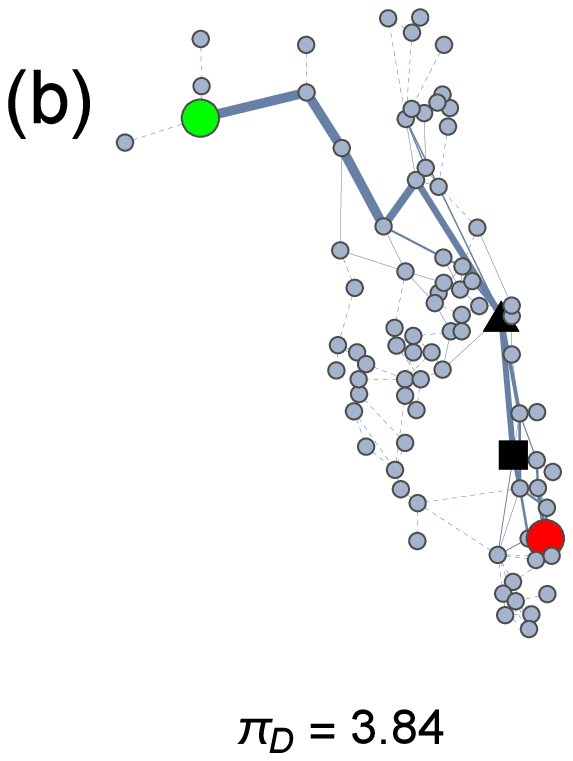}
\hspace{-8em} 
\includegraphics[scale=.85, trim={0cm .3cm .4cm .0cm},clip]{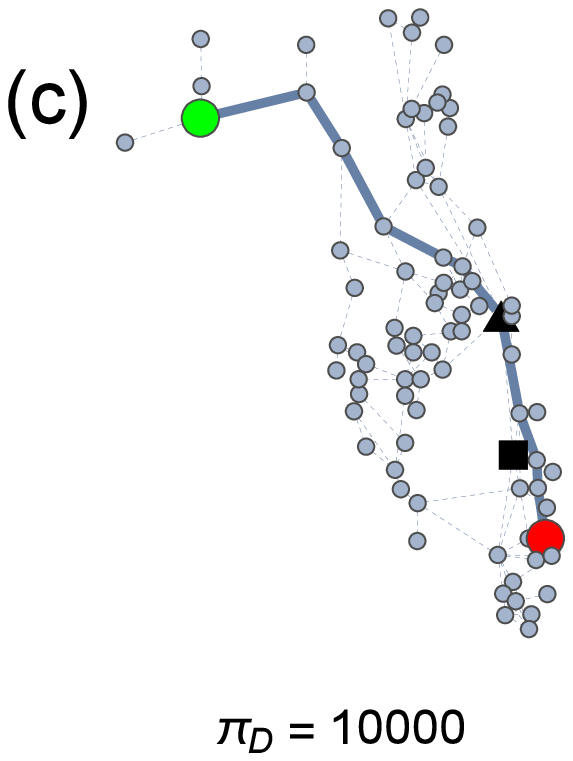}

\caption{\label{fig:floridacond} \em Conditional current flows $\cond$ at extreme and intermediate values of $\para_D$ illustrated on the weighted Florida power-grid network (FLG) \em \cite{xu2014architecture}. One unit of conditional current $\cond$ originates on the source node (green) and is absorbed at the target (red). Line thickness indicates conditional current magnitude, and edges with negligible conditional current ($< 0.001$ units) are shown as dashed lines. The nodes marked with a square and a triangle are referred to in Sec.~\ref{sec:networkresults} and Appendix \ref{sec:nonmon data}. (a) At $\para_D=0$,  $\cond$ is equal to physical current flow in a resistor network with the same topology as FLG: the current fans out  over the network. (b) At $\para_D=3.84$, $\cond$ begins to focus on shorter weighted paths. (c) At $\para_D=10000$, $\cond$ is confined to the shortest weighted path from the source to the target. See the detailed discussion of this network in Sec.~\ref{sec:networkresults}.} 

\end{figure}

Figures \ref{fig:floridacond}(a) and \ref{fig:floridacond}(c)  illustrate low-$\para_D$ and high-$\para_D$   conditional current $\cond$ on a weighted network representing the electrical power grid of the U.S. state of Florida \cite{dale,xu2014architecture}. Similarly, the first and last subfigure of Fig.~\ref{fig:condcurrent} illustrates extremal $\cond$, as applied to a weighted network of social interactions in a group of kangaroos \cite{kangadata,grant1973dominance}. These networks will be explained and studied in further detail in Sec.~\ref{sec:results}; here they are just used to demonstrate the general behavior of conditional current. In the case of low $\para_D$, $\cond$ spreads out exactly like physical current in a resistor network. In the case of high $\para_D$, $\cond$ follows only weighted geodesic paths.

     Employing our  parametrization, Eqs.~(\ref{eq:RCC}) and (\ref{eq:CBT}) undergo the transformation $I\longrightarrow \cond(\para_D)$ as $\para_D$ is increased from zero. This  naturally interpolates between the current-flow measure and the corresponding geodesic measure: between the  resistance-closeness centrality [Eq.~({\ref{eq:RCC}})] (as $\para_D\to 0$)  and the  closeness [Eq.~({\ref{eq:HCL}})] (as $\para_D\to \infty$), and likewise between the current betweenness (or random-walk betweenness)  (as $\para_D\to 0$) [Eq.~({\ref{eq:CBT}})] and the original betweenness [Eq.~({\ref{eq:bet}})] (as $\para_D\to \infty$). To get a sense for why this is so, take $\lim_{\para_D \to 0}$. In this case, when the ``death probability'' is zero, the walk reduces to the standard random walk on a weighted network. Such walks correspond to  current flows as described in Sec.~\ref{sec:origcentralities}: $\lim_{\para_D \to 0}\cond(\para_D)=I$, which reproduces the centralities in Eqs.~(\ref{eq:RCC}) and (\ref{eq:CBT}).
    
    In the other direction, take a random walk with an extremely high $\para_D$ and consider the effects on the the flow of walkers, $\lim_{\para_D \to \infty}\cond(\para_D)$, from source $s$ to target $j$. Almost no such walks succeed in escaping from $s$ to $j$ before succumbing to the death probability $p_D$ from Table \ref{paraDlims}.  Of the walkers that make this escape, the overwhelming majority will have  taken walks along geodesics  because even a single unnecessary step will incur a steep penalty from $\para_D$. Furthermore, Eq.~(\ref{eq:semidegen}) in Appendix \ref{app:degen} shows that every geodesic path will get the same conditional current. Thus,  $\lim_{\para_D \to \infty}\cond_{s i j}$ becomes proportional  to $n_{s i j}$, and the parametrized betweenness reduces to the standard betweenness in Eq.~(\ref{eq:bet}). (The more technical proof of the reduction of the parametrized resistance closeness to the harmonic closeness in Eq.~(\ref{eq:HCL}) is presented at the end of Sec.~\ref{sec:reff}.)

The reduction of the conditional resistance closeness to the closeness and the conditional current betweenness to the betweenness  are confirmed numerically for several example networks, as shown in Sec.~\ref{sec:results}.  We note that the reduction would not be possible without splitting weighted edges into intermediate edges and nodes with connections to the death node. Consider the alternative scheme where we capture edge weights in the random walk by dramatically  increasing the transition probability for highly weighted edges and  decreasing it for weakly weighted edges. In this case, the walker current would not be able to flow along geodesics that contain any long lines. For example, the geodesic in Fig.~\ref{fig:floridacond}(c) contains a very long line incident on the node marked with a triangle. Even though this line is one of the longest (lowest weight) in the network, if the walker were to bypass it the conditional current would no longer flow along a geodesic, and the  reduction  to the closeness centrality would be impossible. 

\subsubsection{$\cond$ at intermediate values of $\para_D$: tuning the preference for geodesic paths}
\label{sec:tuning}

\begin{figure*}
\includegraphics[scale=0.65, trim={3cm 4cm 2cm 4cm},clip]{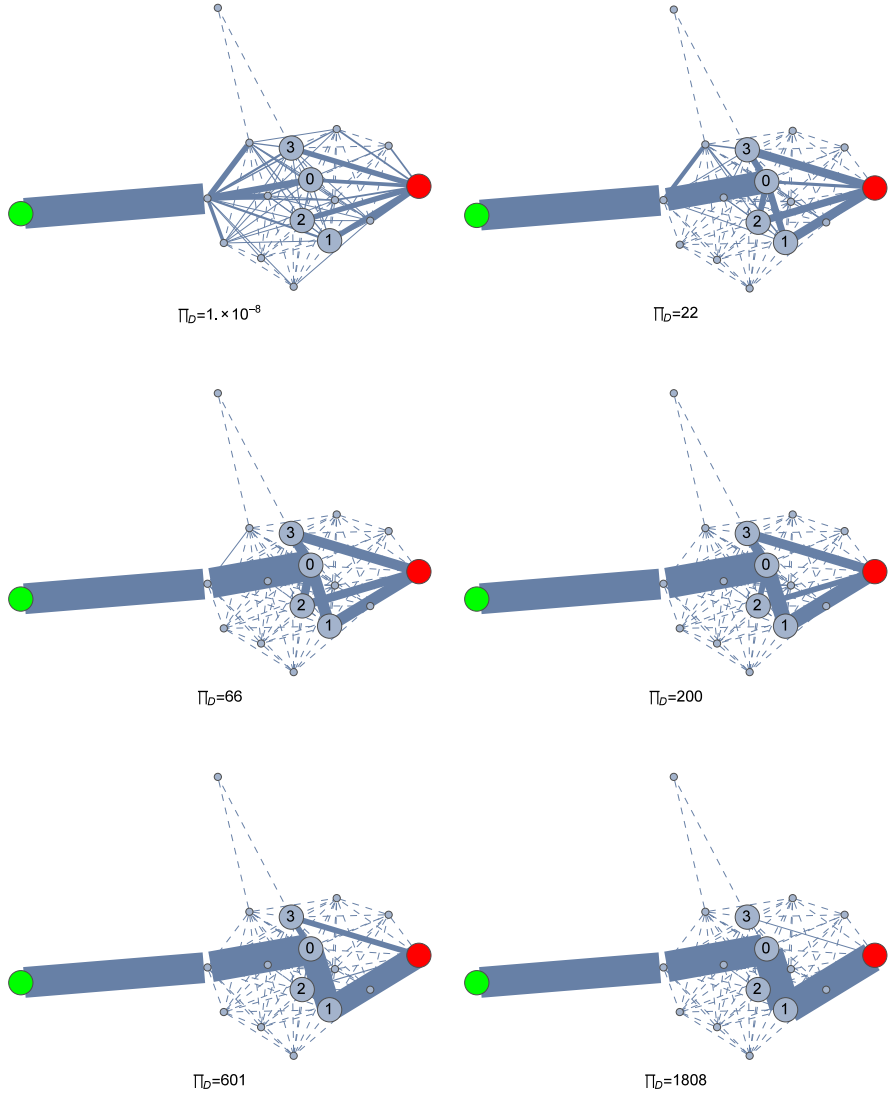}
\caption{\label{fig:condcurrent} \em Conditional current at increasing values of $\para_D$ illustrated on the weighted kangaroo social interaction network \em \cite{kangadata,grant1973dominance}. Symbols have the same meanings as in Fig.~\ref{fig:floridacond}. At values of $\para_D$ near 0,  $\cond$ is equal to the physical current flow in a resistor network, fanning out over all possible paths from source to target. As $\para_D$ approaches $\infty$, $\cond$ follows only the shortest \em weighted \em path. In the intermediate $\para_D$ regime, $\cond$ splits among three approximately equal-length paths, passing through nodes 1, 2, and 3, respectively. All of these paths pass through node 0. As $\para_D$ increases,  more and more $\cond$ flows along the shortest of these paths. 
See further discussion  in Sec.~\ref{sec:networkresults} and Sec.~\ref{sec:pathres}. }
\end{figure*}

Figure \ref{fig:condcurrent} shows the conditional currents $\cond$ on a weighted network representing social interaction in a group of kangaroos \cite{kangadata,grant1973dominance}. It includes the full range of conditional current behavior, including intermediate parameter values. As discussed in Sec.~\ref{sec:shortpaths}, the smaller $\para_D$ the longer the paths the random walker is able to explore. In the figure, at low values of $\para_D$ the conditional current splays out over the network: the walker is able to take advantage of many parallel paths from the start to the target, whether they are short or long. As $\para_D$ increases, fewer and fewer of the long parallel paths can be followed. Finally, at large $\para_D$, the conditional current becomes negligible for all but the geodesic path, and all parallelism is suppressed. In this way, the parameter tunes the conditional current's preference for geodesic (and nearly-geodesic) paths.  The centralities based on this conditional current will probe the network at the level of parallelism specified by the value of the parameter $\para_D$.  (Note that $\para_D$ does not simply  tune the conditional current's preference for {\it short} paths, since the walker's start and target nodes may be an arbitrary distance apart. )


\subsubsection{Calculating $\cond$ vs $\para_D$: conditional current betweenness}
\label{sec:CalcCond}

 For values of $\para_D$  between 0 and $\infty$, the  conditional current $\cond(\para_D)$ of walkers must be calculated  using the theory of absorbing Markov Chains \cite{novotny2001annual,iosifescu2014finite}. Take the random-walk matrix $\mathbf{W}$, whose elements $\mathbf{W}_{m n}$ indicate the single-step transition probability from node $m$ to $n$, and partition it according to a  form which picks out absorbing  and transient  nodes. In the present case, there are two absorbing nodes: the first is a sink that corresponds to the  death probability $p_D$, while the second is the ``target'' node where the conditional current leaves the network: node $j$ in the calculation of $\cond_{ij}(\para_D)$. The walkers begin on node $i$. The canonical form for $\mathbf{W}$  is 
   
   \begin{equation}\label{WCanonical}
\mathbf{W}=
\begin{pmatrix}
\pmb{\mathbb{I}} & \pmb{\mathbb{O}} \\ 
\begin{pmatrix} 
\ket{\mathscr{A}^\mathrm{sink}} &
\ket{\mathscr{A}^\mathrm{target}}
\end{pmatrix} & \mathbf{T}
\end{pmatrix} .
\end{equation}

\noindent
Here, $\mathbf{T}$ is the $(N-1)\times(N-1)$ transient transition matrix, whose elements are transition probabilities between the non-absorbing (transient) states. $\pmb{\mathbb{O}}$ is the $2\times (N-1)$ matrix of zeroes, indicating that walkers cannot make transitions from absorbing states to transient states. Here,  $\pmb{\mathbb{I}}$ is the $2\times 2$ identity matrix, indicating that a walker in an absorbing state is stuck there forever. The two $(N-1)$-dimensional  column vectors $\ket{\mathscr{A}}$ describe absorption transitions to the sink and the target node, respectively. An element  of the $\mathbf{T}$ matrix $\mathbf{T}_{m n}$ is given by $[1-p_D(m)]( \mathbf{A}_{m n}/A_m)$, 
according to the standard definition of random walks on a network with adjacency matrix $\mathbf{A}$  \cite{Doyle06randomwalks},  modified by the walker death probability 
$p_D(m)$ from Eq.~(\ref{eq:dedprob}). (Recall that $A_m$ is the weighted degree of node $m$.) Similarly, $(\mathscr{A}^\mathrm{target})_m$, the $m$th entry of $\ket{\mathscr{A}^\mathrm{target}},$ equals $(1-p_D(m))( \mathbf{A}_{m j}/A_m)$, while  $(\mathscr{A}^\mathrm{sink})_m$ is just $p_D(m)$. 

A key object in the theory of absorbing random walks
 is the \em fundamental matrix \/\em $\mathbf{F}$, given by 
  \begin{equation}
 \mathbf{F} = (\pmb{\mathbb{I}}-\mathbf{T})^{-1}.
 \end{equation}
  $\mathbf{F}_{i n}$ is the  expected number of  times a walker starting on source $i$ can make it to $n$ before being absorbed by the sink. Because the target node $j$ is not transient, $n\neq j$. Define the unbolded variable  $F_{i n}$ similarly to $\mathbf{F}_{i n}$, but now allowing $n$ to take on the value of $j$.  By the properties of the fundamental matrix,
 \begin{equation}\label{eq:fundmat}
 F_{i n}=
 \begin{cases}
 \mathbf{F}_{i n}& n\neq j\\
 \sum_{m\sim n}\mathbf{F}_{i m}\mathscr{A}^\mathrm{target}_m & n=j
 \end{cases},
 \end{equation}
 where the sum is over the neighbors of $n$, and node $j$ is the target of the random walk.

The random-walk formulation can be connected to the current-flow formulation  by extrapolating from the well-known \cite{Doyle06randomwalks} isomorphism for the case $\para_D=0$. In that case, the edge current produced by a unit voltage is proportional to the \em net \em number of walker crossings: the number in the forward direction, minus the number in the reverse direction. The proportionality constant is the inverse of the resistance distance, $(R^\mathrm{eff}_{i\,j})^{-1}$, which describes the total number of walkers released from the node maintained at unit voltage. To generalize the resistance-closeness centrality of Eq.~({\ref{eq:RCC}}) to non-zero values of $\para_D$, $R^\mathrm{eff}_{ij}$ must deviate from its value at $\para_D=0$, so the proportionality constant is unknown. Thus, in the  $\para_D>0$ regime, we must work with  
  ratios of currents so that the constant does not appear.  Recall that, for $\para_D>0$, the current is \em conditional \em on reaching target $j$; we denote this condition as ``$\mid j$''. 
  Equation (\ref{eq:fundmat}) can be used to formulate $\cond_{i (a,b) j}$: the current entering the network at $i$, eventually flowing through the edge $(a,b)$, and finally leaving the network at $j$ (\em i.e.\/\em, not succumbing to $\para_D$):
\begin{equation}\label{conditionalcurrent}
\begin{array}{ccccc}
\frac{\cond_{i  (a,b) j}}{\cond_{i  j}}&=&
\mathbb{ E}\small(\mbox{\# walker crosses from $a$ to $b$ $\mid$   $j$})
&-&\mathbb{E}\small(\mbox{\# walker crosses from $b$ to $a$ $\mid$   $j$})\\
&=&F_{i\;a} \mathbf{T}_{a\;b}F_{b\;j}/F_{i\;j}& - &F_{i\;b} \mathbf{T}_{b\;a}F_{a\;j}/F_{i\;j}
\end{array}.
\end{equation}
Here, every term has an implicit dependence on $\para_D$, and the $\mathbb{E}(* \mid j)$ are conditional expectation values. The above equation is just the ``$\mid j$'' conditional version of a well-known connection between walker paths and electric currents (see, \em e.g.\/\em, \cite{newman2005measure}).  Note that this expression for conditional current satisfies Kirchhoff's Current Law, since the path of any individual walker must do so. 

Equation (\ref{conditionalcurrent}) only provides {\it ratios} of conditional currents $\cond$. However, in the betweenness centrality of Eq.~(\ref{eq:CBT}), the currents $I$ are found in ratios as well. The above can be used to calculate the interpolated betweenness currents  by summing the edge currents \em into \em a given node, as in Sec.~\ref{sec:bet}: $\cond_{s i j}/\cond_{s  j}= \sum_{a:\cond_{s(a,i)j}>0} \cond_{s(a,i)j}/\cond_{s  j}$. This process leads to a parametrized form of the current-betweenness centrality: \em conditional current-betweenness 
centrality \em in Table \ref{curbettab}.

\begin{table}[h]

  \begin{center}

	\caption{\em Current betweenness, betweenness,  and intermediate centralities\/\em. The top-left entry defines current-betweenness centrality (equivalent to random-walk centrality \cite{newman2005measure}), while the top-right entry defines betweenness centrality  in an analogous form. The interpolation between these two centralities in terms of the ``death parameter'' $\para_D$ is described by the top-middle entry. The middle row indicates the corresponding values of $\para_D$. The bottom row describes the type and behavior of the current corresponding to each parameter value. }
	  	\label{curbettab}
    \begin{tabular}{c|c|c}

\textbf{Current Betweenness}$\;$ & $\longleftarrow$ Interpolation$\longrightarrow$ & \textbf{Betweenness} 
\\

[see Eq.~(\ref{eq:CBT})] & \textbf{Conditional Current Betweenness}  & $ $ { [see Eq.~(\ref{eq:bet})]}
\\
    
 $\mathbf{M}^\mathrm{CBT}_{ij}=\sum_s I_{s i  j} /I_{s j}$ &     
 $\mathbf{M}^\mathrm{CBT}_{ij}(\para_D)=\sum_s \cond_{s i  j}(\para_D) /\cond_{s j}(\para_D)$& 
$\mathbf{M}^\mathrm{BET}_{ij}=\sum_s n_{s i j} /g_{s j}$ 
\\[.5em] \hline 

 $\lim_{\para_D \to 0}$ & $\para_D>0$ &$\lim_{\para_D \to \infty}$  
\\\hline
   
$I_{ij}$&
 $ \cond_{ij}(\para_D)$&
$\cond$ only flows on geodesics: 

\\ 
   
(physical current)&  (conditional current) & $\cond_{sj}\propto g_{sj}$ and  $\cond_{s i j}\propto n_{s i j}.$
\\

    \end{tabular}
  \end{center}

\end{table}

\subsubsection{Conditional resistance closeness: calculating the $\para_D$-parametrized effective resistance}
\label{sec:reff}
To naively parametrize the form of the resistance-closeness centrality in Eq.~({\ref{eq:RCC}})  would  require the values $\cond_{i j}(\para_D)$, which cannot be determined from Eq.~(\ref{conditionalcurrent}).   This is because the absorbing random walk outlined above, for $\para_D>0$, does not correspond to a physical current, and thus only current \em ratios \em are determined.

	To bridge the gap, we seek to determine which edge resistances---given the same network topology---would reproduce the calculated conditional current as a physical current: $I=\cond$. 
If we could obtain a set of resistances $\{R^\cond_\nu\}$  that would reproduce the set of conditional currents $\{\cond_\nu\}$ as physical currents, then the corresponding voltage drop 
$V_{i\,j}$ from $i$ to $j$ would simply be equal to   $\sum_{\nu \in \mathcal{P}} \cond_\nu R^\cond_\nu$, where the edge index $\nu$ runs over the  edges  in \em any oriented \em path $\mathcal{P}$ from $i$ to $j$. (Note that in general when $I\neq\cond$, $V_\nu \neq \cond_\nu R^\cond_\nu$.)  

From $V_{i\,j}=I_{i\,j}R^\mathrm{eff}_{i\,j}$, the voltage drop for a unit current is  equal to the  effective resistance. 
Hence, it is convenient to set the total conditional current (from $i$ to $j$) to unity.
\begin{equation} 
\label{eq:effres}
R^{\mathrm{eff}}_{i\,j}=\sum_{\nu \in \mathcal{P}} \cond_\nu R^\cond_\nu. \hspace{2em} (\mathrm{for}\;\cond_{ij}=I_{ij}=1)
\end{equation}
With $I_{i\,j}=\cond_{i\,j}$ set to unity,   the edge current $\cond_\nu$ over edge $\nu=(a,b)$ becomes $\cond_\nu=\cond_{i (a,b) j}/\cond_{i\,j}=\cond_{i (a,b) j}$.
(Even though we are dealing with undirected networks, consistency with what follows forces us to specify edge orientation explicitly, meaning that $\cond_{(a,b)}=-\cond_{(b,a)}$.)

Unfortunately, the values $\{R^\cond_\nu\}$ (and hence, also the value of $R^\mathrm{eff}_{i\,j}$) are under-determined by the currents in Eq.~(\ref{conditionalcurrent}). This can be seen from the following linear condition on $\{R^\cond_\nu\}$ \cite{bollobas1979graph}, which is equivalent to Kirchhoff's Voltage Law :
\begin{equation}\label{nullspace}
\forall r : \sum_{\nu} \mathbf{K}_{r\,\nu}\cond_\nu R^\cond_\nu =0 .
\end{equation}
Here, $\mathbf{K}$ is the reduced cycle matrix, describing the edges that form a maximal 
collection of linearly independent cycles on the network. The index $r$ denotes independent 
oriented cycles, and $\mathbf{K}_{r\,\nu}$ is non-zero only for network edges $\nu$ participating in  cycle $r$. Thus, the possible edge resistances $\{R^\cond_\nu\}$ form the (generally multidimensional) null-space of the matrix $[\mathbf{K}_{r\,\nu}\cond_\nu]$, with the added physical constraint that $R^\cond_\nu \ge 0,\forall \nu$. For a network with $N$ nodes and $M$ edges, the matrix $[\mathbf{K}_{r\,\nu}\cond_\nu]$ has dimensions $(M-N+1)\times M$.  Using this equation, it can be verified that sometimes wildly different resistance distributions can lead to the same current flow on a given network.

Nonetheless, it is possible to compute a uniquely suitable set of resistances $\{R^\cond_\nu\}$, given two  common-sense criteria: (1) Because increasing $\para_D$ serves to  inhibit  current, we constrain the resistances $R^\cond_\nu$ to be larger than or equal to their $\para_D=0$ values; \em i.e.\/\em, $R^\cond_\nu\ge R^\mathrm{orig}_\nu,\forall \nu$. (2) Because any vector in the null-space of $[\mathbf{K}_{r\,\nu}\cond_\nu]$ remains in the null-space after scaling, there is no upper bound on the effective resistance $R^\mathrm{eff}_{i\,j}=V_{i\,j}$, and thus, we define the $\para_D$-parametrized  effective resistance $\mathfrak{R}_{ij}$ to be the {\it minimum value } of $R^\mathrm{eff}_{ij}$. 

To find $\mathfrak{R}_{ij}$, we minimize the expression in Eq.~(\ref{eq:effres}), requiring that a valid solution $\{R^\cond_\nu\}$ must satisfy condition (1) and Eq.~(\ref{nullspace}):
\begin{subequations}\label{linprog}
\begin{alignat*}{2}
  & \underset{\{R^\cond_\nu\}}{\text{minimize}} & & \sum_{\nu \in \mathcal{P}} \cond_\nu R^\cond_\nu \\
   & \text{subject to}& \quad & \sum_{\mathclap{\nu}}\begin{aligned}[t]
                    \mathbf{K}_{r\,\nu}\cond_\nu R^\cond_\nu =0&,& \forall& \; \mathrm{independent\; cycles} \; r \\[3ex]
                    R^\cond_\nu \ge R^\mathrm{orig}_\nu&,& \forall& \; \mathrm{edges} \; \nu \\[3ex]
                \end{aligned}\tag{\ref{linprog}}
\end{alignat*}
\end{subequations}
Note that $\cond$ depends on $i, j,$ and $\para_D$. Even though the first sum is over the edges in the arbitrary path $\mathcal{P}$, the minimization is over all the edges in the network. $\mathfrak{R}_{ij}$ is the minimized objective function.

Finding $\{R^\cond_\nu\}$ that satisfies the above is a standard linear programming problem, which  is guaranteed to be feasible, as shown in  Appendix \ref{app:feasible}. Because the problem is feasible (and clearly bounded), the optimal solution   $\mathfrak{R}_{ij}$  is, in principle, calculable. Techniques such as the simplex method \cite{murty1983linear} and  interior point methods   \cite{boyd2004convex} are generally used to find this optimal solution. As a practical matter, the linear programming algorithms struggle to find solutions when conditional currents $\cond_\nu$ become too small. The difficulty is overcome by removing low-current edges from the network, since they do not contribute to $\mathfrak{R}_{ij}$ anyway. For the networks under consideration, once low-current edges are removed, the convergence is fast and the particular choice of linear solver is not important \cite{noteSolver}.

Finally, the optimized value $\mathfrak{R}$ of the linear programming problem in Eq.~(\ref{linprog}), given conditional currents calculated from Eq.~(\ref{conditionalcurrent}), leads to a parametrized form of the resistance-closeness centrality of  Eq.~({\ref{eq:RCC}}): the \em conditional resistance-closeness centrality \em. See Table \ref{curclostab}.

 We can now understand the behavior of the conditional resistance closeness as $\para_D \rightarrow \infty$. In this limit, conditional current  becomes restricted to the shortest path $\mathcal{P^*}$ from source $i$ to target $j$.
Setting $\mathcal{P}=\mathcal{P^*}$ in Eq.~(\ref{eq:effres}), $R^{\mathrm{eff}}_{i\,j}=\sum_{\nu \in \mathcal{P^*}} \cond_\nu R^\cond_\nu$. Because we are dealing with unit  current, and because there is only one shortest path (see Appendix \ref{sec:randomnoise}),  
\begin{equation}
\label{eq:limitReff}
\lim_{\para_D \to \infty}R^{\mathrm{eff}}_{i\,j}=\sum_{\nu \in \mathcal{P^*}} R^\cond_\nu.
\end{equation}  
The final centrality is calculated using $\mathfrak{R}_{ij}$ . The sum in Eq.~(\ref{eq:limitReff}) is clearly minimized, within our constraints, by setting $R^\cond_\nu=R^\mathrm{orig}_\nu = d_\nu$, for all $\nu \in \mathcal{P^*}$. The current can be forced to match the conditional current by setting $R_\nu^\cond = \infty$ for all $\nu \notin \mathcal{P^*}$, but this does not affect $\mathfrak{R}_{ij}$. Thus, 
\begin{equation}
\label{eq:limitReffmin}
\lim_{\para_D \to \infty}\mathfrak{R}_{ij} =\sum_{\nu \in \mathcal{P^*}} d_\nu  = d_{i j}.
\end{equation}
 Inverting this effective resistance, as in the top-middle entry of Table \ref{curclostab}, results in the formula for harmonic closeness, as in the top-right entry of the table.

\begin{table}[h]

  \begin{center}

	\caption{
		\em Resistance closeness, harmonic closeness,  and intermediate centralities\/\em. The presentation is similar to  that of  Table~\ref{curbettab}.}
	  	\label{curclostab}
    \begin{tabular}{c|c|c} 
    
\textbf{Resistance Closeness}$\;$&$\longleftarrow$ Interpolation$\longrightarrow$ & \textbf{Harmonic Closeness}  
\\
      
\;[see Eq.~(\ref{eq:RCC})]& \;\textbf{Conditional Resistance Closeness}\; &    $ $ {[see Eq.~(\ref{eq:HCL})] }
\\
     
$\mathbf{M}^\mathrm{RCC}_{ij}={1}/{R^\mathrm{eff}_{ij}}= I_{ij}$ &
$\mathbf{M}^\mathrm{RCC}_{ij}(\para_D)=1/{\mathfrak{R}_{ij} (\para_D)}\;\mbox{\;and Eq. }(\ref{linprog})$ &
$\mathbf{M}^\mathrm{HCL}_{ij}= {1}/{d_{ij}}$ 
\\[.5em] \hline 

$\lim_{\para_D \to 0}$ &
$\para_D>0$ & 
$\lim_{\para_D \to \infty}$ 
\\\hline
   
$I_{ij}$  &
$ \cond_{ij}(\para_D)$&
$\cond$ only flows   on
\\ 

(physical current)&
(conditional current)& 
geodesics from $i$ to $j$.
\\

    \end{tabular}
  \end{center}

\end{table}

\subsubsection{The conditional walker-flow centralities} \label{sec:meaningpd}

In summary, we have shown that the parameter $\para_D$ interpolates between the leftmost and rightmost columns in Tables \ref{curbettab} and \ref{curclostab}. The transition is from current-betweenness centrality at $\para_D=0$ to  betweenness centrality as $\para_D$ approaches $\infty$ (Table \ref{curbettab}), and from resistance-closeness centrality at $\para_D=0$ to the harmonic closeness centrality as $\para_D$ approaches $\infty$ (Table \ref{curclostab}). Intermediate parameter values tune the closeness and betweenness centralities' preference for geodesic paths and, as a result, probe the network's behavior at different levels of parallelism. 

The new centralities that interpolate between these limits may be called \em conditional walker-flow \em centralities. The measures in Table \ref{curbettab} are connected to each other by the same random-walk process that connects the measures in Table \ref{curclostab}, reinforcing the idea that  all four centralities in those tables are part of the natural class of {\it walker-flow centralities}. Stephenson's information centrality \cite{stephenson1989rethinking} 
and  Newman's random-walk betweenness \cite{newman2005measure} also belong to 
this class, having been proved equivalent to resistance closeness and current 
betweenness, respectively.

\section{Conditional Walker-Flow Centrality Results}
 \label{sec:results}
\subsection{Results on example networks}
\label{results_a}
We now apply the conditional current centralities developed in the previous section to four example networks, demonstrating the interpolations and limits summarized in Tables \ref{curbettab} and \ref{curclostab}. The  networks are chosen to have different edge densities and types of edge weighting: the kangaroo network is dense and has integer weights, the karate club network has intermediate density and is unweighted, and the power-grid network is sparse with continuous weights. (We also study the unweighted version of the power-grid network for comparison.) The characteristics of the example networks, as well as literature references and the figure numbers of corresponding results, are summarized in Table~\ref{tab:nets}. 

\begin{table}[]
\caption{\em Example network summary\/\em. Networks have $N$ nodes and $M$ edges. The density of a network is defined as the number of edges divided by the number of possible edges: $M/(0.5 N (N-1))$. See text for discussion.  }
\label{tab:nets}
\setlength\tabcolsep {.35em}

\begin{tabular}{clcccc|cc}
Network               & Refs. & \textit{N} & \textit{M}& Density & Weights    & Results                                                         \\ \hline
Kangaroo Group             &      \cite{kangadata,grant1973dominance}     & 17         & 91    & 0.67     & Integer    & Fig. \ref{fig:kanga_cond_bet_lines}                  \\
Zachary Karate Club   &   \cite{zachary}        & 34         & 78 & 0.14        & Unweighted       & Fig. \ref{fig:karate_cond_bet_lines}                 \\
Weighted Power Grid   &     \cite{xu2014architecture,dale}      & 84         & 137   &0.04     & Continuous & Figs. \ref{fig:florida_weighted_bet_lines},   \ref{fig:flflcorr}, and \ref{fig:nonmon}     \\
Unweighted Power Grid &     \cite{xu2014architecture,dale}      & 84         & 137 &0.04       & Unweighted       & Figs. \ref{fig:florida_unweighted_bet_lines},  and  \ref{fig:flflcorr}  
\end{tabular}
\end{table}

\subsubsection{Overview of numerical results}

The  values of the conditional walker-flow centralities---the \em conditional current betweenness \em and the \em conditional resistance closeness\em---are presented in Figs.~\ref{fig:kanga_cond_bet_lines}-\ref{fig:florida_unweighted_bet_lines}. There, each line represents the centrality results of a different node across a range of values of the dimensionless parameter  $\para_D \langle L \rangle$, where $ \langle L \rangle$ is the average edge length (edge resistance) of the
network.

 The large circles in the plots show that the conditional walker-flow centralities correspond to the limiting centralities in Tables \ref{curbettab} and \ref{curclostab}, regardless of weighting type. As an example, consider the conditional current-betweenness centrality for the kangaroo network, Fig.~\ref{fig:kanga_cond_bet_lines}(a).  The circles on the left side of the figure correspond to the current-betweenness centrality. Those on the right correspond to the weighted betweenness centrality, obtained with a weighted version of the algorithm  from \cite{brandes2001faster}.  In the limits, the lines coincide with the circles, showing that the conditional current betweenness  reduces to the current betweenness at low $\para_D$ and to the standard weighted betweenness at high  $\para_D$. Similarly, Fig.~\ref{fig:kanga_cond_bet_lines}(b) shows that  the conditional resistance closeness reduces to the resistance closeness at low  $\para_D$ and to the harmonic closeness at high  $\para_D$.

As discussed in Sec.~\ref{sec:notation}, the figures in this section depict the unnormalized centrality values produced by our algorithms. This enables us to better compare centralities across different values of $\para_D$. In the normalized version, an increase in node $i$'s centrality may create a spurious decrease in the centrality of node $j$, even if the conditional currents through $j$ remain unchanged.

\newcommand{\BetCaptionText}{ The circles on the left and right ends show the values of the current-betweenness centrality and the betweenness centrality, respectively. Note that some nodes on the periphery of the network have a centrality value of zero. The data are thus represented on a semi-logarithmic scale. } 

\newcommand{\BetCaptionTextTwo}{ See the caption to Fig.~\ref{fig:kanga_cond_bet_lines} for explanatory details. All centrality values are unnormalized. }

\newcommand{\ClosCaptionText}{The circles on the left and right ends show the values of the resistance-closeness centrality and the harmonic closeness centrality, respectively. The data are  represented on a log-log scale. } 

\newcommand{\ClosCaptionTextTwo}{See the caption to Fig.~\ref{fig:kanga_cond_clos_lines} for explanatory details. }

 \begin{figure*}

 \begin{overpic}[unit=1mm,scale=.8,trim={0 0cm 0cm 0cm},clip]{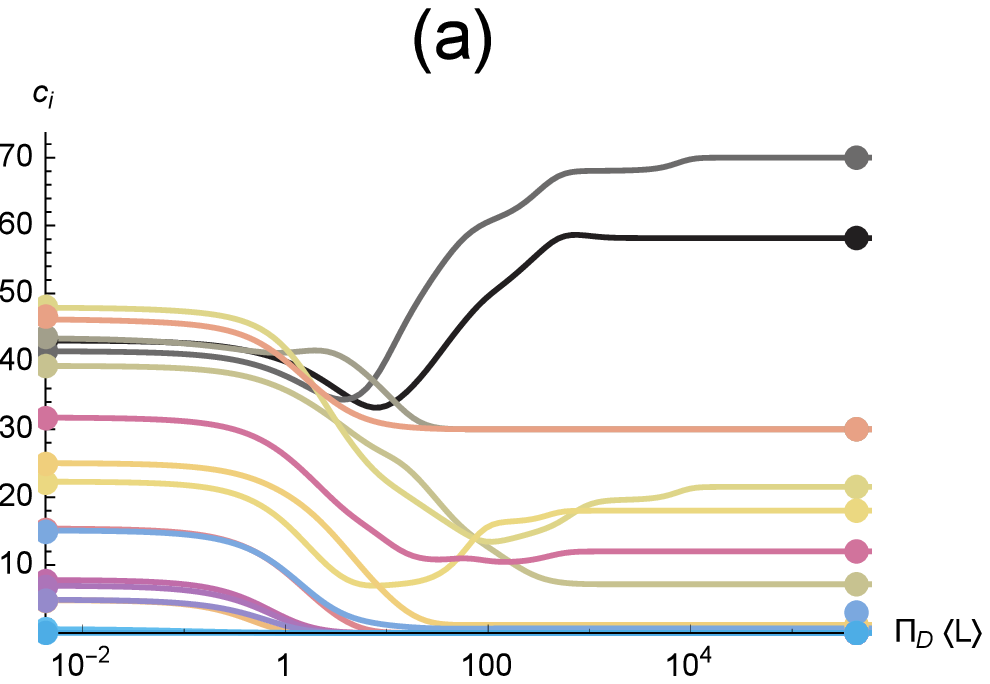}\end{overpic}
 \begin{overpic}[unit=1mm,scale=.8,trim={0 0cm 0cm 0cm},clip]{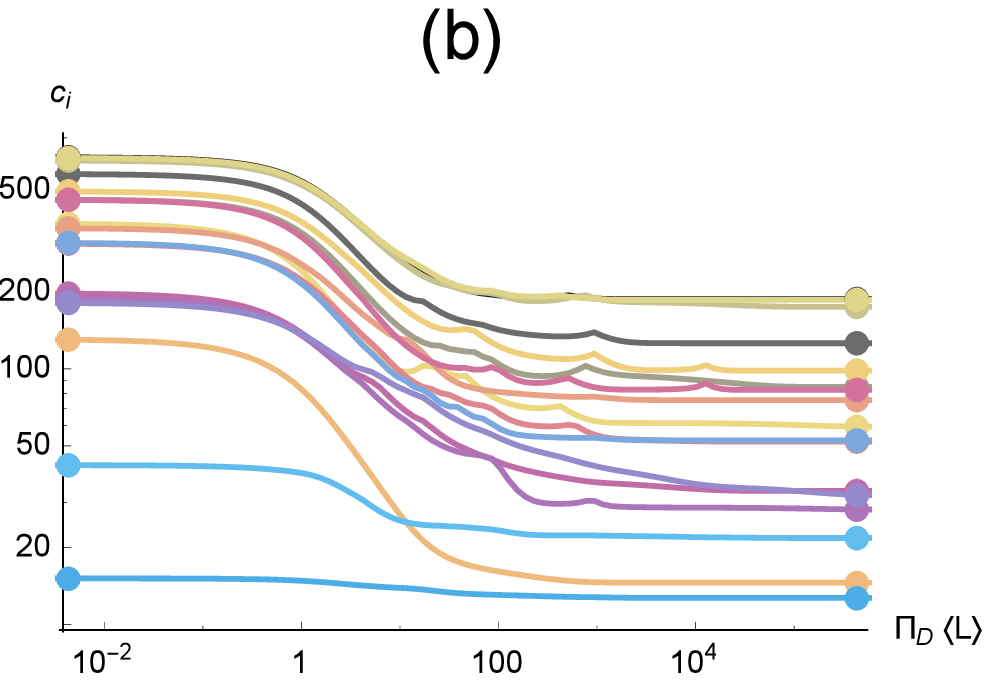}\end{overpic}
\caption{\label{fig:kanga_cond_bet_lines} \em Conditional walker-flow centrality of every node in the kangaroo network\em. Each line represents the {\it unnormalized} centrality of a different node. In this and the following figures, the abscissa is made dimensionless by
multiplying $\para_D$  by $\langle L \rangle$, the average edge length (edge resistance) of the network. Here, $\langle L \rangle \approx 0.432$.  (a) \em Conditional current betweenness\em. \BetCaptionText    (b) \em Conditional resistance closeness\em. \ClosCaptionText   }
\end{figure*}

 \begin{figure*}

 \begin{overpic}[unit=1mm,scale=.8,trim={0 0cm 0cm 0cm},clip]{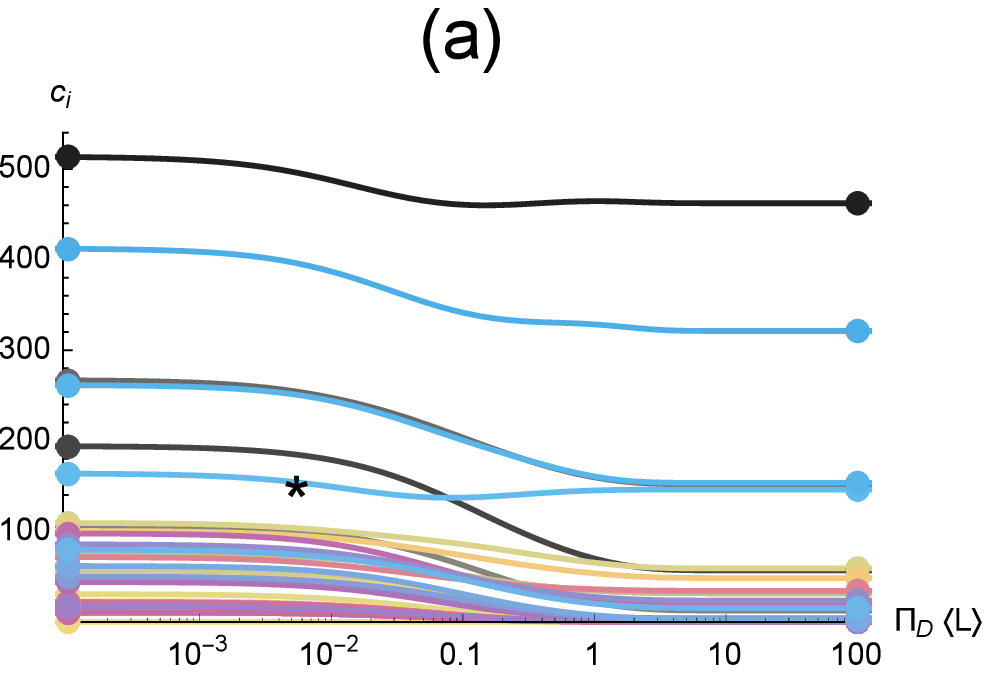}\put(0,62){\bf \large }\end{overpic}
 \begin{overpic}[unit=1mm,scale=.75,trim={0 0cm 0cm 0cm},clip]{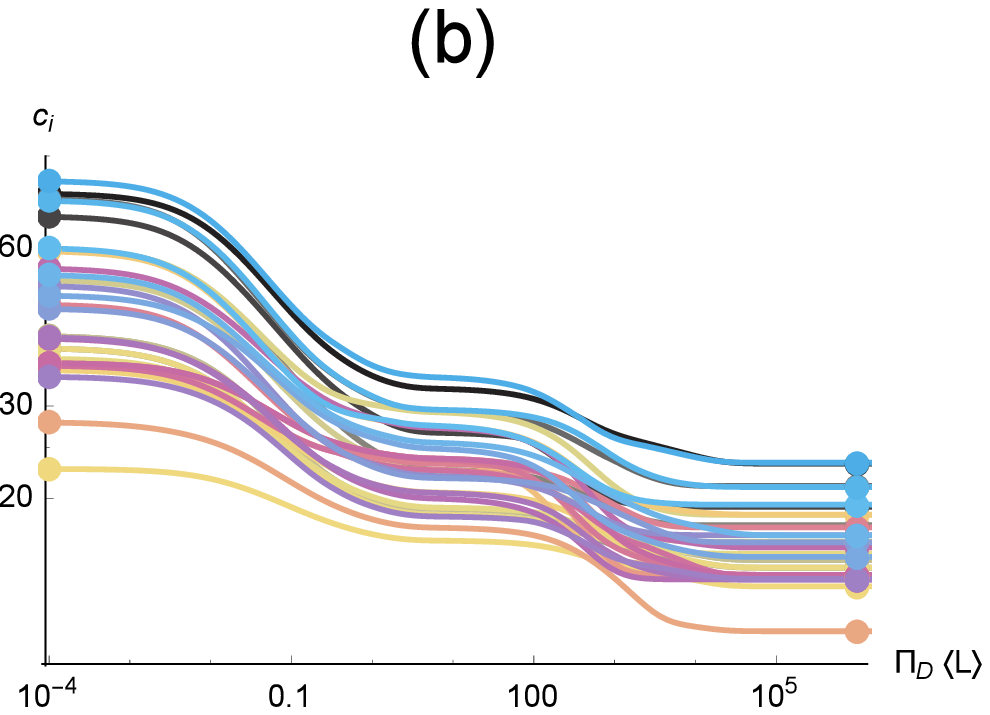}\put(0,62){\bf \large}\end{overpic}
\caption{\label{fig:karate_cond_bet_lines} \em Conditional walker-flow centrality of every node in the karate-club network\em. \BetCaptionTextTwo  Because this network is unweighted, $\langle L \rangle=1$.  (a) \em Conditional current betweenness\em. The curve marked with a star  is referenced in  Sec.~\ref{sec:nonmon} and in Appendix \ref{sec:nonmon data}. (b) \em Conditional resistance closeness\em.  The plateau region between $\para_D\langle L\rangle\approx 1$ and $\para_D\langle L\rangle\approx 100$ occurs because $\para_D$ is large enough to pick out (possibly multiple) shortest paths in the original network  but not yet large enough to resolve the unique shortest path created by the introduction of random noise. See further discussion in Appendix~\ref{sec:randomnoise}. }
\end{figure*}

 \begin{figure*}

 \begin{overpic}[unit=1mm,scale=.8,trim={0 0cm 0cm 0cm},clip]{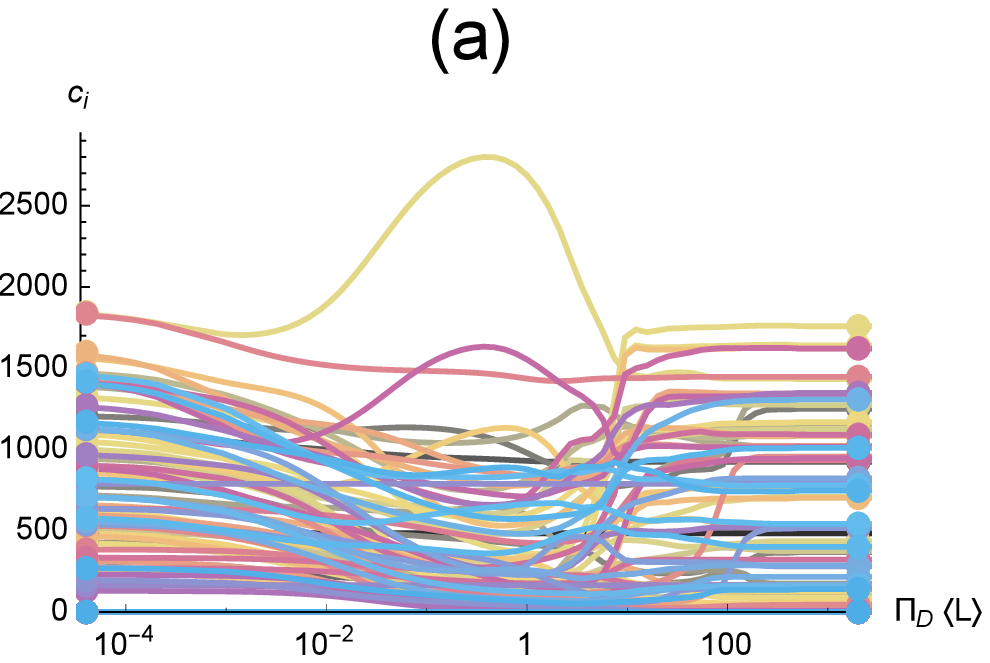}\end{overpic}
 \begin{overpic}[unit=1mm,scale=.75,trim={0 0cm 0cm 0cm},clip]{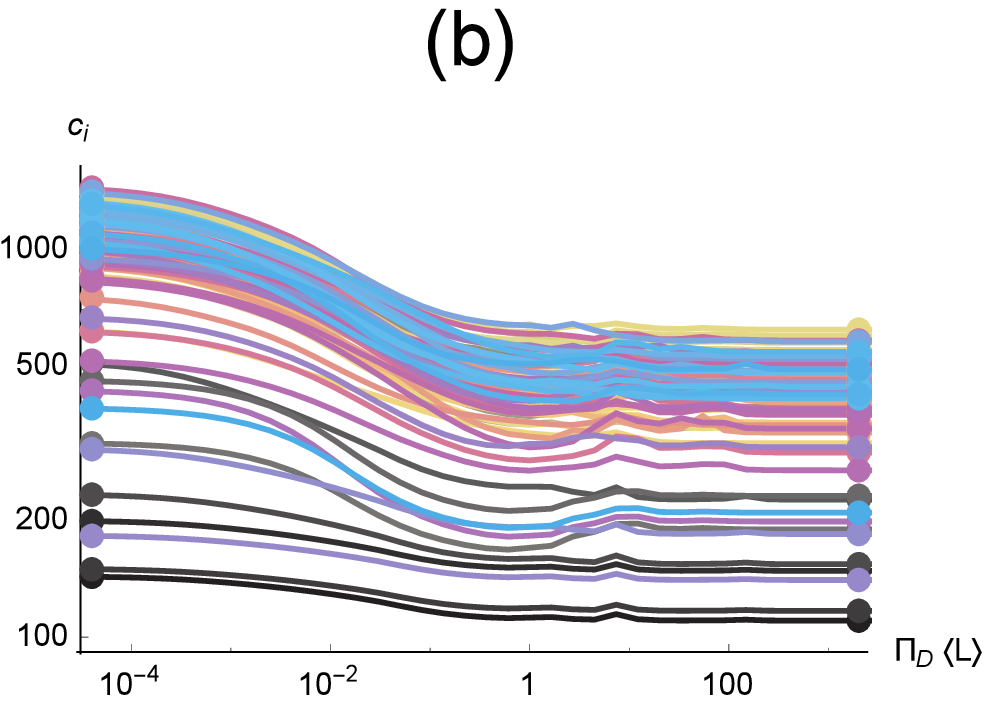}\end{overpic}

\caption{\label{fig:florida_weighted_bet_lines} \em Conditional walker-flow centrality of every node in the weighted power-grid network\em. \BetCaptionTextTwo  In this network, $\langle L \rangle \approx 0.067$.  (a) \em Conditional current betweenness\em. (b) \em Conditional resistance closeness\em. }
\end{figure*}

 \begin{figure*}

 \begin{overpic}[unit=1mm,scale=.8,trim={0 0cm 0cm 0cm},clip]{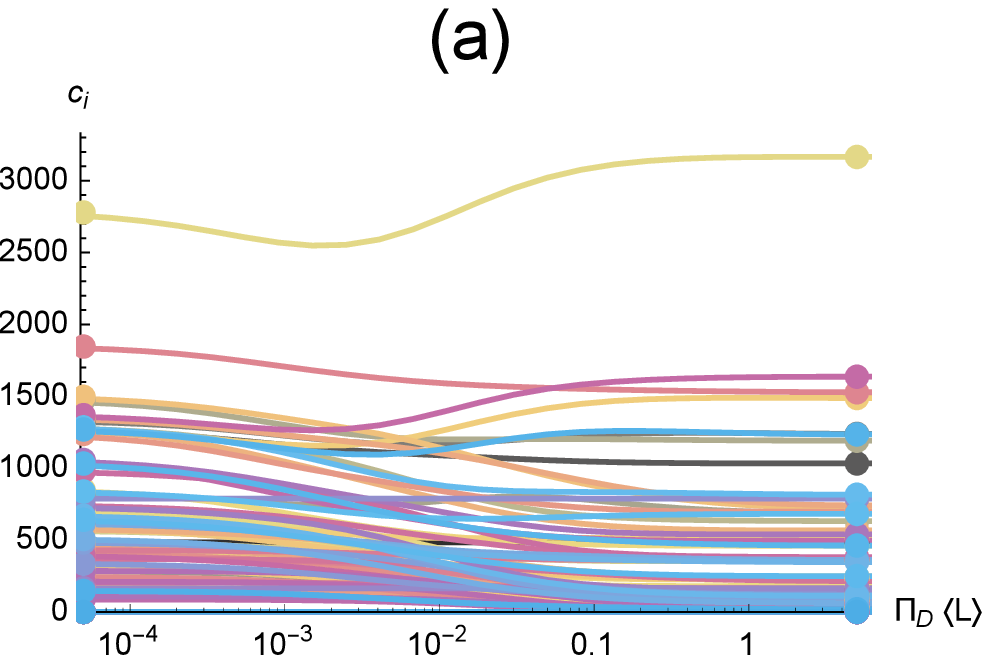}\end{overpic}
 \begin{overpic}[unit=1mm,scale=.75,trim={0 0cm 0cm 0cm},clip]{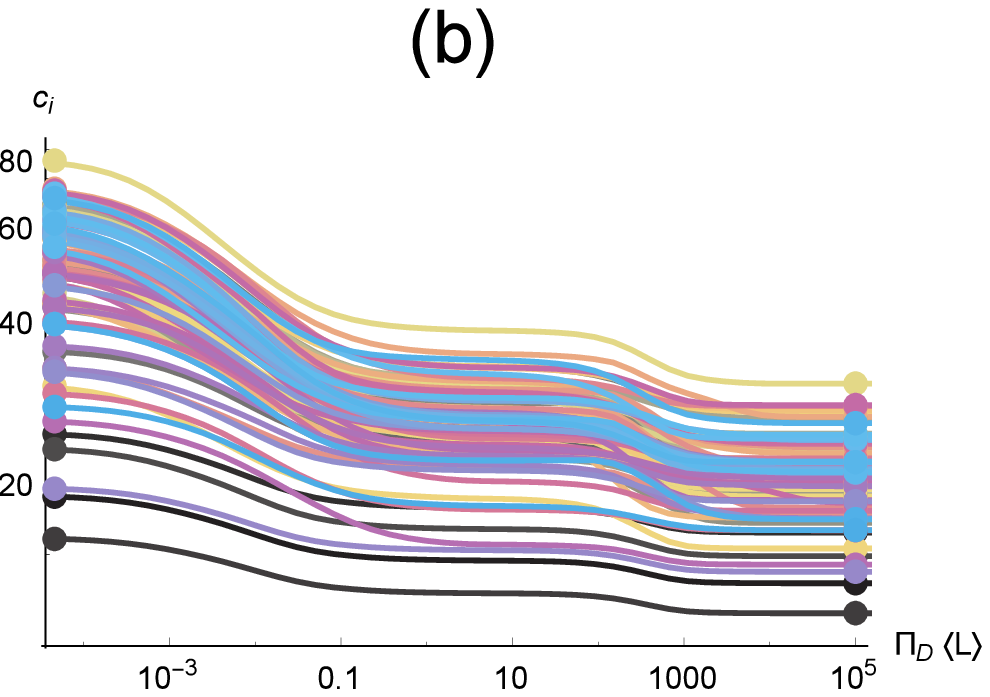}\end{overpic}

\caption{\label{fig:florida_unweighted_bet_lines} \em Conditional walker-flow centrality of every node in the unweighted power-grid network\em. \BetCaptionTextTwo  Because this network is unweighted, $\langle L \rangle=1$. (a) \em Conditional current betweenness\em.  (b) \em Conditional resistance closeness\em. The plateau region is similar to that described in the caption to Fig.~\ref{fig:karate_cond_bet_lines}. }
\end{figure*}

\subsubsection{Numerical results for individual networks}
\label{sec:networkresults}

We next remark on some particulars of the results for the different example networks.

{\it Kangaroo network.}
The first network under consideration is a weighted network of social interactions within a group of 17 kangaroos \cite{kangadata,grant1973dominance}. The nodes represent individual animals, and the 91 weighted edges represent their social interactions. The weights are integers indicating the number of observed interactions. This network is illustrated in Fig.~\ref{fig:condcurrent}.

Figure \ref{fig:kanga_cond_bet_lines} depicts the conditional walker-flow centralities for this network. In part (a), the conditional current betweenness behavior of two nodes stands out. Consider the  nodes with the two highest values of the standard betweenness (the two dots at the top-right of the figure). These correspond to the nodes marked ``0'' and ``1'' in Fig.~\ref{fig:condcurrent}. For a broad range of $\para_D$ values, these  have much higher conditional current betweenness than any other node. At  $\para_D \langle L \rangle \lessapprox 20$, their centrality values become close to those of several  other nodes. We have remarked that $\para_D$ tunes the conditional current-betweenness centrality's preference for geodesic paths (see Sec.~\ref{sec:tuning}). Thus, Fig.~\ref{fig:kanga_cond_bet_lines}(a) shows that at tuning level indicated by  $\para_D \langle L \rangle \lessapprox 20$, the network structure ceases to prioritize the two nodes in question. In fact, they rank only number four and five in current-betweenness centrality. A similar sensitivity to resolution is not observed in the conditional current-closeness centralities in Fig.~\ref{fig:kanga_cond_bet_lines}(b). Such large, non-monotonic variations in the conditional current betweenness are analyzed in Sec.~\ref{sec:nonmon}.

{\it Karate club network.}
The second network under consideration is Zachary's karate club \cite{zachary}. The nodes represent the 34 members of the club. The 78 unweighted edges of the network represent the presence of social interactions between  club members. This is a standard test case in network science. Fig.~\ref{fig:karate_cond_bet_lines} shows that the two nodes representing the club's instructor and administrator have the highest conditional walker-flow centralities across all values of $\para_D$. Thus, unlike the two kangaroo network nodes discussed previously, the two club officials' high centrality rank does not depend on the centrality's preference for geodesic paths. This is consistent with the non-monotonicity analysis of Sec.~\ref{sec:nonmon}, where we explain why unweighted networks are less susceptible to large changes in centrality. In such a network, nodes with a clear centrality lead in the extreme $\para_D$ limits are likely to keep their high centrality rank in the intermediate regime.

{\it Power-grid networks.}
The last two networks under consideration are based on the map of the Florida power grid obtained from \cite{dale} and studied in \cite{xu2014architecture}. The 84 nodes represent high-capacity generators and important substations of the  Florida power grid in 2009. The 137 edges represent power transmission lines. This network is illustrated in Fig.~\ref{fig:floridacond}, and walker-flow centrality results are reported in Figs.~\ref{fig:florida_weighted_bet_lines}-\ref{fig:florida_unweighted_bet_lines}. We analyzed both a weighted (Fig.~\ref{fig:florida_weighted_bet_lines}) and an unweighted (Fig.~\ref{fig:florida_unweighted_bet_lines}) version of this network. The unweighted version only captures the presence or absence of transmission lines. In the weighted version, edge weights are real numbers proportional to the  estimated total conductance of the direct connection between two nodes. Specifically, the edge weight between nodes $a$ and $b$ is equal to the number of  transmission lines divided by the  geographical distance between $a$ and $b$, as in \cite{HAMA11}. 

In both the weighted and unweighted cases, a single node (marked with a triangle in Fig.~\ref{fig:floridacond}) stands out as having the highest centrality across a broad range of parameter values. This node corresponds to an electrical substation with one of the largest degrees in the network. Standard betweenness centrality tends to pick out bottlenecks, and while the node in question does find itself in a bottleneck region of the network, it also has unusually long connections, which link geographically different regions of the graph.
In fact, this node lies at the intersection of multiple 
communities in high-modularity partitions of the power grid network 
by different methods \cite{HAMA10,HAMA11}.
In addition, another node  (marked with a square in Fig.~\ref{fig:floridacond}) exhibits striking behavior in the intermediate $\para_D$ regime in Fig.~\ref{fig:florida_weighted_bet_lines}(a). This node, having the second-highest centrality  at $\para_D \left<L\right> \approx 0.5 $,  is not assigned high centrality in either the betweenness or the current-betweenness limits. Thus, it is only important at moderate levels of parallelism. Both of the salient nodes in the weighted power-grid network will be studied further in Sec.~\ref{sec:nonmon}, which discusses non-monotonic behavior in the conditional current-betweenness centrality.

\subsection{Degenerate and nearly-degenerate paths} \label{sec:degen}

In addition to providing insight about specific networks, the numerical results presented in this section help us understand the workings of conditional current more clearly. In particular, we discuss the properties of the conditional current 
in the presence of degenerate and nearly-degenerate paths in the network.

 \subsubsection{Degenerate and semi-degenerate paths}
 \label{sec:degen1}
 The conditional walker-flow centralities exhibit noteworthy behavior in the presence of degenerate and semi-degenerate paths. (We consider two paths semi-degenerate if they have the same weighted length but different unweighted lengths.) In the case of degenerate paths, at large $\para_D$  the conditional current betweenness reproduces the potentially huge combinatorial weighting that is a consequence of the definition of the standard betweenness centrality. Because the conditional current calculations do not explicitly consider the combinatorics of possible paths, it is noteworthy that the conditional betweenness can still reduce to the standard betweenness in such edge cases. In the case of semi-degenerate paths, convergence to the betweenness centrality sometimes requires a slight modification to the walk matrix $\mathbf{W}$ used to calculate $\cond$ in Eqs.~(\ref{WCanonical}-\ref{conditionalcurrent}). See further details in Appendix \ref{app:degen}.

\subsubsection{Nearly degenerate paths: $\para_D$ controls path-length resolution}
 \label{sec:pathres}

In Sec.~\ref{sec:tuning}, we demonstrated that as we decrease $\para_D$   from $\infty$ to $0$, the conditional current $\cond$ will prefer geodesic paths less and less. Our numerical studies lead to an illuminating interpretation of this behavior: $\para_D$ tunes the conditional walker-flow centralities' ability to distinguish between paths of similar weighted length. In this view, the centralities always prefer nearly geodesic paths, and $\para_D$ controls which path lengths count as ``nearly'' geodesic. At high $\para_D$, only the true geodesics qualify. As $\para_D$ decreases,  longer and longer parallel paths from the start to the target will be indistinguishable from geodesics.

This resolution-tuning effect is illustrated in Fig.~\ref{fig:condcurrent}, where $\cond$ flows from a specific start node (green) to a specific target node (red). There, at $\para_D=66$, almost all of $\cond$ passes through three similarly long paths, each of which goes through node 0. The shortest path goes through node 1, with a weighted length of 1.481. The paths through 2 and 3 have weighted lengths of 1.486 and 1.483, respectively. For comparison, the path that goes directly from 0 to the target node has a weighted length of 1.6.  At $\para_D=66$, the centralities can resolve length differences between the direct 0-to-target path and the other three paths. However, it cannot yet resolve the smaller differences between the paths through 1, 2, and 3, so these three paths have nearly equal values of $\cond$. As the parameter value increases to $\para_D=601$, the centralities begin to distinguish between these three paths, and $\cond$ through node 2 is nearly eliminated. As $\para_D$ grows even larger, all of $\cond$ will pass through the node-1 path, which is the shortest start-target path in the network.

 \begin{figure*}
\includegraphics[scale=1.0, trim={0 0cm 0cm 0cm},clip]{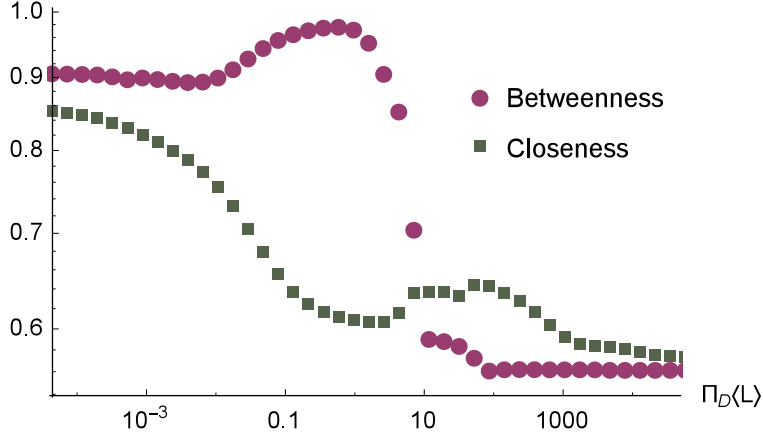}
\caption{\label{fig:flflcorr} \em Pearson  correlations of weighted with unweighted network versions of walker-flow centralities on the Florida power-grid network\/\em. The conditional current-betweenness centrality is represented by circles, while the conditional resistance-closeness 
centrality is represented by squares. Both correlations tend to get larger as $\para_D$, and hence the path-length resolution level, gets smaller. When the centralities are less sensitive to differences in edge weights, the differences between the weighted and unweighted networks are diminished. The conditional current-betweenness correlation maximum occurs at the same value of $\para_D\langle L \rangle$ that produces the large bump in maximum centrality in Fig.~\ref{fig:florida_weighted_bet_lines} because, in the unweighted power-grid network, there is a large gap between the maximum centrality and the other nodes' centralities (see Fig.~\ref{fig:florida_unweighted_bet_lines}).}
\end{figure*}

The resolution-tuning interpretation of $\para_D$ explains the similarity of the current betweenness and resistance closeness on weighted and unweighted versions of the same network. In Fig.~\ref{fig:flflcorr} we present the Pearson correlations of the conditional walker-flow centralities on the weighted Florida power-grid network with those on the unweighted network, across a large range of $\para_D$ values.  (For two centrality measures $c$ and $c^\prime$ on a given network, the  correlation is calculated using $\sum_i (c_i-\langle c \rangle)(c^\prime_i -\langle c^\prime \rangle )/(N \sigma_c \sigma_{c^\prime})$, where the sum runs over all nodes $i$.) For both the conditional current betweenness and the conditional resistance closeness, the correlations tend to increase for smaller values of $\para_D$. At smaller $\para_D$ the centralities are less sensitive to differences in edge weights, so the differences between the weighted and unweighted networks are diminished, and the correlations become large.  In the intermediate range, betweenness is almost independent of the weighting (high correlation), while closeness is quite strongly affected (low correlation). 

The resolution-tuning effect also has practical implications for helping the conditional resistance-closeness centrality converge to the harmonic closeness, in the case of unweighted networks. The method involves adding a small amount of random noise to the edge weights of the network---not enough to substantially affect centrality values at low and intermediate values of $\para_D$. This method is explained more fully in Appendix \ref{sec:randomnoise}. Note that we do not need to add the random noise when calculating the conditional current-betweenness centrality, since in that case, degenerate paths must be included for the centrality to correctly reduce to the betweenness centrality.

\subsection{Non-monotonicity \label{sec:nonmon}}

\subsubsection{Complex behavior in the conditional walker-flow centralities}

 Figs.~\ref{fig:kanga_cond_bet_lines}(a), \ref{fig:karate_cond_bet_lines}(a),  \ref{fig:florida_weighted_bet_lines}(a), and  \ref{fig:florida_unweighted_bet_lines}(a) show that the conditional current betweenness can exhibit complex behavior over the full range of $\para_D$, including non-monotonicity, pronounced maxima and minima, and line crossings that indicate changes in the centrality rankings.  The general reason is that, being based on a conserved walker current, betweenness is a limited resource. 
 Though the conditional current centrality curves are complex, they are produced by a simple process: the tuning behaviors described in Sec.~\ref{sec:tuning} . The tuning is generally monotonic in the sense that increasing $\para_D$ takes conditional current from indirect paths and re-routes it along more direct paths (see Fig.~\ref{fig:condcurrent}). Sec.~\ref{sec:pathres} provides an alternate interpretation of the same tuning process, where the underlying monotonicity can be seen even more clearly. From that perspective, as $\para_D$ is reduced from $\infty$ the walker-flow centralities monotonically reduce the length resolution with which they identify shortest paths. The complex, non-monotonic behaviors seen in the figures emerge from the aggregate of many such simple processes and from the uneven distribution of the conserved walker current. 
 
The behavior of the conditional resistance-closeness centrality [Figs.~\ref{fig:kanga_cond_bet_lines}(b), \ref{fig:karate_cond_bet_lines}(b),  \ref{fig:florida_weighted_bet_lines}(b), and  \ref{fig:florida_unweighted_bet_lines}(b)] is much more regular than that of the conditional  current betweenness. Generally, the conditional resistance closeness  decreases with increasing $\para_D$. This is because increasing $\para_D$  restricts parallel paths of conditional current as in Fig.~\ref{fig:condcurrent}, which leads to higher resistance values in the linear programming procedure of Sec.~\ref{sec:reff}, and thus to lower closeness. The main exceptions to this trend are the small centrality spikes seen in Figs.~\ref{fig:kanga_cond_bet_lines}(b) and \ref{fig:florida_weighted_bet_lines}(b), which are discussed in Appendix~\ref{sec:appSemiDeg}.

\subsubsection{How non-monotonicity in the conditional current betweenness arises in bottlenecks}

 Much of the complex behavior of the conditional current-betweenness centrality can be described as non-monotonicity. Non-monotonic behavior is evident in Figs.~\ref{fig:kanga_cond_bet_lines}(a), \ref{fig:karate_cond_bet_lines}(a),  \ref{fig:florida_weighted_bet_lines}(a), and  \ref{fig:florida_unweighted_bet_lines}(a). The figures show that  the most important nodes often exhibit such non-monotonic behavior---for example the two high-centrality power-grid  nodes discussed in Sec.~\ref{sec:networkresults}. Thus, understanding the non-monotonicity becomes crucial.

 \begin{figure*}
  \begin{overpic}[unit=1mm,scale=.8,trim={0 -.5cm 0cm 0cm},clip]{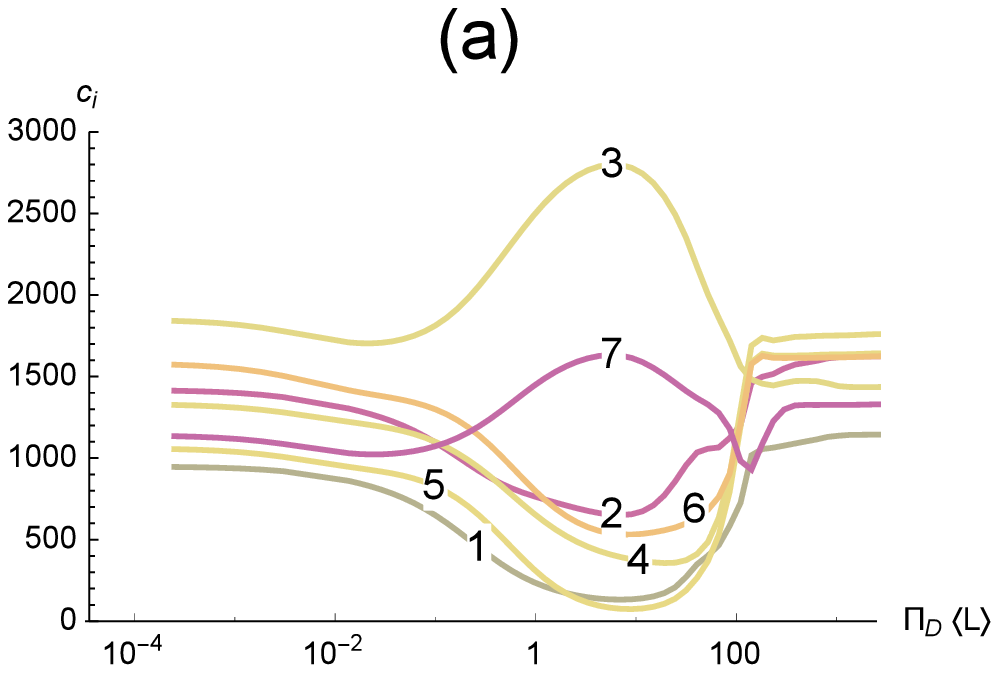}\end{overpic}
 \hspace{0em}
 \begin{overpic}[unit=1mm,scale=.545,trim={0 0cm 0cm 0cm},clip]{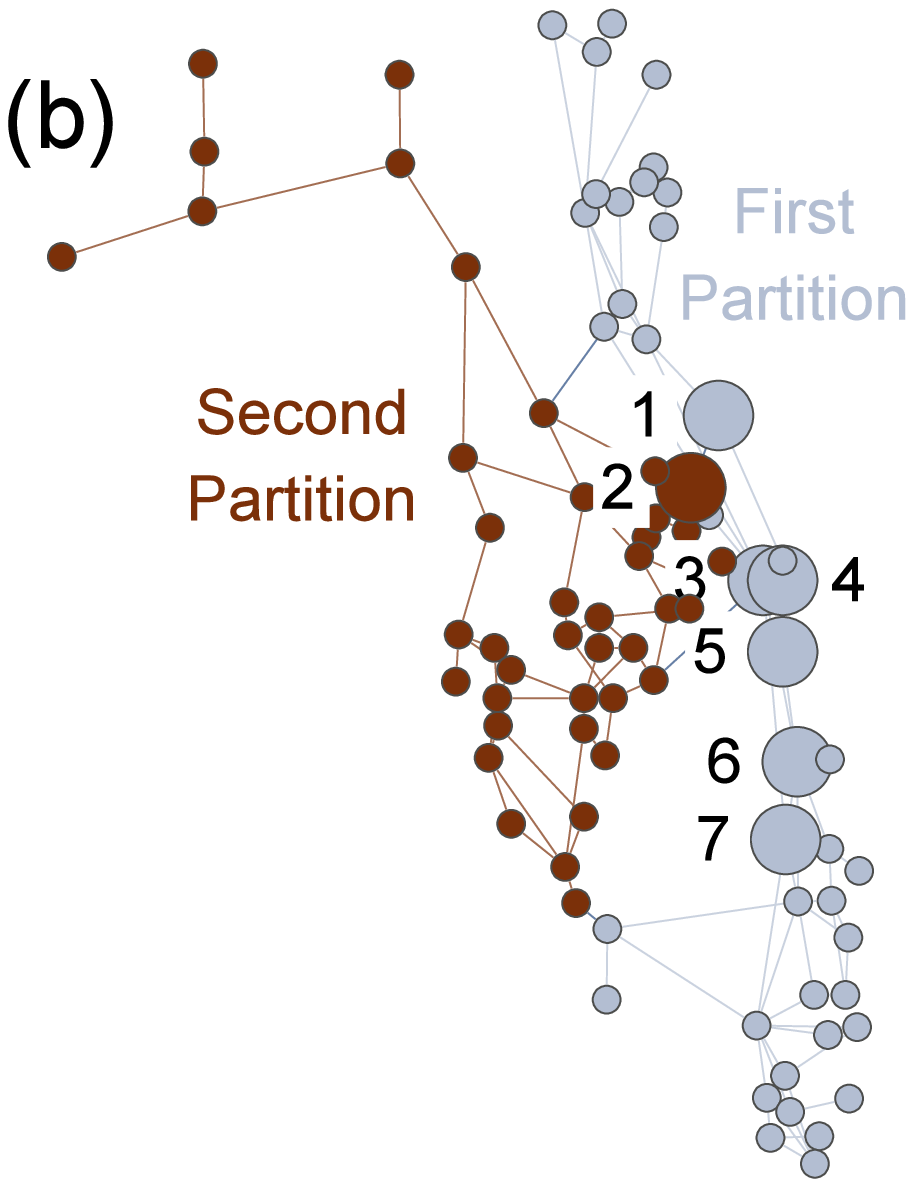}\end{overpic}
\caption{\label{fig:nonmon} \em   Non-monotonicity and partitioning of the weighted Florida power-grid network\em. (a) The conditional current-betweenness centrality curves from Fig.~\ref{fig:florida_weighted_bet_lines}(a) that show pronounced non-monotonicity. (b) Node colors indicate membership in the partitions found using the Kernighan--Lin algorithm. The large, numbered nodes correspond to the curves in (a). See further details in Appendix \ref{sec:nonmon data}.  }
\end{figure*}

It is noteworthy, then, that in our numerical examples, every single node with relatively pronounced non-monotonicity (relative to the network's other nodes) lies at a bottleneck. For example, consider Fig.~\ref{fig:nonmon}, which describes the weighted power-grid network. Part (a)  reproduces the curves exhibiting pronounced non-monotonicity in Fig.~\ref{fig:florida_weighted_bet_lines}(a). Part (b) shows the network nodes divided into two equally-sized partitions using the Kernighan--Lin algorithm \cite{newman2004detecting}. This is a greedy algorithm for finding partitions that minimize the sum of the (weighted) lengths of inter-partition edges, meaning that these edges are bottlenecks between the two halves of the network. (Since the algorithm uses a random start, we took the best partitioning obtained in 1000 runs.) All the most  non-monotonic nodes are bottleneck nodes: they  either lie at the boundary between partitions, or are directly connected to such nodes \cite{noteOnDownstream}. This includes nodes 3 and 7, which are the two high-centrality nodes mentioned earlier (depicted as a triangle and square, respectively, in Fig.~\ref{fig:floridacond}).

This result is very general. In the three other example networks, the nodes with the highest non-monotonicity are always found at---or adjacent to---the partition boundary.  This can even occur for non-central nodes and ones with low absolute non-monotonicity. For example, the node corresponding to the starred curve in Fig.~\ref{fig:karate_cond_bet_lines} lies on the partition boundary of the karate club network. Furthermore, the exact partitioning method is not   important: we obtain the same behavior with partitions derived from the first nontrivial Laplacian eigenvector  \cite{HAMA10}.  In Appendix \ref{sec:nonmon data}, we quantify the non-monotonicity of centrality curves and provide a complete listing of the most non-monotonic nodes in the example data.

 \begin{figure*}
\includegraphics[scale=1.0, trim={0 0cm 0cm 0cm},clip]{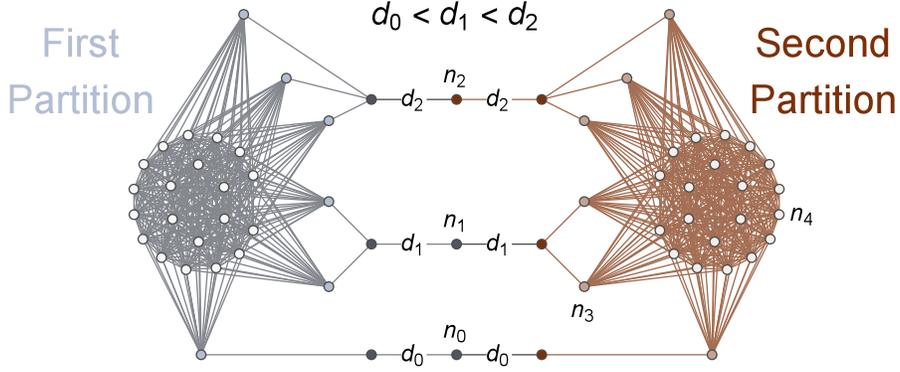}
\caption{\label{fig:artificial} \em   Example network leading to non-monotonic conditional current-betweenness\em. Hue indicates membership in the partitions found using the Kernighan--Lin algorithm. Node lightness/darkness encodes the connection pattern: the lightest nodes form two complete subnetworks of size $N_C$. The  nodes of intermediate lightness have connections to all nodes in the nearest complete subnetwork. The darkest nodes have only a few connections, and thus form bottlenecks.  Edge lengths are labeled with $d$'s. Unlabeled edges all have the same length $d_\mathrm{other}$. }
\end{figure*}

The non-monotonic behavior of nodes on the partition boundary can be understood with reference to the idealized network of Fig.~\ref{fig:artificial}.  There, as in Fig.~\ref{fig:nonmon}, we have two (nearly) equally sized partitions separated by a few bottleneck nodes. Most of each partition is made up of a  complete subnetwork of size $N_C$. Because $N_C$ is large, most of the conditional current betweenness is due to current flowing between the complete subnetworks. The non-monotonicity in node $n_1$ is caused by a trade-off between number of paths  and path length. The most direct paths between subnetworks take only 6 steps. Of these, there are 4 paths through $n_1$ and 9 paths through $n_2$ for every path through $n_0$. This biases the conditional current betweenness to assign  more centrality to node $n_2$ and less to $n_0$. However,  the lengths of all these direct paths are slightly different because $d_0 \lessapprox d_1 < d_2 <d_\mathrm{other}$, which introduces an opposite bias. The effect is that, as $\para_D$ is decreased from $\infty$ to $0$, the centrality of $n_1$ first rises, then falls. 

\begin{figure*}
 \begin{overpic}[unit=1mm,scale=.8,trim={0 0cm 0cm 0cm},clip]{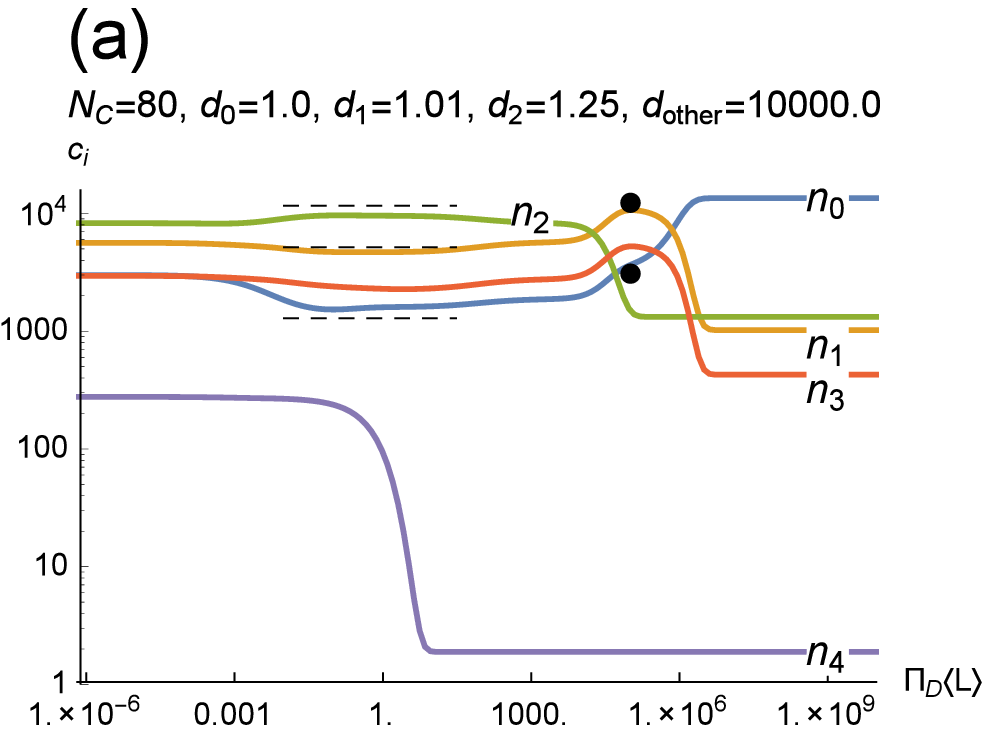}\end{overpic}
 \hspace{0em}
 \begin{overpic}[unit=1mm,scale=.8,trim={0 0cm 0cm 0cm},clip]{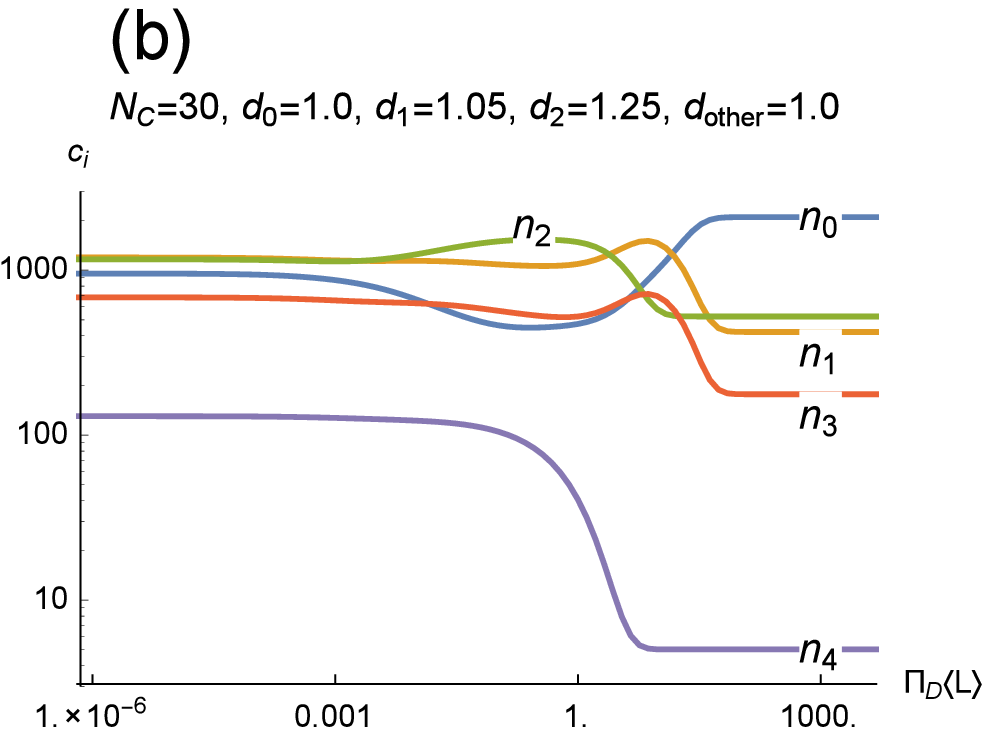}\end{overpic}
\caption{\label{fig:artificial_res} \em Conditional current betweenness for the network in Fig.~\ref{fig:artificial}\em. (a) The vertical coordinates of the dashed lines are in the ratio 1:4:9, equal to the number of short paths from left cluster to right cluster passing through $n_0$, $n_1$, and $n_2$, respectively. For the same reason, the vertical coordinates of the two circles are in the ratio 1:4. The horizontal coordinates of the circles correspond to the smallest $\para_D$ where the centrality of $n_2$ is minimal.  The large values of $N_C$ and $d_\mathrm{other}$ are chosen so that the data approach this ratio in the intermediate $\para_D$ regime. The curves thus exhibit non-monotonic behavior, by design. See the text for discussion. (b) Even at more realistic values of $N_C$ and $d_\mathrm{other}$, the non-monotonic behavior is qualitatively similar.}
\end{figure*}

This can be understood more specifically using the resolution tuning concept  (Sec.~\ref{sec:pathres}) and the observation that, at high $\para_D$, the centrality reproduces the path counting of the standard betweenness (Sec.~\ref{sec:degen1} and Appendix~\ref{app:degen1}). The details of these considerations are borne out by the numerical results in Fig.~\ref{fig:artificial_res}(a). At very high values of $\para_D$ (right-most plateaus in the figure), current  flows mostly over the shortest weighted path, which goes through $n_0$. As a result, $n_1$ and $n_2$ have lower betweenness. As $\para_D$ is decreased, the centrality reaches a resolution level where $d_0$ and $d_1$ are indistinguishable, but both are still distinguishable from $d_2$ and $d_\mathrm{other}$, so the betweenness of $n_2$ is still low. Here, because there are 4 times as many paths through $n_1$ as $n_0$, the centrality of the former is about 4 times as big. (See the circles in the figure, which are placed  on the horizontal axis just before $n_2$'s centrality begins to rise. The vertical coordinates are in the ratio of 4:1.) As $\para_D$ is decreased even further, $d_1$, $d_2$, and $d_3$ are all indistinguishable from each other, but still distinguishable from $d_\mathrm{other}$. Hence, the betweenness values for $n_0$, $n_1$, and $n_2$ are close to the ratio 1:4:9 (as indicated by the dashed lines). The centrality rank of $n_1$ goes from third, to first, to second. Because of current conservation, this means that the centrality curves exhibit extrema \cite{noteOnMaximum} and are thus non-monotonic.

In Fig.~\ref{fig:artificial_res}(a), we chose  large values of $N_C$ and $d_\mathrm{other}$ to match the above theoretical betweenness ratios  (encoded by the circles and dashed lines). However, as shown in part (b) of the figure, the qualitative features of (a)---especially the non-monotonicity---persist even with more realistic parameters. We  propose that the cause of non-monotonicity in the real example networks is qualitatively the same as that in the idealized network of Fig.~\ref{fig:artificial}. That is, non-monotonicity is caused by the competition of different-length and differently-connected paths between large network communities. In realistic networks, there may be more than two large communities and more than three paths between them, resulting in more complex non-monotonic behavior than is seen in Fig.~\ref{fig:artificial_res}. Nonetheless, the real networks  studied here accord with the non-monotonicity model of Fig.~\ref{fig:artificial}, including  the placement of non-monotonic nodes on partition boundaries, as well as the reduced occurrence of non-monotonicity in non-weighted graphs (Figs.~\ref{fig:karate_cond_bet_lines} and \ref{fig:florida_unweighted_bet_lines}), which are less likely to have small differences in path-length. 

These considerations can shed light on the behaviors of the high-centrality nodes described in Sec.~\ref{sec:networkresults}. In our examples, large changes in centrality are almost always consequences of non-monotonicity, so unweighted networks are  more likely to have stable centralities. This is exactly what happens in the (unweighted) karate club network (Fig.~\ref{fig:karate_cond_bet_lines}), where the two highest-centrality nodes maintain their rank across the entire range of $\para_D$.

Figure \ref{fig:artificial_res} also gives  insight into two types of nodes that are not high-centrality bottlenecks. The first, $n_3$, does not have the highest centrality rank at any value of $\para_D$. Furthermore, it does not lie on the network's partition boundary. However, it is directly connected to $n_1$, which does lie on the boundary. As a result, $n_3$  receives half of that node's  walker current, and therefore inherits  its non-monotonicity as well. This is analogous to nodes 4, 5, 6, and 7 in Fig.~\ref{fig:nonmon}, which inherit the non-monotonicity from  node 3, which has much higher centrality than any other in the intermediate $\para_D$ regime. Second,  $n_4$ is a node that lies {\it within} a  community, far from the partition boundary. Therefore, it has an entirely monotonic centrality curve, much like the monotonic nodes of the kangaroo network and weighted power-grid network, seen in Figs.~\ref{fig:kanga_cond_bet_lines}(a) and \ref{fig:florida_weighted_bet_lines}(a), respectively. 

\section{Conclusion} \label{sec:conc}
We have shown that the class of walker-flow centralities includes important and commonly known centrality measures.
	The walker-flow centralities that are most frequently encountered in the literature admit
		 a natural parametrization scheme, based on the walker death parameter $\para_D$, which interpolates between the measures in the left and right columns of Tables \ref{curbettab} and \ref{curclostab}. Our conditional current-betweenness centrality interpolates from random-walk betweenness (equivalently, current betweenness) at $\para_D=0$ (no walker death), to standard betweenness as $\para_D \to \infty$ (walker death likely). Our conditional resistance-closeness centrality interpolates from harmonic information centrality (equivalently, resistance closeness)  at $\para_D=0$, to harmonic closeness as $\para_D \to \infty$. We believe our absorbing walker-flow method is the first to interpolate simultaneously across both the betweenness and the closeness continua.
		 
The interpolating centralities are also meaningful at intermediate values of $\para_D$. As $\para_D$ increases from zero, the centralities prioritize geodesic and nearly-geodesic paths. Therefore, parallelism in the network becomes less and less important. Finally, as $\para_D \to \infty$, parallel paths between any pair of nodes are ignored  (unless they have degenerate lengths). From this perspective, $\para_D$ can be seen as a tuning parameter that sets the centralities' preference for geodesic paths and thus the level of parallelism. 
This can be informative of the structure of specific networks, since some centrality features only appear in limited $\para_D$ regimes. For example, in Sec.~\ref{sec:networkresults}, we discuss how a particular node in the Florida power-grid network becomes important only in the intermediate $\para_D$ regime.  Generally, non-monotonic behavior of conditional current betweenness in the intermediate $\para_D$ regimes is found in nodes that lie close to a community boundary, which experience a competition between (a) lying on many paths and (b) lying on short paths.

We have argued that considering the entire range of $\para_D$ can be revealing about a network's structure, but it can also be useful to pick out a single ``best'' parameter value for a given network. In previous work on the Florida power-grid network \cite{xu2014architecture,GURF15}, we found a strong correlation between  the known generation capacities of power plants and the values of a centrality based on Estrada's communicability \cite{estrada2008communicability,estrada2009communicability}. The best-fit parameter in the communicability centrality can be viewed as a measure of a length scale inherent in the network. In future reports, we will describe how several different centrality measures  also reveal the same length scale. We similarly plan to use centrality-matching to reveal ``best'' values of $\para_D$.


The present work has focused on  general features of the centralities' numerical results. These reveal a second effect of the interpolation parameter: in addition to controlling geodesic-path preference, it controls the centralities' path-length resolution ability. At low values of $\para_D$, the centralities cannot distinguish between geodesic and nearly geodesic paths. This implies that standard current betweenness and resistance closeness centralities, obtained at $\para_D=0$, are not  sensitive to small variations in edge weights in complex networks, as seen in Fig.~\ref{fig:flflcorr}. Our method provides versions of betweenness and closeness centralities with an adjustable resolution level.

\begin{acknowledgements}

We thank Y. Murase, K.~S. Olsen, G. M. Buend\'\i a and S. Gallagher for useful comments. We also thank the anonymous reviewers for particularly careful and insightful criticism. 

Supported in part by U.S.  National Science Foundation Grant No. DMR-1104829. 

Work at the University of Oslo was  supported by the Research Council of Norway 
through the Center of Excellence funding scheme, Project No. 262644.
\end{acknowledgements}

\appendix
\section{Derivation of Eqs. (\ref{eq:tranprob}) and (\ref{eq:dedprob})}
\label{app:cont}
   Consider the absorbing random walk on a chain of $n_\mathrm{edge}-1$ intermediary nodes  depicted in Fig.~\ref{fig:longline}(a). The situation describes a random walker attempting to cross a long edge $(a,b)$ with constant death probability at every intermediary node. The walker begins on the first intermediary node to the right of $a$ and can absorb on $a$ (transmission failed), $b$ (transmission succeeded), and the ``death'' node (walker died).     Here, the difference between transmission failure and walker death is that, in the former case, the walker can try again: a new transmission attempt will start on some edge $(a,c)$, where $c$ is a neighbor of $a$ in the original network. Standard random-walk dynamics require that the death probability at every intermediary node be $p=w_D/(w_D + 2 w)$, while the probability of moving along each of the two intermediary edges is $w/(w_D + 2 w)$. [See Eq.~(\ref{eq:intermed}) and the following sentence for the definitions of $w$ and $w_D$.]
   
The probability of successful transmission  in a \em single \em attempt, $p_T(a,b)$, is found using standard methods \cite{logan2013applied}. We  solve the following linear difference relation for $p_{T;k}$, the probability  of transmission over $(a,b)$ given a start on  intermediary node $k$:
\begin{equation}
 \label{eq:difrel}
p_{T;k} = \frac{1-p}{2}p_{T;k-1}+\frac{1-p}{2}p_{T;k+1}
 \end{equation} 
 Let $k=0$ correspond to node $a$ and $k=n_\mathrm{edge}$ correspond to node $b$. The boundary conditions become $ p_{T;0}=0$ and $p_{T;n_\mathrm{edge}}=1$. This leads to
 \begin{equation}
 \label{eq:1tranprob}
 p_T = p_{T;1}= \frac{2}{1-p}\frac{  \sqrt{2 p-p^2}}{\left(\frac{1 +\sqrt{2 p -p^2}}{1-p}\right)^{n_\mathrm{edge}}-\left(\frac{1 -\sqrt{2 p -p^2}}{1-p}\right)^{\vphantom{E^E}n_\mathrm{edge}}},
 \end{equation}
 where $p$ is a function of $w$ and $w_D$.

To obtain the continuum limit, $n_\mathrm{edge}$ will increase to infinity. Therefore, $w$ and $w_D$ must be described in terms of quantities per unit length. Analogy with the lossy transmission line model from power engineering \cite{steer2010microwave} suggests these quantities to be the ground conductance per unit length, $G$,  and the line resistance per unit length, $R$. The correspondence between electrical networks and random walks \cite{Doyle06randomwalks} then implies that $w_D=G\Delta x$ and $w=(R \Delta x)^{-1}$, where $\Delta x = d_{(a,b)}/n_\mathrm{edge}$.

Expansion in terms of $\Delta x$ results in

\noindent
 \begin{equation}
 \label{eq:1tranproblow}
 p_T(a,b) = \frac{ \sqrt{G R}\, \Delta x}{\sinh({ d_{(a,b)}\sqrt{G R})}} + \mathcal{O}(\Delta x^2).
 \end{equation}
Reversing the boundary conditions for Eq.~(\ref{eq:difrel}) results in $p_R(a,b)$, the probability that the walker will return to $a$ before reaching $b$:
  \begin{equation}
 \label{eq:1retproblow}
 p_R(a,b) = 1-\frac{ \sqrt{G R}\, \Delta x}{\tanh({ d_{(a,b)}\sqrt{G R})}} + \mathcal{O}(\Delta x^2). 
 \end{equation}

As remarked earlier, $p_T(a,b)$ and $p_R(a,b)$ describe only a \em single \em attempt at transmission over the edge $(a,b)$.   The \em final \em transmission probability $p_{(a,b)}$ can include failed attempts to reach any nearest neighbor of $a$; so long as the walker returns to $a$ rather than dying, it can try again. What matters is that the ultimately successful transmission occurs over $(a,b)$. This reasoning is captured in the recursive equation 
 \begin{equation}
  \label{eq:finalprob}
 p_{(a,b)}=k_a^{-1}\left(\sum_{l\sim a} p_R(a,l) p_{(a,b)} + p_T(a,b) \right).
 \end{equation}
Here, the sum is over nearest neighbors of $a$  and  the inverse unweighted degree factor $k_a^{-1}$  comes from the random choice of the first edge the walker attempts to cross.  

Solving the linear equation (\ref{eq:finalprob}) for $p_{(a,b)}$ and substituting the lowest-order terms from Eqs. (\ref{eq:1tranproblow}) and  (\ref{eq:1retproblow}) results in
\begin{equation}
\label{eq:totprobapp}
p_{(a,b)}=\frac{[\sinh(\sqrt{GR}\,d_{(a,b)})]^{-1}}{ \sum_{l\sim a} [\tanh(\sqrt{GR}\,d_{(a,l)})]^{-1}}.
\end{equation}
Note that the dependence on the granularity parameter $\Delta x$ has canceled out. This cancellation further justifies the use of the physically-motivated parameters $G$ and $R$ in the per-step death probability $p$: the cancellation does not occur if we instead choose a constant death probability per unit length.

\begin{figure*}
\includegraphics[scale=.9, trim={0 0cm 0cm 0cm},clip]{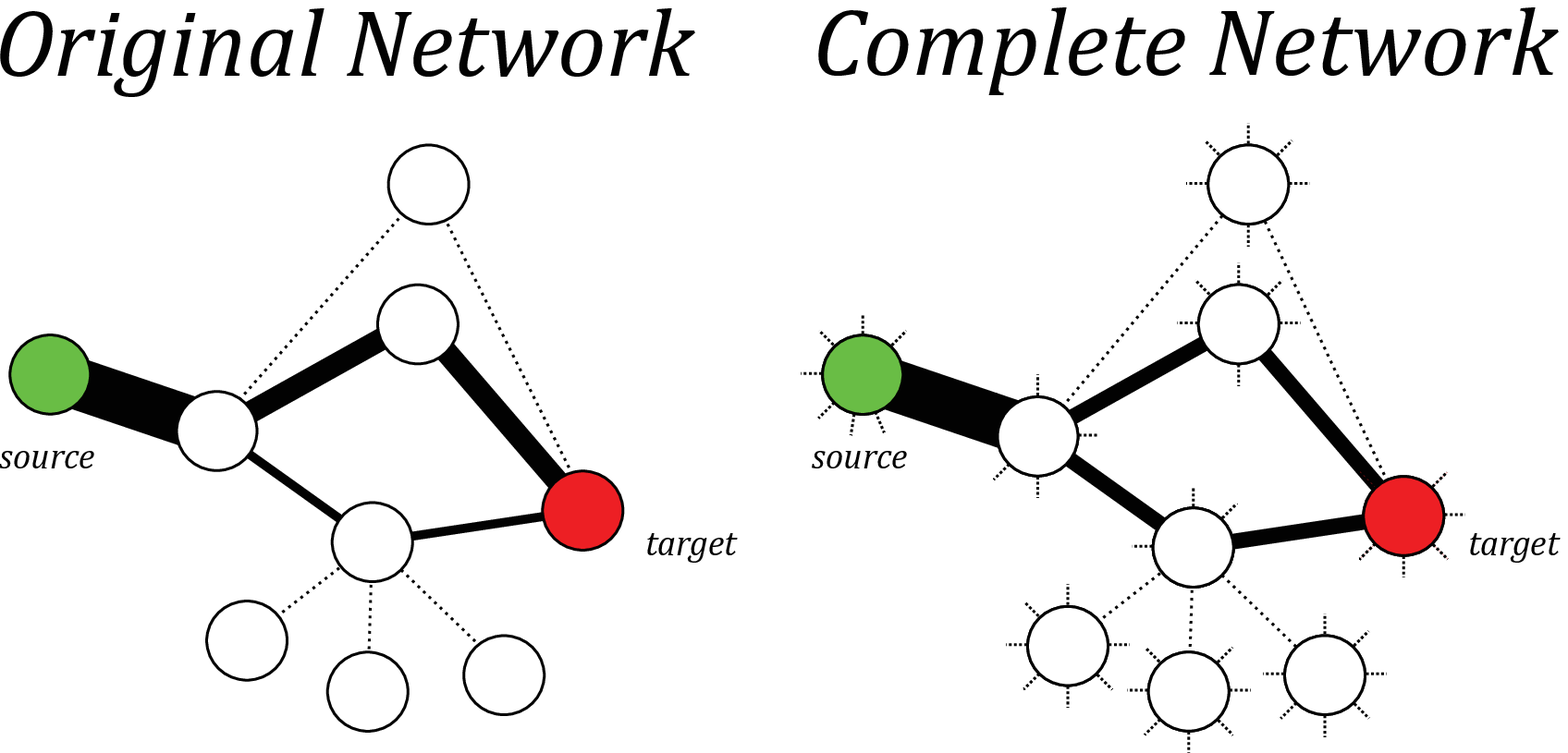}
\caption{\label{fig:degen} \em Conditional current flow in the case of degenerate shortest paths in an unweighted network\/\em. The conserved walker current from ``source'' to ``target'' in a simple example graph is illustrated for large $\para_D$ ($\para_D=1000)$. Edge current magnitude is proportional to line thickness, and  infinitesimal current is depicted with dotted lines. Conditional current $\cond$  flows only on shortest paths from ``source'' to ``target''. If (left  side) transmission probabilities are given by Eq.~(\ref{eq:totprobapp}), then less current will flow on geodesics that contain higher-degree nodes. When transmission probabilities are given by Eq.~(\ref{eq:totprobappdone}) (right side), all degenerate geodesics carry equal currents because all nodes have degree $N-1$. In this case the network is described by a complete graph, but the edges not present in the original network have infinite length and, therefore, no conditional currents. }
\end{figure*}

A final consideration is that Eq.~(\ref{eq:totprobapp}) leads to unwanted behavior in the case of unweighted networks with degenerate (equal length) paths. Figure  \ref{fig:degen} (left) shows the conditional current $\cond$ in a simple example-network for large values of $\para_D=\sqrt{G R}$. The figure illustrates that while $\cond$ is restricted to geodesics, it is smaller for paths that include higher-degree nodes. The solution is to replace all non-edges in the network with edges of infinite length. In effect, this gives all nodes the same unweighted degree of $N-1$. As a result, degenerate geodesics will share equal conditional currents, as  shown in Fig.~\ref{fig:degen} (right). (However, we continue to use $k_a$ to refer to the {\it unweighted\/} degree of node $a$: 
$k_a= \sum_{l\sim a}1$.) With this change, Eq.~(\ref{eq:totprobapp}) becomes 
\begin{equation}
\label{eq:totprobappdone}
p_{(a,b)}=\frac{[\sinh(\sqrt{GR}\,d_{(a,b)})]^{-1}}{N-1 -k_a+ \sum_{l\sim a} [\tanh(\sqrt{GR}\,d_{(a,l)})]^{-1}},
\end{equation}
which leads to Eqs. (\ref{eq:tranprob}) and (\ref{eq:dedprob}).

\section{Feasibility of the linear programming problem for $\mathfrak{R}$}
\label{app:feasible}

To show that the linear programming problem of Eq.~(\ref{linprog}) is \em feasible \em is to show that there exists a  solution $\{R^\cond_\nu\}$ that does not necessarily minimize $R^\mathrm{eff}$. If a node potential mapping $\{V^\cond_l\}$ can  be found to reproduce the \em conditional \em currents  as physical currents, $I=\cond$, then  $\sum_{r} \mathbf{K}_{r\,\nu}\cond_\nu R^\cond_\nu =0$ is trivially satisfied for all independent cycles $r$ because $\cond_\nu R^\cond_\nu$ is edge $\nu$'s potential drop $V^\cond_\nu$, and the sum of potential drops around a cycle must be zero. Indeed, the condition in question is just a re-statement of Kirchhoff's Voltage Law. 

A \em directed acyclic graph \em always admits  a topological ordering $\mathcal{O}$ on the nodes, such that any directed edge $\nu=(a,b)$ satisfies $\mathcal{O}_a >\mathcal{O}_b $ (edges  point from higher to lower order). Below, we  prove that the conditional current $\cond$ results in  a directed acyclic graph. The topological ordering obtained from the graph of $\cond$s can be converted into a consistent potential mapping by assigning $V^\cond_a>V^\cond_b$ whenever $\mathcal{O}_a >\mathcal{O}_b$. The value of $R^\cond_\nu$ is then chosen to satisfy $V^\cond_\nu =V^\cond_a-V^\cond_b= \cond_\nu R^\cond_\nu$. Finally, the potential of every node can  be scaled to ensure that $R^\cond_\nu \ge R^\mathrm{orig}_\nu$ for all $\nu$, and Eq.~(\ref{linprog}) is proven feasible.

The conditional current mapping clearly defines a directed graph. We show that the resulting graph is acyclic through contradiction. Assume that  nodes $k$ through $k+m-1$ form a directed cycle of $m$ edges, such that $\cond$ flows from $l$ to $l+1$ for $l\in[k,k+m-1]$. (Because this is a cycle, nodes $l$ and $l+m$ are equivalent.) The previous statement, in light of Eq.~(\ref{conditionalcurrent}), becomes

\begin{equation}\label{conditionalcurrent2}
\begin{array}{ccccc}
F_{i\;l} \mathbf{T}_{l\;l+1} F_{l+1\;j}   & > &F_{i\;l+1} \mathbf{T}_{l+1\;l}F_{l\;j}   &&\\
&\Updownarrow&&\\
F_{i\;l} \frac{[\sinh(\sqrt{GR}\,d_{(l,l+1)})]^{-1}}
{g(l)} F_{l+1\;j} 
& > 
&F_{i\;l+1} 
\frac
{[\sinh(\sqrt{GR}\,d_{(l+1,l)})]^{-1}}
{g(l+1)} F_{l\;j} &&\\.
\end{array}
\end{equation}
for all $l\in[k,k+m-1]$. Here, $\mathbf{T}$ is substituted from Eq.~(\ref{eq:tranprob}), from which we define $g(l)=N-1-k_l + \sum_\mu[\tanh (\sqrt{G R}d_\mu)]^{-1}$, where the sum runs over edges incident on node $l$.  Noting that $d_{(l,l+1)}=d_{(l+1,l)}$, the above can be rewritten as $f(l)>f(l+1) $ where $f(l)= F_{i\; l}\left( g(l)  F_{l\; j} \right)^{-1}$. The inequalities form a chain: $f(l)>f(l+1)>\cdots>f(l+m)=f(l)$, which  is a contradiction. Therefore, $\cond$ always results in a directed acyclic graph.

\section{Degenerate and semi-degenerate paths}
\label{app:degen}

 \subsection{Degenerate paths}
 \label{app:degen1}
 \begin{figure*}
\includegraphics[scale=.7, trim={0 0cm 0cm 0cm},clip]{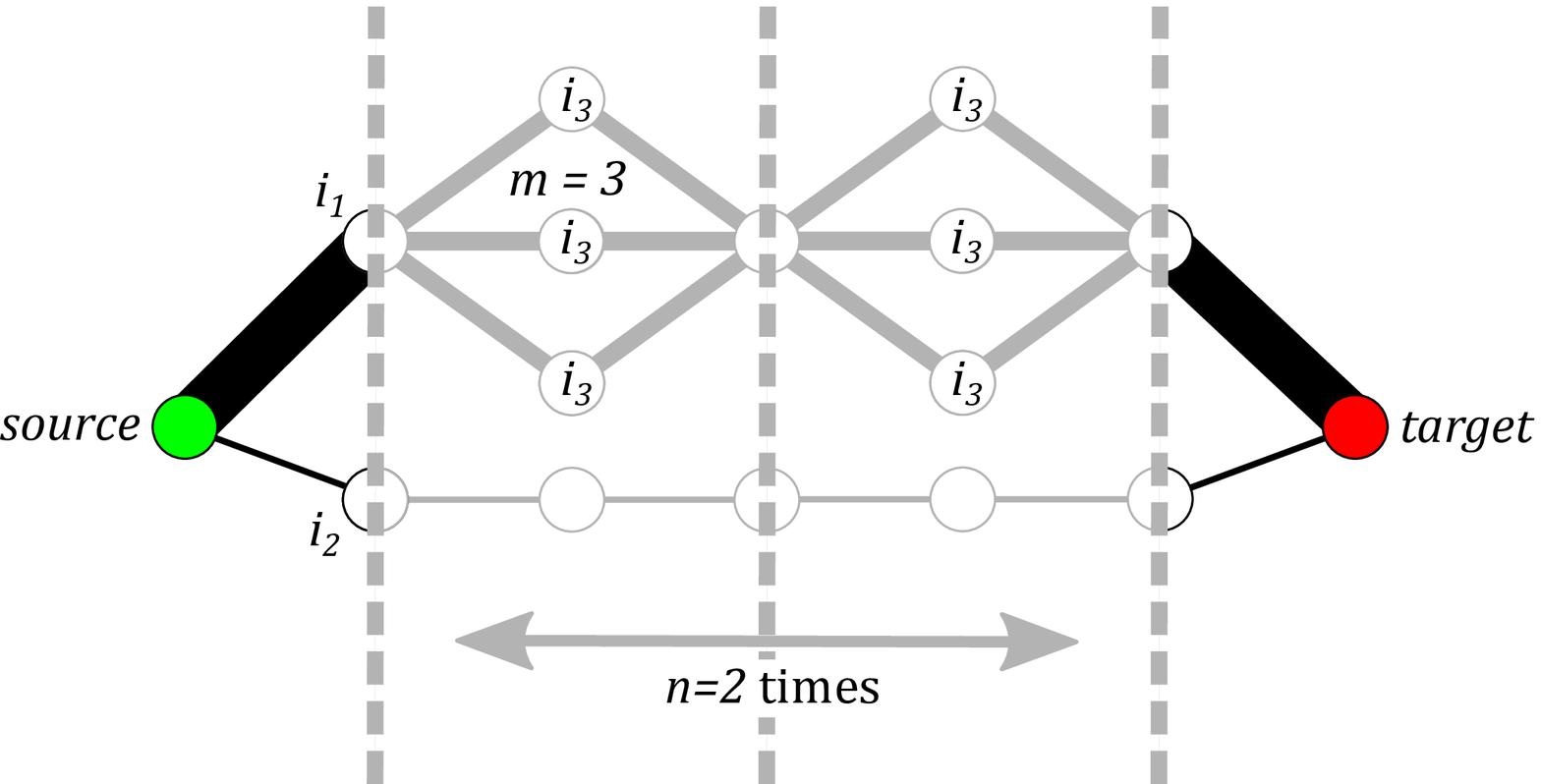}
\caption{\label{fig:degen_braids} \em Example unweighted network with many degenerate paths from source to target\/\em. The graph distance between the  source and target nodes is $2(n+1)$. There are $m^n$ times as many geodesics of this length passing through node $i_1$ as there are through node $i_2$. Because $n_{\mathrm{source},i,\mathrm{target}}$ in the betweenness formula [Eq.~(\ref{eq:bet})] counts the total number of geodesics passing through $i$, the contribution to $i_1$'s betweenness centrality from this (source, target) pair is $m^n$ times the contribution to  $i_2$'s betweenness. Our conditional current-betweenness centrality reproduces this result at large values of $\para_D$. Here, $m=3$ and $n=2$ is illustrated. Line thicknesses are proportional to the number of geodesic paths along an edge, as well as the conditional current at large $\para_D$. Node $i_3$ can be taken to be any of the $m n$ nodes in that position, and is discussed further in the text.
}
\end{figure*}

In the case of many degenerate paths, the standard betweenness centrality [Eq.~(\ref{eq:bet})] can exponentially prefer some nodes over others, even when they both lie on geodesics. Consider the example network in Fig.~\ref{fig:degen_braids}. There, geodesics between the source and target nodes have graph distance $2(n+1)$. However, there are $m^n$ times as many geodesics  passing through node $i_1$ as there are through node $i_2$. Because $n_{\mathrm{source},i,\mathrm{target}}$ in the betweenness formula  counts the total number of geodesics passing through $i$, the contribution to $i_1$'s betweenness centrality from this (source, target) pair is $m^n$ times the contribution to  $i_2$'s betweenness. 

The conditional current-betweenness centrality reproduces this behavior at large $\para_D$, without having been explicitly designed to do so. Because all the nodes in the network lie on geodesics, no nodes will have zero conditional current $\cond$. However, $\cond$ through $i_1$  is $m^n$ times as large as $\cond$ through $i_2$, so the relative contributions to current betweenness are the same as they are in standard betweenness. In that case, by symmetry and conditional current conservation, $\cond$ through $i_3$ is  $m^{(n-1)}$ times as large as $\cond$ through $i_2$, where $i_3$ can be any of the $m n$ nodes compatible with the position of $i_3$ in the figure. In the other extreme, at low $\para_D$, the conditional current is more evenly shared. At $\para_D=0$, the conditional current becomes identical with the physical current on the corresponding resistor network. In the large $n$ limit, this means that $\cond$ through $i_3$ is identical to $\cond$ through $i_2$, while $\cond$ through $i_1$ is $m$ times as large.

\subsection{Semi-degenerate paths} \label{sec:appSemiDeg}
 
 Consider two paths of the same {\it weighted} length $d_\mathrm{path}$ from source $i$ to target $j$, and calculate $\cond$ in the high $\para_D$ limit. If the two paths also have the same {\it un}weighted length, $\cond$ will be  equal on the two paths. However, if the paths have different unweighted lengths (are {\it semi-degenerate\/}), $\cond$ will not be equal. This can be seen from the formula for transition probability along edge $\nu$ [Eq.~(\ref{eq:tranprob})] which, in the high $\para_D$ limit,  reduces to $p_\nu= \exp(-\para_D d_\nu)/(N-1)$. In this limit, the conditional current $\cond_\mathrm{path}$ along a (weighted) shortest path is proportional to the product of edge transition probabilities along the path. Therefore, 
 \begin{equation}
 \label{eq:semidegen}
 \cond_\mathrm{path} \propto \exp(-\para_D d_\mathrm{path})/(N-1)^{n_\mathrm{path}},
 \end{equation}
  where ${n_\mathrm{path}}$ is the number of non-fictitious nodes along the path. 
  
Eq.~(\ref{eq:semidegen}) means that, in the high $\para_D$ limit, while conditional current will flow along a path if and only if it is a weighted shortest path, {\it more} conditional current will flow along the paths that involve the fewest nodes. Occasionally, this can lead to conditional current betweenness failing to converge to betweenness in the high $\para_D$ limit. The only example of this in our numerical studies can be seen for the integer-weighted kangaroo network in Fig.~\ref{fig:kanga_cond_bet_lines}(a), where in the bottom right corner, one data-point indicating non-zero betweenness does not match up with the corresponding conditional current betweenness curve, which goes to zero. 

In principle, this convergence problem for semi-degenerate paths can only occur in weighted networks (in unweighted networks $d_\mathrm{path}=n_\mathrm{path}$). Furthermore, it cannot occur for continuously weighted networks  because it is overwhelmingly unlikely that two different paths would have precisely the same weighted length. For the same reason, the convergence of the conditional resistance distance is unaffected, since in this case the addition of a small amount of random noise effectively creates a continuously weighted network. Of all realistic networks, the problem primarily occurs in networks that have integer edge lengths, up to a constant factor. One way around this difficulty is to introduce {\it macroscopic} intermediary nodes such that, after  the addition of the new nodes, every edge has the same length. [Unlike the microscopic intermediary nodes of Sec.~\ref{sec:rootgr}, the macroscopic nodes {\it are } included in the walk matrix $\mathbf{W}$ of Eq.~(\ref{WCanonical}). However, since they are not real network nodes, they are not included in the centrality matrix $\mathbf{M}$ of Eq.~(\ref{centrality}).] This effectively transforms the integer-weighted network into an unweighted network, which removes any issues with semi-degenerate paths. However, finding a single version of our conditional current that gives correct results for all types of weighted networks is a priority for future research.
 
The conditional walker-flow centralities also prefer shorter unweighted paths in the case of merely approximate semi-degeneracy, though this does not affect convergence to the limiting centralities (betweenness, current betweenness, closeness, and resistance closeness). Consider a network with only two paths from $i$ to $j$; path 1 has a slightly longer weighted length than path 2, but a shorter unweighted length. The two paths are thus approximately semi-degenerate. When $\para_D$ is low enough that the difference between the two weighted lengths cannot be resolved, path 1 will carry more conditional current $\cond$. As the centrality's resolution increases with $\para_D$, more and more of the conditional current will flow along path 2. At some value of $\para_D$, $\cond$ will be equal across the two paths. At this point, the effective resistance $\mathfrak{R}_{ij}$  will be lowest  because $\cond$ mimics current flow for two resistors in parallel. In networks with more than two paths, a similar phenomenon causes the small spikes in nodes' resistance closeness, as can be seen in Figs.~\ref{fig:kanga_cond_bet_lines}(b) and \ref{fig:florida_weighted_bet_lines}(b). 

\subsection{Random noise ensures reduction of conditional resistance-closeness to closeness}
\label{sec:randomnoise}

Table~\ref{curclostab} shows that, for the conditional resistance-closeness to reduce to the harmonic closeness centrality at high $\para_D$,   $\mathfrak{R}_{ij}$  must equal $d_{ij}$ in that parameter regime.
Since unweighted networks generally have multiple equal length (degenerate) paths between a given source $i$ and target $j$,  the linear programming method assigns a  value of effective resistance $\mathfrak{R}_{ij}$ lower than that of the graph distance $d_{ij}$---parallel paths lower the resistance. We  thus add a small amount of random noise to every edge weight, changing the network from unweighted ({\it i.e.\/}, unit edge weights) to weighted. This creates a single shortest path from $i$ to $j$, whose length is approximately $d_{ij}$. Therefore, at large values of $\para_D$,  we find $\mathfrak{R}_{ij}\approx d_{ij}$.  The amount of random noise is too small to be resolved at anything but very large values of $\para_D$, so it does not affect our results when $\para_D$ is not large. At  large values of $\para_D$, the noise is resolved, and the centrality reduces to closeness centrality.

The resolution-tuning effect of $\para_D$ is evident in the plateau regions  in Figs.~\ref{fig:karate_cond_bet_lines}(b) and \ref{fig:florida_unweighted_bet_lines}(b), for example between $\para_D \langle L \rangle \approx 1$ and $\para_D \langle L \rangle \approx 100$ in Fig.~\ref{fig:florida_unweighted_bet_lines}(b).  In such plot regions, where most of the curves are approximately constant, even as $\para_D$ increases the differences in path lengths are not large enough to be resolved by the centrality. The end of the plateau in Fig.~\ref{fig:florida_unweighted_bet_lines}(b) corresponds to the value of $\para_D$ at which the centrality is capable of resolving the  random noise. 

Without the addition of random noise, the plateaus would extend to arbitrarily large values of $\para_D$. The resulting centrality can be viewed as an alternative closeness measure, where only shortest paths contribute, but the presence of degenerate paths is taken into account and makes the source and target ``closer''. This is because the alternative closeness considers flows rather than single travelers. The harmonic closeness does not distinguish between situations in which there is a unique shortest path of length $l$ and where there are many degenerate shortest paths of length $l$.

\section{Non-monotonicity data tables}
\label{sec:nonmon data}
In Sec.~\ref{sec:nonmon}, we explained how non-monotonicity in the conditional current betweenness arises in nodes that experience a competition between (a) lying on many paths and (b) lying on short paths.  To quantify the non-monotonicity of centrality curves $f(\para_D)$, we rely on the {\it Lack of Monotonicity } (LOM) index  \cite{davydov_zitikis_2017}, a functional given by
\begin{equation}\label{eq:lom}
\mathrm{LOM}(f(x))= 2 \min\left( \int_0^\infty (f'(x))^+ \mathrm{d}x,\left| \int_0^\infty  (f'(x))^- \mathrm{d}x \right| \right).
\end{equation}
The $+$ and $-$ superscripts pick out, respectively, the positive and negative part of a function $g$. Specifically, $(g(x))^+=\max(g(x),0) $ and $(g(x))^-= \min(g(x),0)$. A monotonic function results in a LOM  of 0, while heavily non-monotonic functions have large LOM.

 Tables \ref{tab1}-\ref{tab4} list the nodes in the four example networks in order of decreasing LOM  for the conditional current betweenness, along with the nodes' relationships to the partition boundary. For every network, we list up to the 10 highest-LOM nodes---so long as the LOM is above zero. The nodes with the top-ranked LOM index invariably appear on the partition boundary. The nodes in the top 5\% of LOM rankings appear either on the boundary or adjacent to it. 
 
 All the partition boundaries in the tables are found using the Kernighan--Lin algorithm. As discussed in Sec.~\ref{sec:nonmon}, the presence of high LOM nodes on or near partition boundaries is reproduced with an alternative partitioning scheme based on the smallest non-trivial eigenvector of the graph Laplacian (the Fiedler vector). The alternative partitioning is a modified form of the standard Laplacian partitioning  \cite{newman2010networks}: a constant  is added to every element of the Fiedler  vector to ensure partitions with equal (or approximately equal) numbers of nodes. The relationship between non-monotonic nodes and boundaries is not observed for partitioning methods that do not produce two equally-sized partitions---such as the unmodified Laplacian method and the multi-community islanding method from \cite{HAMA11}. This may be because the conditional current betweenness (see Table~\ref{curbettab}) assigns equal weight to all source/target pairs $(s,j)$. Bottlenecks between two equally-sized partitions will experience the most conditional current flow once all pairs are accounted for, similar to the bottlenecks in Fig.~\ref{fig:artificial}.

\begin{table}[h]
\begin{center}

	\caption{{\it Non-monotonicity data for the kangaroo network.} Of the 9 nodes not appearing in the table, all have zero LOM. Of these, 2 are not boundary nodes. (Because of the tight-knit network structure, all of the network's nodes are either on or adjacent to the partition boundary. )
		}
	  	\label{tab1}
\begingroup
\renewcommand{\arraystretch}{0.6}
\begin{tabular}{cccc}

 \rule{0pt}{3ex}\text{LOM Index }\, & \text{Partition}\, & \text{Boundary Node}\, & \text{Boundary Neighbor}
 \\ \hline
\rule{0pt}{3ex}22.09 & 2 & \text{YES} & \text{NO} \\
 20.70 & 1 & \text{YES} & \text{NO} \\
 16.17 & 1 & \text{YES} & \text{NO} \\
 14.27 & 2 & \text{YES} & \text{NO} \\
 3.43 & 2 & \text{YES} & \text{NO} \\
 0.69 & 1 & \text{YES} & \text{NO} \\
 0.10 & 1 & \text{YES} & \text{NO} \\
 0.01 & 1 & \text{YES} & \text{NO} \\

\end{tabular}\endgroup\end{center}\end{table}

\begin{table}[h]
\begin{center}
	\caption{{\it Non-monotonicity data for the karate club network.} Of the 32 nodes not appearing in the table, all have zero LOM. Of these, 21 are not boundary nodes. (Because of the tight-knit network structure, all but one of the network's nodes are either on or adjacent to the partition boundary.) The  highest-LOM node is denoted with a star in Fig.~\ref{fig:karate_cond_bet_lines}.
		}
	  	\label{tab2}
\begingroup
\renewcommand{\arraystretch}{0.6}
\begin{tabular}{cccc}
\rule{0pt}{3ex}\text{LOM Index}\, & \text{Partition}\, & \text{Boundary Node}\, & \text{Boundary Neighbor} 
\\ \hline
\rule{0pt}{3ex}17.92 & 2 & \text{YES} & \text{NO} \\
9.20 & 1 & \text{YES} & \text{NO} \\
\end{tabular}\endgroup\end{center}\end{table}

\begin{table}[h]
\begin{center}
	\caption{{\it Non-monotonicity data for the weighted power-grid network.} Of the 74 nodes not appearing in the table, 51 have non-zero LOM (mean 472.96), and 28 of these are neither boundary nodes nor boundary neighbors. Of the 23 nodes with zero LOM, 21 are neither boundary nodes nor boundary neighbors.  See Fig.~\ref{fig:nonmon} for the locations and centrality curves of the highest-LOM nodes. Nodes 3 and 7 are marked in Fig.~\ref{fig:floridacond} with a triangle and square, respectively.
		}
	  	\label{tab3}
\begingroup
\renewcommand{\arraystretch}{0.6}
\begin{tabular}{ccccc}
\rule{0pt}{3ex}\text{LOM Index}\, & \text{Fig. 10 Index}\, & \text{Partition}\, & \text{Boundary Node}\, &
   \text{Boundary Neighbor}\\ \hline
\rule{0pt}{3ex}2250.13 & 3 & 1 & \text{YES} & \text{NO} \\
 2118.49 & 6 & 1 & \text{NO} & \text{YES} \\
 1995.26 & 5 & 1 & \text{NO} & \text{YES} \\
 1978.92 & 4 & 1 & \text{NO} & \text{YES} \\
 1639.39 & 7 & 1 & \text{NO} & \text{YES} \\
 1635.02 & 1 & 1 & \text{YES} & \text{NO} \\
 1533.49 & 2 & 2 & \text{YES} & \text{NO} \\
 1492.88 & \text{} & 2 & \text{NO} & \text{NO} \\
 1452.09 & \text{} & 1 & \text{NO} & \text{NO} \\
 1308.06 & \text{} & 2 & \text{NO} & \text{NO} \\
\end{tabular}\endgroup\end{center}\end{table}

\begin{table}[h]
\begin{center}
	\caption{{\it Non-monotonicity data for the unweighted power-grid network.} Of the 74 nodes not appearing in the table, 14 have non-zero LOM (mean 1.44), and 7 of these are neither boundary nodes nor boundary neighbors. Of the 60 nodes with zero LOM, 43 are neither boundary nodes nor boundary neighbors. The ``Fig. 10 Index'' column shows that the highest-LOM nodes in the unweighted power-grid network are very different from those in the weighted version.
		}
	  	\label{tab4}
\begingroup
\renewcommand{\arraystretch}{0.6}
\begin{tabular}{ccccc}
\rule{0pt}{3ex}\text{LOM Index}\, & \text{Fig. 10 Index}\, & \text{Partition}\, & \text{Boundary Node}\, &
   \text{Boundary Neighbor}\\ \hline
\rule{0pt}{3ex}419.13 & 2 & 1 & \text{YES} & \text{NO} \\
 329.90 & \text{} & 1 & \text{YES} & \text{NO} \\
 190.01 & 7 & 1 & \text{NO} & \text{YES} \\
 185.57 & \text{} & 2 & \text{YES} & \text{NO} \\
 162.74 & \text{} & 1 & \text{NO} & \text{YES} \\
 135.60 & \text{} & 1 & \text{NO} & \text{YES} \\
 69.02& \text{} & 2 & \text{NO} & \text{YES} \\
 28.56 & \text{} & 1 & \text{NO} & \text{NO} \\
 6.78 & \text{} & 2 & \text{YES} & \text{NO} \\
 5.93 & \text{} & 1 & \text{NO} & \text{NO} \\
\end{tabular}\endgroup\end{center}\end{table}

\clearpage


\begin{thebibliography}{52}%
\makeatletter
\providecommand \@ifxundefined [1]{%
 \@ifx{#1\undefined}
}%
\providecommand \@ifnum [1]{%
 \ifnum #1\expandafter \@firstoftwo
 \else \expandafter \@secondoftwo
 \fi
}%
\providecommand \@ifx [1]{%
 \ifx #1\expandafter \@firstoftwo
 \else \expandafter \@secondoftwo
 \fi
}%
\providecommand \natexlab [1]{#1}%
\providecommand \enquote  [1]{``#1''}%
\providecommand \bibnamefont  [1]{#1}%
\providecommand \bibfnamefont [1]{#1}%
\providecommand \citenamefont [1]{#1}%
\providecommand \href@noop [0]{\@secondoftwo}%
\providecommand \href [0]{\begingroup \@sanitize@url \@href}%
\providecommand \@href[1]{\@@startlink{#1}\@@href}%
\providecommand \@@href[1]{\endgroup#1\@@endlink}%
\providecommand \@sanitize@url [0]{\catcode `\\12\catcode `\$12\catcode
  `\&12\catcode `\#12\catcode `\^12\catcode `\_12\catcode `\%12\relax}%
\providecommand \@@startlink[1]{}%
\providecommand \@@endlink[0]{}%
\providecommand \url  [0]{\begingroup\@sanitize@url \@url }%
\providecommand \@url [1]{\endgroup\@href {#1}{\urlprefix }}%
\providecommand \urlprefix  [0]{URL }%
\providecommand \Eprint [0]{\href }%
\providecommand \doibase [0]{http://dx.doi.org/}%
\providecommand \selectlanguage [0]{\@gobble}%
\providecommand \bibinfo  [0]{\@secondoftwo}%
\providecommand \bibfield  [0]{\@secondoftwo}%
\providecommand \translation [1]{[#1]}%
\providecommand \BibitemOpen [0]{}%
\providecommand \bibitemStop [0]{}%
\providecommand \bibitemNoStop [0]{.\EOS\space}%
\providecommand \EOS [0]{\spacefactor3000\relax}%
\providecommand \BibitemShut  [1]{\csname bibitem#1\endcsname}%
\let\auto@bib@innerbib\@empty
\bibitem [{\citenamefont {Euler}(1736)}]{EULER1736}%
  \BibitemOpen
  \bibfield  {author} {\bibinfo {author} {\bibfnamefont {L.}~\bibnamefont
  {Euler}},\ }\bibfield  {title} {\enquote {\bibinfo {title} {Solutio
  problematis ad geometriam situs pertinentis},}\ }\href@noop {} {\bibfield
  {journal} {\bibinfo  {journal} {Commentarii Academiae Scientarum Imperialis
  Petropolitanae}\ }\textbf {\bibinfo {volume} {8}},\ \bibinfo {pages}
  {128--140} (\bibinfo {year} {1736})}\BibitemShut {NoStop}%
\bibitem [{\citenamefont {Euler}(1956)}]{EULER1956}%
  \BibitemOpen
  \bibfield  {author} {\bibinfo {author} {\bibfnamefont {L.}~\bibnamefont
  {Euler}},\ }\bibfield  {title} {\enquote {\bibinfo {title} {The seven bridges
  of {K}{\"o}nigsberg},}\ }in\ \href@noop {} {\emph {\bibinfo {booktitle}
  {World of Mathematics, Vol. 1}}},\ \bibinfo {editor} {edited by\ \bibinfo
  {editor} {\bibfnamefont {J.~R.}\ \bibnamefont {Newman}}}\ (\bibinfo
  {publisher} {George Allen and Unwin},\ \bibinfo {address} {London},\ \bibinfo
  {year} {1956})\ pp.\ \bibinfo {pages} {573--580},\ \bibinfo {note}
  {({E}nglish translation of \cite{EULER1736})}\BibitemShut {NoStop}%
\bibitem [{\citenamefont {Bollob{\'a}s}(1998)}]{bollobas1979graph}%
  \BibitemOpen
  \bibfield  {author} {\bibinfo {author} {\bibfnamefont {B.}~\bibnamefont
  {Bollob{\'a}s}},\ }\href@noop {} {\emph {\bibinfo {title} {Modern Graph
  Theory}}}\ (\bibinfo  {publisher} {Springer Science+Business Media},\
  \bibinfo {address} {New York},\ \bibinfo {year} {1998})\BibitemShut {NoStop}%
\bibitem [{\citenamefont {Moreno}(1934)}]{MORENO1934}%
  \BibitemOpen
  \bibfield  {author} {\bibinfo {author} {\bibfnamefont {J.~L.}\ \bibnamefont
  {Moreno}},\ }\href@noop {} {\emph {\bibinfo {title} {Who Shall Survive?}}}\
  (\bibinfo  {publisher} {Beakon House},\ \bibinfo {address} {Beakon, NY},\
  \bibinfo {year} {1934})\BibitemShut {NoStop}%
\bibitem [{\citenamefont {Opsahl}\ \emph {et~al.}(2010)\citenamefont {Opsahl},
  \citenamefont {Agneessens},\ and\ \citenamefont {Skvoretz}}]{opsahl2010node}%
  \BibitemOpen
  \bibfield  {author} {\bibinfo {author} {\bibfnamefont {T.}\ \bibnamefont
  {Opsahl}}, \bibinfo {author} {\bibfnamefont {F.}\ \bibnamefont
  {Agneessens}}, \ and\ \bibinfo {author} {\bibfnamefont {J.}\ \bibnamefont
  {Skvoretz}},\ }\bibfield  {title} {\enquote {\bibinfo {title} {Node
  centrality in weighted networks: Generalizing degree and shortest paths},}\
  }\href@noop {} {\bibfield  {journal} {\bibinfo  {journal} {Soc. Networks}\
  }\textbf {\bibinfo {volume} {32}},\ \bibinfo {pages} {245--251} (\bibinfo
  {year} {2010})}\BibitemShut {NoStop}%
\bibitem [{\citenamefont {Kirchhoff}(1847)}]{KIRCHHOFF1847}%
  \BibitemOpen
  \bibfield  {author} {\bibinfo {author} {\bibfnamefont {G.}~\bibnamefont
  {Kirchhoff}},\ }\bibfield  {title} {\enquote {\bibinfo {title} {{Ue}ber die
  {A}ufl{\"o}sung der {G}leichungen, auf welche man bei der {U}ntersuchung der
  {l}inearen {V}ertheilung galvanischer {S}tr{\"o}me gef{\"u}hrt wird},}\
  }\href@noop {} {\bibfield  {journal} {\bibinfo  {journal} {Ann. Phys. Chem.}\
  }\textbf {\bibinfo {volume} {72}},\ \bibinfo {pages} {497--508} (\bibinfo
  {year} {1847})}\BibitemShut {NoStop}%
\bibitem [{\citenamefont {O'Toole}(1958)}]{KIRCHHOFF1958}%
  \BibitemOpen
  \bibfield  {author} {\bibinfo {author} {\bibfnamefont {J.~B.}\ \bibnamefont
  {O'Toole}},\ }\bibfield  {title} {\enquote {\bibinfo {title} {On the solution
  of the equations obtained from the investigation of the linear distribution
  of galvanic currents},}\ }\href@noop {} {\bibfield  {journal} {\bibinfo
  {journal} {IRE Trans. Circuit Theory}\ }\textbf {\bibinfo {volume} {5}},\
  \bibinfo {pages} {4--7} (\bibinfo {year} {1958})},\ \bibinfo {note}
  {({E}nglish translation of \cite{KIRCHHOFF1847})}\BibitemShut {NoStop}%
\bibitem [{\citenamefont {Albert}\ and\ \citenamefont
  {Barab{\'a}si}(2002)}]{ALBE02}%
  \BibitemOpen
  \bibfield  {author} {\bibinfo {author} {\bibfnamefont {R.}~\bibnamefont
  {Albert}}\ and\ \bibinfo {author} {\bibfnamefont {A.-L.}\ \bibnamefont
  {Barab{\'a}si}},\ }\bibfield  {title} {\enquote {\bibinfo {title}
  {Statistical mechanics of complex networks},}\ }\href@noop {} {\bibfield
  {journal} {\bibinfo  {journal} {Rev. Mod. Phys.}\ }\textbf {\bibinfo {volume}
  {74}},\ \bibinfo {pages} {47--97} (\bibinfo {year} {2002})}\BibitemShut
  {NoStop}%
\bibitem [{\citenamefont {Caldarelli}(2007)}]{CALD07}%
  \BibitemOpen
  \bibfield  {author} {\bibinfo {author} {\bibfnamefont {G.}~\bibnamefont
  {Caldarelli}},\ }\href@noop {} {\emph {\bibinfo {title} {Scale-Free
  Networks}}}\ (\bibinfo  {publisher} {Oxford University Press},\ \bibinfo
  {address} {Oxford, UK},\ \bibinfo {year} {2007})\BibitemShut {NoStop}%
\bibitem [{\citenamefont {Dorogovtsev}\ \emph {et~al.}(2008)\citenamefont
  {Dorogovtsev}, \citenamefont {Goltsev},\ and\ \citenamefont
  {Mendes}}]{DORO08}%
  \BibitemOpen
  \bibfield  {author} {\bibinfo {author} {\bibfnamefont {S.~N.}\ \bibnamefont
  {Dorogovtsev}}, \bibinfo {author} {\bibfnamefont {A.~V.}\ \bibnamefont
  {Goltsev}}, \ and\ \bibinfo {author} {\bibfnamefont {J.~F.~F.}\ \bibnamefont
  {Mendes}},\ }\bibfield  {title} {\enquote {\bibinfo {title} {Critical
  phenomena in complex networks},}\ }\href@noop {} {\bibfield  {journal}
  {\bibinfo  {journal} {Rev. Mod. Phys.}\ }\textbf {\bibinfo {volume} {80}},\
  \bibinfo {pages} {1275--1335} (\bibinfo {year} {2008})}\BibitemShut {NoStop}%
\bibitem [{\citenamefont {Newman}(2010)}]{newman2010networks}%
  \BibitemOpen
  \bibfield  {author} {\bibinfo {author} {\bibfnamefont {M.~E.~J.}\
  \bibnamefont {Newman}},\ }\href@noop {} {\emph {\bibinfo {title} {Networks:
  an Introduction}}}\ (\bibinfo  {publisher} {Oxford University Press},\
  \bibinfo {address} {Oxford, UK},\ \bibinfo {year} {2010})\BibitemShut
  {NoStop}%
\bibitem [{\citenamefont {Estrada}(2011)}]{ESTR11}%
  \BibitemOpen
  \bibfield  {author} {\bibinfo {author} {\bibfnamefont {E.}~\bibnamefont
  {Estrada}},\ }\href@noop {} {\emph {\bibinfo {title} {The Structure of
  Complex Networks.Theory and Applications}}}\ (\bibinfo  {publisher} {Oxford
  University Press},\ \bibinfo {address} {Oxford, UK},\ \bibinfo {year}
  {2011})\BibitemShut {NoStop}%
\bibitem [{\citenamefont {Blunt}\ \emph {et~al.}(2013)\citenamefont {Blunt},
  \citenamefont {Bijelic}, \citenamefont {Dong}, \citenamefont {Gharbi},
  \citenamefont {Iglauer}, \citenamefont {Mostaghimi}, \citenamefont
  {Paluszny},\ and\ \citenamefont {Pentland}}]{BLUN13}%
  \BibitemOpen
  \bibfield  {author} {\bibinfo {author} {\bibfnamefont {M.~J.}\ \bibnamefont
  {Blunt}}, \bibinfo {author} {\bibfnamefont {B.}~\bibnamefont {Bijelic}},
  \bibinfo {author} {\bibfnamefont {H.}~\bibnamefont {Dong}}, \bibinfo {author}
  {\bibfnamefont {O.}~\bibnamefont {Gharbi}}, \bibinfo {author} {\bibfnamefont
  {S.}~\bibnamefont {Iglauer}}, \bibinfo {author} {\bibfnamefont
  {P.}~\bibnamefont {Mostaghimi}}, \bibinfo {author} {\bibfnamefont
  {A.}~\bibnamefont {Paluszny}}, \ and\ \bibinfo {author} {\bibfnamefont
  {C.}~\bibnamefont {Pentland}},\ }\bibfield  {title} {\enquote {\bibinfo
  {title} {Pore-scale imaging and modelling},}\ }\href@noop {} {\bibfield
  {journal} {\bibinfo  {journal} {Adv. Water Resour.}\ }\textbf {\bibinfo
  {volume} {51}},\ \bibinfo {pages} {197--216} (\bibinfo {year}
  {2013})}\BibitemShut {NoStop}%
\bibitem [{\citenamefont {Savani}\ \emph {et~al.}(2017)\citenamefont {Savani},
  \citenamefont {Bedeaux}, \citenamefont {Kjelstrup}, \citenamefont {Vassvik},
  \citenamefont {Sinha},\ and\ \citenamefont {Hansen}}]{SAVA17}%
  \BibitemOpen
  \bibfield  {author} {\bibinfo {author} {\bibfnamefont {I.}~\bibnamefont
  {Savani}}, \bibinfo {author} {\bibfnamefont {D.}~\bibnamefont {Bedeaux}},
  \bibinfo {author} {\bibfnamefont {S.}~\bibnamefont {Kjelstrup}}, \bibinfo
  {author} {\bibfnamefont {M.}~\bibnamefont {Vassvik}}, \bibinfo {author}
  {\bibfnamefont {S.}~\bibnamefont {Sinha}}, \ and\ \bibinfo {author}
  {\bibfnamefont {A.}~\bibnamefont {Hansen}},\ }\bibfield  {title} {\enquote
  {\bibinfo {title} {Ensemble distribution for immiscible two-phase flow in
  porous media},}\ }\href@noop {} {\bibfield  {journal} {\bibinfo  {journal}
  {Phys. Rev. E}\ }\textbf {\bibinfo {volume} {95}},\ \bibinfo {pages} {023116}
  (\bibinfo {year} {2017})}\BibitemShut {NoStop}%
\bibitem [{\citenamefont {Shamboek}\ \emph {et~al.}(2019)\citenamefont
  {Shamboek}, \citenamefont {Iedema},\ and\ \citenamefont {Kryven}}]{SHAM19}%
  \BibitemOpen
  \bibfield  {author} {\bibinfo {author} {\bibfnamefont {V.}~\bibnamefont
  {Shamboek}}, \bibinfo {author} {\bibfnamefont {P.~D.}\ \bibnamefont
  {Iedema}}, \ and\ \bibinfo {author} {\bibfnamefont {I.}~\bibnamefont
  {Kryven}},\ }\bibfield  {title} {\enquote {\bibinfo {title} {Process of
  irreversible step-growth polymerization},}\ }\href@noop {} {\bibfield
  {journal} {\bibinfo  {journal} {Sci. Rep.}\ }\textbf {\bibinfo {volume}
  {9}},\ \bibinfo {pages} {2276} (\bibinfo {year} {2019})}\BibitemShut
  {NoStop}%
\bibitem [{\citenamefont {Rossberg}(2013)}]{ROSS13}%
  \BibitemOpen
  \bibfield  {author} {\bibinfo {author} {\bibfnamefont {A.~G.}\ \bibnamefont
  {Rossberg}},\ }\href@noop {} {\emph {\bibinfo {title} {Food Webs and
  Biodiversity}}}\ (\bibinfo  {publisher} {Wiley Blackwell},\ \bibinfo
  {address} {Oxford, UK},\ \bibinfo {year} {2013})\BibitemShut {NoStop}%
\bibitem [{\citenamefont {Verma}\ \emph {et~al.}(2014)\citenamefont {Verma},
  \citenamefont {Ara{\'u}jo},\ and\ \citenamefont {Herrmann}}]{VERM14}%
  \BibitemOpen
  \bibfield  {author} {\bibinfo {author} {\bibfnamefont {T.}~\bibnamefont
  {Verma}}, \bibinfo {author} {\bibfnamefont {N.~A.~M.}\ \bibnamefont
  {Ara{\'u}jo}}, \ and\ \bibinfo {author} {\bibfnamefont {H.~J.}\ \bibnamefont
  {Herrmann}},\ }\bibfield  {title} {\enquote {\bibinfo {title} {Revealing the
  structure of the world airline network},}\ }\href@noop {} {\bibfield
  {journal} {\bibinfo  {journal} {Sci. Rep.}\ }\textbf {\bibinfo {volume}
  {4}},\ \bibinfo {pages} {5638} (\bibinfo {year} {2014})}\BibitemShut
  {NoStop}%
\bibitem [{\citenamefont {Xu}\ \emph {et~al.}(2014)\citenamefont {Xu},
  \citenamefont {Gurfinkel},\ and\ \citenamefont
  {Rikvold}}]{xu2014architecture}%
  \BibitemOpen
  \bibfield  {author} {\bibinfo {author} {\bibfnamefont {Y.}~\bibnamefont
  {Xu}}, \bibinfo {author} {\bibfnamefont {A.~J.}\ \bibnamefont {Gurfinkel}}, \
  and\ \bibinfo {author} {\bibfnamefont {P.~A.}\ \bibnamefont {Rikvold}},\
  }\bibfield  {title} {\enquote {\bibinfo {title} {Architecture of the
  {F}lorida power grid as a complex network},}\ }\href@noop {} {\bibfield
  {journal} {\bibinfo  {journal} {Physica A}\ }\textbf {\bibinfo {volume}
  {401}},\ \bibinfo {pages} {130--140} (\bibinfo {year} {2014})}\BibitemShut
  {NoStop}%
\bibitem [{\citenamefont {Gurfinkel}\ \emph {et~al.}(2015)\citenamefont
  {Gurfinkel}, \citenamefont {Silva},\ and\ \citenamefont {Rikvold}}]{GURF15}%
  \BibitemOpen
  \bibfield  {author} {\bibinfo {author} {\bibfnamefont {A.~J.}\ \bibnamefont
  {Gurfinkel}}, \bibinfo {author} {\bibfnamefont {D.~A.}\ \bibnamefont
  {Silva}}, \ and\ \bibinfo {author} {\bibfnamefont {P.~A.}\ \bibnamefont
  {Rikvold}},\ }\bibfield  {title} {\enquote {\bibinfo {title} {Centrality
  fingerprints for power grid network growth models},}\ }\href@noop {}
  {\bibfield  {journal} {\bibinfo  {journal} {Phys. Proc.}\ }\textbf {\bibinfo
  {volume} {68}},\ \bibinfo {pages} {52--55} (\bibinfo {year}
  {2015})}\BibitemShut {NoStop}%
\bibitem [{\citenamefont {Kutner}\ \emph {et~al.}(2019)\citenamefont {Kutner},
  \citenamefont {Ausloos}, \citenamefont {Grech}, \citenamefont {Di~Matteo},
  \citenamefont {Schinckus},\ and\ \citenamefont {Stanley}}]{KUTN19}%
  \BibitemOpen
  \bibfield  {author} {\bibinfo {author} {\bibfnamefont {R.}~\bibnamefont
  {Kutner}}, \bibinfo {author} {\bibfnamefont {M.}~\bibnamefont {Ausloos}},
  \bibinfo {author} {\bibfnamefont {D.}~\bibnamefont {Grech}}, \bibinfo
  {author} {\bibfnamefont {T.}~\bibnamefont {Di~Matteo}}, \bibinfo {author}
  {\bibfnamefont {C.}~\bibnamefont {Schinckus}}, \ and\ \bibinfo {author}
  {\bibfnamefont {H.~E.}\ \bibnamefont {Stanley}},\ }\bibfield  {title}
  {\enquote {\bibinfo {title} {Econophysics and sociophysics: Their milestones
  \& challenges},}\ }\href@noop {} {\bibfield  {journal} {\bibinfo  {journal}
  {Physica A}\ }\textbf {\bibinfo {volume} {516}},\ \bibinfo {pages} {240--253}
  (\bibinfo {year} {2019})}\BibitemShut {NoStop}%
\bibitem [{\citenamefont {Tiropanis}\ \emph {et~al.}(2015)\citenamefont
  {Tiropanis}, \citenamefont {Hall}, \citenamefont {Crowcroft}, \citenamefont
  {Contracctor},\ and\ \citenamefont {Tassiulas}}]{TIRO15}%
  \BibitemOpen
  \bibfield  {author} {\bibinfo {author} {\bibfnamefont {T.}~\bibnamefont
  {Tiropanis}}, \bibinfo {author} {\bibfnamefont {W.}~\bibnamefont {Hall}},
  \bibinfo {author} {\bibfnamefont {J.}~\bibnamefont {Crowcroft}}, \bibinfo
  {author} {\bibfnamefont {N.}~\bibnamefont {Contractor}}, \ and\ \bibinfo
  {author} {\bibfnamefont {L.}~\bibnamefont {Tassiulas}},\ }\bibfield  {title}
  {\enquote {\bibinfo {title} {Network science, {W}eb science, and {I}nternet
  science},}\ }\href@noop {} {\bibfield  {journal} {\bibinfo  {journal}
  {Commun. ACM}\ }\textbf {\bibinfo {volume} {58}} (8),\ \bibinfo {pages} {76--82}
  (\bibinfo {year} {2015})}\BibitemShut {NoStop}%
\bibitem [{\citenamefont {Page}\ \emph {et~al.}(1999)\citenamefont {Page},
  \citenamefont {Brin}, \citenamefont {Motwani},\ and\ \citenamefont
  {Winograd}}]{page1999pagerank}%
  \BibitemOpen
  \bibfield  {author} {\bibinfo {author} {\bibfnamefont {L.}~\bibnamefont
  {Page}}, \bibinfo {author} {\bibfnamefont {S.}~\bibnamefont {Brin}}, \bibinfo
  {author} {\bibfnamefont {R.}~\bibnamefont {Motwani}}, \ and\ \bibinfo
  {author} {\bibfnamefont {T.}~\bibnamefont {Winograd}},\ }\href@noop {} {\emph
  {\bibinfo {title} {The PageRank Citation Ranking: Bringing Order to the
  Web.}}},\ \bibinfo {type} {Tech. Rep.}\ (\bibinfo  {institution} {Stanford
  InfoLab},\ \bibinfo {year} {1999})\BibitemShut {NoStop}%
\bibitem [{\citenamefont {Jeong}\ \emph {et~al.}(2001)\citenamefont {Jeong},
  \citenamefont {Mason}, \citenamefont {Barab{\'a}si},\ and\ \citenamefont
  {Oltvai}}]{jeong2001lethality}%
  \BibitemOpen
  \bibfield  {author} {\bibinfo {author} {\bibfnamefont {H.}~\bibnamefont
  {Jeong}}, \bibinfo {author} {\bibfnamefont {S.~P.}\ \bibnamefont {Mason}},
  \bibinfo {author} {\bibfnamefont {A.~L.}\ \bibnamefont {Barab{\'a}si}}, \
  and\ \bibinfo {author} {\bibfnamefont {Z.~N.}\ \bibnamefont {Oltvai}},\
  }\bibfield  {title} {\enquote {\bibinfo {title} {Lethality and centrality in
  protein networks},}\ }\href@noop {} {\bibfield  {journal} {\bibinfo
  {journal} {Nature}\ }\textbf {\bibinfo {volume} {411}},\ \bibinfo {pages}
  {41} (\bibinfo {year} {2001})}\BibitemShut {NoStop}%
\bibitem [{\citenamefont {Estrada}\ and\ \citenamefont
  {Hatano}(2008)}]{estrada2008communicability}%
  \BibitemOpen
  \bibfield  {author} {\bibinfo {author} {\bibfnamefont {E.}~\bibnamefont
  {Estrada}}\ and\ \bibinfo {author} {\bibfnamefont {N.}~\bibnamefont
  {Hatano}},\ }\bibfield  {title} {\enquote {\bibinfo {title} {Communicability
  in complex networks},}\ }\href@noop {} {\bibfield  {journal} {\bibinfo
  {journal} {Phys. Rev. E}\ }\textbf {\bibinfo {volume} {77}},\ \bibinfo
  {pages} {036111} (\bibinfo {year} {2008})}\BibitemShut {NoStop}%
\bibitem [{\citenamefont {Estrada}\ \emph {et~al.}(2009)\citenamefont
  {Estrada}, \citenamefont {Higham},\ and\ \citenamefont
  {Hatano}}]{estrada2009communicability}%
  \BibitemOpen
  \bibfield  {author} {\bibinfo {author} {\bibfnamefont {E.}~\bibnamefont
  {Estrada}}, \bibinfo {author} {\bibfnamefont {D.~J.}\ \bibnamefont {Higham}},
  \ and\ \bibinfo {author} {\bibfnamefont {N.}~\bibnamefont {Hatano}},\
  }\bibfield  {title} {\enquote {\bibinfo {title} {Communicability betweenness
  in complex networks},}\ }\href@noop {} {\bibfield  {journal} {\bibinfo
  {journal} {Physica A}\ }\textbf {\bibinfo {volume} {388}},\ \bibinfo {pages}
  {764--774} (\bibinfo {year} {2009})}\BibitemShut {NoStop}%
\bibitem [{\citenamefont {Doyle}\ and\ \citenamefont
  {Snell}(1984)}]{Doyle06randomwalks}%
  \BibitemOpen
  \bibfield  {author} {\bibinfo {author} {\bibfnamefont {P.~G.}\ \bibnamefont
  {Doyle}}\ and\ \bibinfo {author} {\bibfnamefont {J.~L.}\ \bibnamefont
  {Snell}},\ }\href@noop {} {\emph {\bibinfo {title} {Random Walks and Electric
  networks}}}\ (\bibinfo  {publisher} {Mathematical Association of America},\
  \bibinfo {year} {1984})\ \bibinfo {note} {. Since 2006, available under the
  GNU FDL at
  \url{https://math.dartmouth.edu/~doyle/docs/walks/walks.pdf}}\BibitemShut
  {NoStop}%
\bibitem [{\citenamefont {Newman}(2005)}]{newman2005measure}%
  \BibitemOpen
  \bibfield  {author} {\bibinfo {author} {\bibfnamefont {M.~E.~J.}\
  \bibnamefont {Newman}},\ }\bibfield  {title} {\enquote {\bibinfo {title} {A
  measure of betweenness centrality based on random walks},}\ }\href@noop {}
  {\bibfield  {journal} {\bibinfo  {journal} {Soc. Networks}\ }\textbf
  {\bibinfo {volume} {27}},\ \bibinfo {pages} {39--54} (\bibinfo {year}
  {2005})}\BibitemShut {NoStop}%
\bibitem [{\citenamefont {Brandes}\ and\ \citenamefont
  {Fleischer}(2005)}]{brandes2005centrality}%
  \BibitemOpen
  \bibfield  {author} {\bibinfo {author} {\bibfnamefont {U.}~\bibnamefont
  {Brandes}}\ and\ \bibinfo {author} {\bibfnamefont {D.}~\bibnamefont
  {Fleischer}},\ }\bibfield  {title} {\enquote {\bibinfo {title} {Centrality
  measures based on current flow},}\ }in\ \href@noop {} {\emph {\bibinfo
  {booktitle} {Annual symposium on theoretical aspects of computer science}}}\
  (\bibinfo {organization} {Springer},\ \bibinfo {address}
  {Berlin-Heidelberg},\ \bibinfo {year} {2005})\ pp.\ \bibinfo {pages}
  {533--544}\BibitemShut {NoStop}%
\bibitem [{\citenamefont {Klein}\ and\ \citenamefont
  {Randi{\'c}}(1993)}]{klein1993resistance}%
  \BibitemOpen
  \bibfield  {author} {\bibinfo {author} {\bibfnamefont {D.~J.}\ \bibnamefont
  {Klein}}\ and\ \bibinfo {author} {\bibfnamefont {M.}~\bibnamefont
  {Randi{\'c}}},\ }\bibfield  {title} {\enquote {\bibinfo {title} {Resistance
  distance},}\ }\href@noop {} {\bibfield  {journal} {\bibinfo  {journal} {J.
  Math. Chem.}\ }\textbf {\bibinfo {volume} {12}},\ \bibinfo {pages} {81--95}
  (\bibinfo {year} {1993})}\BibitemShut {NoStop}%
  \bibitem [{\citenamefont {Latora}\ and\ \citenamefont
  {Marchiori}(2001)}]{latora2001efficient}%
  \BibitemOpen
  \bibfield  {author} {\bibinfo {author} {\bibfnamefont {V.}\ \bibnamefont
  {Latora}}\ and\ \bibinfo {author} {\bibfnamefont {M.}\ \bibnamefont
  {Marchiori}},\ }\bibfield  {title} {\enquote {\bibinfo {title} {Efficient
  behavior of small-world networks},}\ }\href@noop {} {\bibfield  {journal}
  {\bibinfo  {journal} {Phys. Rev. Lett.}\ }\textbf {\bibinfo {volume}
  {87}},\ \bibinfo {pages} {198701} (\bibinfo {year} {2001})}\BibitemShut
  {NoStop}%
\bibitem [{\citenamefont {Rochat}(2009)}]{Rochat2009}%
  \BibitemOpen
  \bibfield  {author} {\bibinfo {author} {\bibfnamefont {Y.}~\bibnamefont
  {Rochat}},\ }\href@noop {} {\enquote {\bibinfo {title} {Closeness centrality
  extended to unconnected graphs: The harmonic centrality index},}\ }  \bibinfo {type} {Tech. Rep.}\ (\bibinfo
  {institution} {Institute of Applied Mathematics, University of Lausanne},\
  \bibinfo {year} {2009}.) \url{https://infoscience.epfl.ch/record/200525/files/%5bEN%5dASNA09.pdf}
 \BibitemShut {NoStop}%
\bibitem [{\citenamefont {Dekker}(2005)}]{Dekker2005}%
  \BibitemOpen
  \bibfield  {author} {\bibinfo {author} {\bibfnamefont {A.}~\bibnamefont
  {Dekker}},\ }\bibfield  {title} {\enquote {\bibinfo {title} {Conceptual
  distance in social network analysis},}\ }\href@noop {} {\bibfield  {journal}
  {\bibinfo  {journal} {J. Social Structure}\ }\textbf {\bibinfo
  {volume} {6}} (\bibinfo {year} {2005})}\BibitemShut {NoStop}%
\bibitem [{\citenamefont {Anthonisse}(1971)}]{Anthonisse1971}%
  \BibitemOpen
  \bibfield  {author} {\bibinfo {author} {\bibfnamefont {J.~M.}\ \bibnamefont
  {Anthonisse}},\ }\bibfield  {title} {\enquote {\bibinfo {title} {The rush in
  a directed graph},}\ }\href@noop {} {\bibfield  {journal} {\bibinfo
  {journal} {Stichting Mathematisch Centrum. Mathematische Besliskunde}\ }
  (\bibinfo {year} {1971}) }\url{https://ir.cwi.nl/pub/9791}\BibitemShut {NoStop}%
\bibitem [{\citenamefont {Freeman}(1977)}]{Freeman1977}%
  \BibitemOpen
  \bibfield  {author} {\bibinfo {author} {\bibfnamefont {L.~C.}\ \bibnamefont
  {Freeman}},\ }\bibfield  {title} {\enquote {\bibinfo {title} {A set of
  measures of centrality based on betweenness},}\ }\href@noop {} {\bibfield
  {journal} {\bibinfo  {journal} {Sociometry {\bf 40}}\ ,\ \bibinfo {pages} {35--41}}
  (\bibinfo {year} {1977})}\BibitemShut {NoStop}%
\bibitem [{\citenamefont {Steer}(2010)}]{steer2010microwave}%
  \BibitemOpen
  \bibfield  {author} {\bibinfo {author} {\bibfnamefont {M.~B.}\ \bibnamefont
  {Steer}},\ }\href@noop {} {\emph {\bibinfo {title} {Microwave and RF design:
  a systems approach}}}\ (\bibinfo  {publisher} {SciTech Publishing, Inc.},\
  \bibinfo {address} {Raleigh, NC},\ \bibinfo {year} {2010})\BibitemShut{NoStop}%
\bibitem [{kan(2017)}]{kangadata}%
  \BibitemOpen
  \href@noop {} {\enquote {\bibinfo {title} {Kangaroo network dataset},}\
  }\bibinfo {howpublished} {KONECT
  \url{http://konect.uni-koblenz.de/networks/moreno_kangaroo}} (\bibinfo {year}
  {2017})\BibitemShut {NoStop}%
\bibitem [{\citenamefont {Grant}(1973)}]{grant1973dominance}%
  \BibitemOpen
  \bibfield  {author} {\bibinfo {author} {\bibfnamefont {T.~R.}\ \bibnamefont
  {Grant}},\ }\bibfield  {title} {\enquote {\bibinfo {title} {Dominance and
  association among members of a captive and a free-ranging group of grey
  kangaroos (Macropus giganteus)},}\ }\href@noop {} {\bibfield  {journal}
  {\bibinfo  {journal} {Anim. Behav.}\ }\textbf {\bibinfo {volume} {21}},\
  \bibinfo {pages} {449--456} (\bibinfo {year} {1973})}\BibitemShut {NoStop}%
\bibitem [{\citenamefont {Zachary}(1977)}]{zachary}%
  \BibitemOpen
  \bibfield  {author} {\bibinfo {author} {\bibfnamefont {W.~W.}\ \bibnamefont
  {Zachary}},\ }\bibfield  {title} {\enquote {\bibinfo {title} {An information
  flow model for conflict and fission in small groups},}\ }\href@noop {}
  {\bibfield  {journal} {\bibinfo  {journal} {J. Anthrop. Res.}\ }\textbf
  {\bibinfo {volume} {33}},\ \bibinfo {pages} {452--473} (\bibinfo {year}
  {1977})}\BibitemShut {NoStop}%
\bibitem [{\citenamefont {Dale}\ \emph {et~al.}(2009)\citenamefont {Dale},
  \citenamefont {Alqutami}, \citenamefont {Baldwin}, \citenamefont {Faruque},
  \citenamefont {Langston}, \citenamefont {McLaren}, \citenamefont {Meeker},
  \citenamefont {Steurer},\ and\ \citenamefont {Schoder}}]{dale}%
  \BibitemOpen
  \bibfield  {author} {\bibinfo {author} {\bibfnamefont {S.}~\bibnamefont
  {Dale}}, \bibinfo {author} {\bibfnamefont {T.}~\bibnamefont {Alqutami}},
  \bibinfo {author} {\bibfnamefont {T.}~\bibnamefont {Baldwin}}, \bibinfo
  {author} {\bibfnamefont {O.}~\bibnamefont {Faruque}}, \bibinfo {author}
  {\bibfnamefont {J.}~\bibnamefont {Langston}}, \bibinfo {author}
  {\bibfnamefont {P.}~\bibnamefont {McLaren}}, \bibinfo {author} {\bibfnamefont
  {P.}~\bibnamefont {Meeker}}, \bibinfo {author} {\bibfnamefont
  {M.}~\bibnamefont {Steurer}}, \ and\ \bibinfo {author} {\bibfnamefont
  {K.}~\bibnamefont {Schoder}},\ }\href@noop {} {\emph {\bibinfo {title}
  {Progress Report for the Institute for Energy Systems, Economics and
  Sustainability and the {F}lorida Energy Systems Consortium}}},\ \bibinfo
  {type} {Tech. Rep.}\ (\bibinfo  {institution} {Florida State University},\
  \bibinfo {address} {Tallahassee, FL},\ \bibinfo {year} {2009})\BibitemShut
  {NoStop}%
\bibitem [{\citenamefont {Bozzo}\ and\ \citenamefont
  {Franceschet}(2013)}]{bozzo}%
  \BibitemOpen
  \bibfield  {author} {\bibinfo {author} {\bibfnamefont {E.}~\bibnamefont
  {Bozzo}}\ and\ \bibinfo {author} {\bibfnamefont {M.}~\bibnamefont
  {Franceschet}},\ }\bibfield  {title} {\enquote {\bibinfo {title} {Resistance
  distance, closeness, and betweenness},}\ }\href@noop {} {\bibfield  {journal}
  {\bibinfo  {journal} {Soc. Networks}\ }\textbf {\bibinfo {volume} {35}},\
  \bibinfo {pages} {460--469} (\bibinfo {year} {2013})}\BibitemShut {NoStop}%
\bibitem [{\citenamefont {Tizghadam}\ and\ \citenamefont
  {Leon-Garcia}(2010)}]{tizghadam}%
  \BibitemOpen
  \bibfield  {author} {\bibinfo {author} {\bibfnamefont {A.}~\bibnamefont
  {Tizghadam}}\ and\ \bibinfo {author} {\bibfnamefont {A.}~\bibnamefont
  {Leon-Garcia}},\ }\bibfield  {title} {\enquote {\bibinfo {title} {Autonomic
  traffic engineering for network robustness},}\ }\href@noop {} {\bibfield
  {journal} {\bibinfo  {journal} {IEEE journal on selected areas in
  communications}\ }\textbf {\bibinfo {volume} {28}},\ \bibinfo {pages}
  {39--50} (\bibinfo {year} {2010})}\BibitemShut {NoStop}%
\bibitem [{\citenamefont {Alamgir}\ and\ \citenamefont
  {Luxburg}(2011)}]{alamgir2011phase}%
  \BibitemOpen
  \bibfield  {author} {\bibinfo {author} {\bibfnamefont {M.}~\bibnamefont
  {Alamgir}}\ and\ \bibinfo {author} {\bibfnamefont {U.~V.}\ \bibnamefont
  {Luxburg}},\ }\bibfield  {title} {\enquote {\bibinfo {title} {Phase
  transition in the family of p-resistances},}\ }in\ \href@noop {} {\emph
  {\bibinfo {booktitle} {Advances in Neural Information Processing Systems 24 (NIPS 2011)}}}\
  (\bibinfo {year} {2011})\ pp.\ \bibinfo {pages} {379--387}\BibitemShut
  {NoStop}%
\bibitem [{\citenamefont {Avrachenkov}\ \emph {et~al.}(2013)\citenamefont
  {Avrachenkov}, \citenamefont {Litvak}, \citenamefont {Medyanikov},\ and\
  \citenamefont {Sokol}}]{avrachenkov2013alpha}%
  \BibitemOpen
  \bibfield  {author} {\bibinfo {author} {\bibfnamefont {K.}~\bibnamefont
  {Avrachenkov}}, \bibinfo {author} {\bibfnamefont {N.}~\bibnamefont {Litvak}},
  \bibinfo {author} {\bibfnamefont {V.}~\bibnamefont {Medyanikov}}, \ and\
  \bibinfo {author} {\bibfnamefont {M.}~\bibnamefont {Sokol}},\ }\bibfield
  {title} {\enquote {\bibinfo {title} {Alpha current flow betweenness
  centrality},}\ }in\ \href@noop {} {\emph {\bibinfo {booktitle} {International
  Workshop on Algorithms and Models for the Web-Graph}}}\ (\bibinfo
  {organization} {Springer},\ \bibinfo {address} {Berlin-Heidelberg},\ \bibinfo
  {year} {2013})\ pp.\ \bibinfo {pages} {106--117}\BibitemShut {NoStop}%
\bibitem [{\citenamefont {Avrachenkov}\ \emph {et~al.}(2015)\citenamefont
  {Avrachenkov}, \citenamefont {Mazalov},\ and\ \citenamefont
  {Tsynguev}}]{avrachenkov2015beta}%
  \BibitemOpen
  \bibfield  {author} {\bibinfo {author} {\bibfnamefont {K.~E.}\ \bibnamefont
  {Avrachenkov}}, \bibinfo {author} {\bibfnamefont {V.~V.}\ \bibnamefont
  {Mazalov}}, \ and\ \bibinfo {author} {\bibfnamefont {B.~T.}\ \bibnamefont
  {Tsynguev}},\ }\bibfield  {title} {\enquote {\bibinfo {title} {Beta current
  flow centrality for weighted networks},}\ }in\ \href@noop {} {\emph {\bibinfo
  {booktitle} {International Conference on Computational Social Networks}}}\
  (\bibinfo {organization} {Springer},\ \bibinfo {address}
  {Berlin-Heidelberg},\ \bibinfo {year} {2015})\ pp.\ \bibinfo {pages}
  {216--227}\BibitemShut {NoStop}%
\bibitem [{\citenamefont {Kivim{\"a}ki}\ \emph {et~al.}(2016)\citenamefont
  {Kivim{\"a}ki}, \citenamefont {Lebichot}, \citenamefont {Saram{\"a}ki},\ and\
  \citenamefont {Saerens}}]{kivimaki2016two}%
  \BibitemOpen
  \bibfield  {author} {\bibinfo {author} {\bibfnamefont {I.}~\bibnamefont
  {Kivim{\"a}ki}}, \bibinfo {author} {\bibfnamefont {B.}~\bibnamefont
  {Lebichot}}, \bibinfo {author} {\bibfnamefont {J.}~\bibnamefont
  {Saram{\"a}ki}}, \ and\ \bibinfo {author} {\bibfnamefont {M.}~\bibnamefont
  {Saerens}},\ }\bibfield  {title} {\enquote {\bibinfo {title} {Two betweenness
  centrality measures based on randomized shortest paths},}\ }\href@noop {}
  {\bibfield  {journal} {\bibinfo  {journal} {Sci. Rep.}\ }\textbf {\bibinfo
  {volume} {6}},\ \bibinfo {pages} {19668} (\bibinfo {year}
  {2016})}\BibitemShut {NoStop}%
\bibitem [{\citenamefont {Kivim{\"a}ki}\ \emph {et~al.}(2014)\citenamefont
  {Kivim{\"a}ki}, \citenamefont {Shimbo},\ and\ \citenamefont
  {Saerens}}]{kivimaki2014developments}%
  \BibitemOpen
  \bibfield  {author} {\bibinfo {author} {\bibfnamefont {I.}~\bibnamefont
  {Kivim{\"a}ki}}, \bibinfo {author} {\bibfnamefont {M.}~\bibnamefont
  {Shimbo}}, \ and\ \bibinfo {author} {\bibfnamefont {M.}~\bibnamefont
  {Saerens}},\ }\bibfield  {title} {\enquote {\bibinfo {title} {Developments in
  the theory of randomized shortest paths with a comparison of graph node
  distances},}\ }\href@noop {} {\bibfield  {journal} {\bibinfo  {journal}
  {Physica A}\ }\textbf {\bibinfo {volume} {393}},\ \bibinfo {pages} {600--616}
  (\bibinfo {year} {2014})}\BibitemShut {NoStop}%
\bibitem [{\citenamefont {Bavaud}\ and\ \citenamefont
  {Guex}(2012)}]{bavaud2012interpolating}%
  \BibitemOpen
  \bibfield  {author} {\bibinfo {author} {\bibfnamefont {F.}~\bibnamefont
  {Bavaud}}\ and\ \bibinfo {author} {\bibfnamefont {G.}~\bibnamefont {Guex}},\
  }\bibfield  {title} {\enquote {\bibinfo {title} {Interpolating between random
  walks and shortest paths: a path functional approach},}\ }in\ \href@noop {}
  {\emph {\bibinfo {booktitle} {International Conference on Social
  Informatics}}}\ (\bibinfo {organization} {Springer},\ \bibinfo {address}
  {Berlin-Heidelberg},\ \bibinfo {year} {2012})\ pp.\ \bibinfo {pages}
  {68--81}\BibitemShut {NoStop}%
\bibitem [{\citenamefont {Fran{\c{c}}oisse}\ \emph {et~al.}(2017)\citenamefont
  {Fran{\c{c}}oisse}, \citenamefont {Kivim{\"a}ki}, \citenamefont {Mantrach},
  \citenamefont {Rossi},\ and\ \citenamefont {Saerens}}]{franccoisse2017bag}%
  \BibitemOpen
  \bibfield  {author} {\bibinfo {author} {\bibfnamefont {K.}~\bibnamefont
  {Fran{\c{c}}oisse}}, \bibinfo {author} {\bibfnamefont {I.}~\bibnamefont
  {Kivim{\"a}ki}}, \bibinfo {author} {\bibfnamefont {A.}~\bibnamefont
  {Mantrach}}, \bibinfo {author} {\bibfnamefont {F.}~\bibnamefont {Rossi}}, \
  and\ \bibinfo {author} {\bibfnamefont {M.}~\bibnamefont {Saerens}},\
  }\bibfield  {title} {\enquote {\bibinfo {title} {A bag-of-paths framework for
  network data analysis},}\ }\href@noop {} {\bibfield  {journal} {\bibinfo
  {journal} {Neural Networks}\ }\textbf {\bibinfo {volume} {90}},\ \bibinfo
  {pages} {90--111} (\bibinfo {year} {2017})}\BibitemShut {NoStop}%
\bibitem [{\citenamefont {Newman}(2001)}]{newman2001scientific}%
  \BibitemOpen
  \bibfield  {author} {\bibinfo {author} {\bibfnamefont {M.~E.~J.}\
  \bibnamefont {Newman}},\ }\bibfield  {title} {\enquote {\bibinfo {title}
  {Scientific collaboration networks. {II}. {S}hortest paths, weighted
  networks, and centrality},}\ }\href@noop {} {\bibfield  {journal} {\bibinfo
  {journal} {Phys. Rev. E}\ }\textbf {\bibinfo {volume} {64}},\ \bibinfo
  {pages} {016132} (\bibinfo {year} {2001})}\BibitemShut {NoStop}%
\bibitem [{\citenamefont {Bavelas}(1950)}]{bavelas1950communication}%
  \BibitemOpen
  \bibfield  {author} {\bibinfo {author} {\bibfnamefont {A.}\ \bibnamefont
  {Bavelas}},\ }\bibfield  {title} {\enquote {\bibinfo {title} {Communication
  patterns in task-oriented groups},}\ }\href@noop {} {\bibfield  {journal}
  {\bibinfo  {journal} {J. Acoust. Soc. Am.}\
  }\textbf {\bibinfo {volume} {22}},\ \bibinfo {pages} {725--730} (\bibinfo
  {year} {1950})}\BibitemShut {NoStop}%
\bibitem [{\citenamefont {Stephenson}\ and\ \citenamefont
  {Zelen}(1989)}]{stephenson1989rethinking}%
  \BibitemOpen
  \bibfield  {author} {\bibinfo {author} {\bibfnamefont {K.}~\bibnamefont
  {Stephenson}}\ and\ \bibinfo {author} {\bibfnamefont {M.}~\bibnamefont
  {Zelen}},\ }\bibfield  {title} {\enquote {\bibinfo {title} {Rethinking
  centrality: Methods and examples},}\ }\href@noop {} {\bibfield  {journal}
  {\bibinfo  {journal} {Soc. Networks}\ }\textbf {\bibinfo {volume} {11}},\
  \bibinfo {pages} {1--37} (\bibinfo {year} {1989})}\BibitemShut {NoStop}%
\bibitem{RExplained}
 The proportionality factor is simply $R$, so we have chosen units where $R=1$. However, in this paper we keep the $R$ dependence explicit to connect with the engineering literature.
\bibitem [{\citenamefont {Novotny}(2001)}]{novotny2001annual}%
  \BibitemOpen
  \bibfield  {author} {\bibinfo {author} {\bibfnamefont {M.~A.}\ \bibnamefont
  {Novotny}},\ }\bibfield  {title} {\enquote {\bibinfo {title} {A tutorial on advanced dynamic Monte Carlo methods for systems with discrete state spaces},}\ }\href@noop {} in\ \href@noop {} \emph {\bibinfo {booktitle} {Annual Reviews of Computational Physics IX}} pp.\ \bibinfo {pages} {153--210}, edited by D. Stauffer (\bibinfo {year} {World Scientific, Singapore, 2001})\BibitemShut
  {NoStop}%
\bibitem [{\citenamefont {Iosifescu}(1980)}]{iosifescu2014finite}%
  \BibitemOpen
  \bibfield  {author} {\bibinfo {author} {\bibfnamefont {M.}~\bibnamefont
  {Iosifescu}},\ }\href@noop {} {\emph {\bibinfo {title} {Finite Markov
  processes and their applications}}}\ (\bibinfo  {publisher} {Wiley},\
  \bibinfo {address} {New York},\ \bibinfo {year} {1980})\BibitemShut {NoStop}%
\bibitem [{\citenamefont {Murty}(1983)}]{murty1983linear}%
  \BibitemOpen
  \bibfield  {author} {\bibinfo {author} {\bibfnamefont {K.~G.}\ \bibnamefont
  {Murty}},\ }\href@noop {} {\emph {\bibinfo {title} {Linear Programming}}},\
  Vol.~\bibinfo {volume} {60}\ (\bibinfo  {publisher} {Wiley},\ \bibinfo
  {address} {New York},\ \bibinfo {year} {1983})\BibitemShut {NoStop}%
\bibitem [{\citenamefont {Boyd}\ and\ \citenamefont
  {Vandenberghe}(2004)}]{boyd2004convex}%
  \BibitemOpen
  \bibfield  {author} {\bibinfo {author} {\bibfnamefont {S.}\ \bibnamefont
  {Boyd}}\ and\ \bibinfo {author} {\bibfnamefont {L.}\ \bibnamefont
  {Vandenberghe}},\ }\href@noop {} {\emph {\bibinfo {title} {Convex
  optimization}}}\ (\bibinfo  {publisher} {Cambridge University Press},\
  \bibinfo {year} {2004})\BibitemShut {NoStop}%
\bibitem{noteSolver}
  Our results were obtained using the linear programming routines built into {\it Mathematica}, 
which employ either simplex or interior-point methods based on internal criteria.
\bibitem [{\citenamefont {Brandes}(2001)}]{brandes2001faster}%
  \BibitemOpen
  \bibfield  {author} {\bibinfo {author} {\bibfnamefont {U.}~\bibnamefont
  {Brandes}},\ }\bibfield  {title} {\enquote {\bibinfo {title} {A faster
  algorithm for betweenness centrality},}\ }\href@noop {} {\bibfield  {journal}
  {\bibinfo  {journal} {J. Math. Sociol.}\ }\textbf {\bibinfo {volume} {25}},\
  \bibinfo {pages} {163--177} (\bibinfo {year} {2001})}\BibitemShut {NoStop}%
\bibitem [{\citenamefont {Abou~Hamad}\ \emph {et~al.}(2011)\citenamefont
  {Abou~Hamad}, \citenamefont {Rikvold},\ and\ \citenamefont
  {Poroseva}}]{HAMA11}%
  \BibitemOpen
  \bibfield  {author} {\bibinfo {author} {\bibfnamefont {I.}~\bibnamefont
  {Abou~Hamad}}, \bibinfo {author} {\bibfnamefont {P.~A.}\ \bibnamefont
  {Rikvold}}, \ and\ \bibinfo {author} {\bibfnamefont {S.~V.}\ \bibnamefont
  {Poroseva}},\ }\bibfield  {title} {\enquote {\bibinfo {title} {Floridian
  high-voltage power-grid network partitioning and cluster optimization using
  simulated annealing},}\ }\href@noop {} {\bibfield  {journal} {\bibinfo
  {journal} {Phys. Proc.}\ }\textbf {\bibinfo {volume} {15}},\ \bibinfo {pages}
  {2--6} (\bibinfo {year} {2011})}\BibitemShut {NoStop}%
\bibitem [{\citenamefont {Abou~Hamad}\ \emph {et~al.}(2010)\citenamefont
  {Abou~Hamad}, \citenamefont {Israels}, \citenamefont {Rikvold},\ and\
  \citenamefont {Poroseva}}]{HAMA10}%
  \BibitemOpen
  \bibfield  {author} {\bibinfo {author} {\bibfnamefont {I.}~\bibnamefont
  {Abou~Hamad}}, \bibinfo {author} {\bibfnamefont {B.}~\bibnamefont {Israels}},
  \bibinfo {author} {\bibfnamefont {P.~A.}\ \bibnamefont {Rikvold}}, \ and\
  \bibinfo {author} {\bibfnamefont {S.~V.}\ \bibnamefont {Poroseva}},\
  }\bibfield  {title} {\enquote {\bibinfo {title} {Spectral matrix methods for
  partitioning power grids: Applications to the {I}talian and {F}loridian
  high-voltage networks},}\ }\href@noop {} {\bibfield  {journal} {\bibinfo
  {journal} {Phys. Proc.}\ }\textbf {\bibinfo {volume} {4}},\ \bibinfo {pages}
  {125--129} (\bibinfo {year} {2010})}\BibitemShut {NoStop}%
\bibitem [{\citenamefont {Newman}(2004)}]{newman2004detecting}%
  \BibitemOpen
  \bibfield  {author} {\bibinfo {author} {\bibfnamefont {M.~E.~J.}\ \bibnamefont
  {Newman}},\ }\bibfield  {title} {\enquote {\bibinfo {title} {Detecting
  community structure in networks},}\ }\href@noop {} {\bibfield  {journal}
  {\bibinfo  {journal} {The European Physical Journal B}\ }\textbf {\bibinfo
  {volume} {38}},\ \bibinfo {pages} {321--330} (\bibinfo {year}
  {2004})}\BibitemShut {NoStop}%
\bibitem{noteOnDownstream}
   Here we are concerned with current flowing from one partition to the other. So 
   Node 5 in Fig.~\ref{fig:nonmon} is effectively directly connected to a partition boundary because it necessarily inherits all current originating from the other partition from node 4, which is directly connected to the partition boundary. This is the only such special case in the present study.

\bibitem{noteOnMaximum}
  Though the discussion centered around Fig.~\ref{fig:artificial} only seems to explain maxima in the centrality plots, it can also be explain minima when one remembers that betweenness calculations are based on conserved walker flows. Because of the conservation of the walker current used to calculate the conditional current betweenness, a maximum for one node may lead to a minimum for another.

\bibitem [{\citenamefont {Logan}(2013)}]{logan2013applied}%
  \BibitemOpen
  \bibfield  {author} {\bibinfo {author} {\bibfnamefont {J.~D.}\ \bibnamefont
  {Logan}},\ }\href@noop {} {\emph {\bibinfo {title} {Applied Mathematics}}}\
  (\bibinfo  {publisher} {John Wiley \& Sons},\ \bibinfo {address} {New York},\
  \bibinfo {year} {2013})\BibitemShut {NoStop}%
  \bibitem [{\citenamefont {Davydov}\ and\ \citenamefont
  {Zitikis}(2017)}]{davydov_zitikis_2017}%
  \BibitemOpen
  \bibfield  {author} {\bibinfo {author} {\bibfnamefont {Y.}\ \bibnamefont
  {Davydov}}\ and\ \bibinfo {author} {\bibfnamefont {R.}\ \bibnamefont
  {Zitikis}},\ }\bibfield  {title} {\enquote {\bibinfo {title} {Quantifying
  non-monotonicity of functions and the lack of positivity in signed
  measures},}\ }\href {\doibase 10.15559/17-VMSTA84} {\bibfield  {journal}
  {\bibinfo  {journal} {Modern Stochastics: Theory and Applications}\ }\textbf
  {\bibinfo {volume} {4}},\ \bibinfo {pages} {219--231} (\bibinfo {year}
  {2017})}\BibitemShut {NoStop}%
\end{thebibliography}
\end{document}